\documentclass[iop,twocolappendix,numberedappendix,appendixfloats,revtex4]{emulateapj}

\usepackage{hyperref, graphicx, natbib, multirow, captcont, amsmath}

\newcommand{\um}{$\mu$m}
\newcommand{\cleansample}{5,786}
\newcommand{\cleansampleE}{two}
\newcommand{\fullsample}{20,502}
\newcommand{\fullsampleE}{584}
\DeclareRobustCommand{\rchi}{{\mathpalette\irchi\relax}}
\newcommand{\irchi}[2]{\raisebox{\depth}{$#1\chi$}} % inner command, used by \rchi

\slugcomment{}
\shorttitle{Planetary Collisions: Low-Mass Stars with Extreme Mid-Infrared Excesses}
\shortauthors{Theissen \& West}

\begin{document}

\title{Collisions of Terrestrial Worlds: The Occurrence of Extreme Mid-Infrared Excesses around Low-Mass Field Stars}

\author{Christopher A. Theissen\altaffilmark{1} and Andrew A. West}
\affil{Department of Astronomy, Boston University, 725 Commonwealth Avenue, Boston, MA 02215, USA}
\altaffiltext{1}{Visiting Graduate Student, UC San Diego}

\email{ctheisse@bu.edu}

\begin{abstract}

We present the results of an investigation into the occurrence and properties (stellar age and mass trends) of low-mass field stars exhibiting extreme mid-infrared (MIR) excesses ($L_\mathrm{IR} / L_\ast \gtrsim 0.01$). 
Stars for the analysis were initially selected from the Motion Verified Red Stars (MoVeRS) catalog of photometric stars with SDSS, 2MASS, and \emph{WISE} photometry and significant proper motions. 
We identify \fullsampleE\ stars exhibiting extreme MIR excesses, selected based on an empirical relationship for main sequence $W1-W3$ colors.
For a small subset of the sample, we show, using spectroscopic tracers of stellar age (H$\alpha$ and Li \textsc{i}) and luminosity class, that the parent sample is likely comprised of field dwarfs ($\gtrsim 1$ Gyr).
We also develop the Low-mass Kinematics (\emph{LoKi}) galactic model to estimate the completeness of the extreme MIR excess sample.
Using Galactic height as a proxy for stellar age, the completeness corrected analysis indicates a distinct age dependence for field stars exhibiting extreme MIR excesses.
We also find a trend with stellar mass (using $r-z$ color as a proxy).
Our findings are consistent with the detected extreme MIR excesses originating from dust created in a short-lived collisional cascade ($\lesssim 100,000$ years) during a giant impact between two large planetismals or terrestrial planets.
These stars with extreme MIR excesses also provide support for planetary collisions being the dominant mechanism in creating the observed \emph{Kepler} dichotomy (the need for more than a single mode, typically two, to explain the variety of planetary system architectures \emph{Kepler} has observed), rather than different formation mechanisms.

\end{abstract}


\section{Introduction}\label{intro}

	The ability to study circumstellar environments has greatly improved around stars has greatly improved over the past decade, due in part to new technologies that provide higher sensitivity and greater resolution at infrared (IR) and radio wavelengths. Examples of facilities that have contributed to this advance include, but are not limited to the \emph{Spitzer Space Telescope} \citep{werner:2004:1}, the Atacama Large Millimeter Array (ALMA), the \emph{Wide-field Infrared Survey Explorer} \citep[\emph{WISE};][]{wright:2010:1868}, and the \emph{Herschel Space Observatory} \citep{pilbratt:2010:l1}. In recent years, observations at these facilities have led to the discovery of stars exhibiting large amounts of excess mid-IR (MIR) flux ($L_\mathrm{IR} / L_\ast \gtrsim 10^{-2}$), termed ``extreme debris disks" \citep{meng:2012:l17,meng:2015:77} or ``extreme IR excesses" \citep{balog:2009:1989}. Typically found around stars with ages between 10--130 Myr \citep{meng:2012:l17,meng:2015:77}, these systems are believed to have hosted collisions between terrestrial planets or large planetismals \citep{meng:2014:1032}.
	
	The majority of stars exhibiting extreme MIR excesses have been found with ages coinciding with the final stages of terrestrial planet formation \citep[10--200 Myr;][]{meng:2015:77}. However, until recently, there was one known system that did not fall into the same category, BD +20 307, a $\sim$1 Gyr old spectroscopic binary composed of two late F-type stars \citep{weinberger:2008:l41} exhibiting a significant MIR excess \citep[$L_\mathrm{IR} / L_\ast \approx 0.033$;][]{song:2005:363, weinberger:2011:72}. An in-depth study of the disk mineralogy for BD +20 307 found that the best explanation for the observed large MIR excess and low level of crystallinity was a giant impact between two large terrestrial bodies, similar to the Moon forming event in our solar system \citep{weinberger:2011:72}. However, such collisions are expected to occur much earlier during planetary system formation (as stated above), and the lifetime for the observable collisional cascade is expected to be short \citep[$\sim$100,000 years;][]{melis:2010:l57}. It is also possible that the close binary nature of BD +20 307 may have played a role in this late-time collision. The potential for impacts between terrestrial bodies on timescales $\gtrsim$ 1 Gyr is particularly important for low-mass stars ($M_\ast \lesssim 0.8 M_\odot$), which are known to host multiple terrestrial planets \citep[$\sim$3 planets per star on average;][]{ballard:2016:66}, all orbiting closely to their host stars due to the proximity of the snow-line \citep[$\lesssim 0.3$ AU;][]{ogihara:2009:824}. 
	
	Low-mass stars make ideal laboratories for studying the occurrence of extreme MIR excesses, and investigating the hypothesis of planetary collisions as their origin. In addition to the observational evidence suggesting an abundance of close-in terrestrial planets surrounding them, low-mass stars are ubiquitous, making up more than 70\% of the stellar population \citep[][]{bochanski:2010:2679}. Until recently, all of the aforementioned observed extreme MIR excesses have been found around solar-type (FGK-spectral type) stars. However, no explanation has been put forward to explain the dearth of low-mass stars exhibiting similar extreme MIR excesses. In particular, the relative frequency of low-mass stars to solar-type stars should make it more likely to find extreme MIR excesses around low-mass stars, barring any observational limitations. 
	
	Simulations of planet formation around Sun-like stars indicate that impacts are quite common during the period of terrestrial planet formation \citep{quintana:2016:126}. \citet{quintana:2016:126} noted that highly energetic giant impacts (similar to the Moon-forming event) occur far more rarely than smaller collisions, but are a necessity to build a system analogous to our present day solar system. One interesting finding by \citet{quintana:2016:404} is that by removing giant planets from their dynamical simulations, giant impacts can occur much later in the system's evolution (100 Myr to a few Gyrs versus 10--100 Myr). This may have strong implications for planetary systems around low-mass stars, which do not typically form giant planets \citep[e.g.,][]{johnson:2010:905, bonfils:2013:a109}. Efforts are currently underway to extend these models to low-mass stars, however, initial circumstellar disk conditions are not as well constrained observationally at the bottom of the main sequence.
	
	A number of studies have undertaken searches for low-mass stars exhibiting signs of disks and/or M(IR) excesses \citep[e.g.,][]{plavchan:2005:1161, plavchan:2009:1068, avenhaus:2012:a105, wu:2013:29}. \citet{plavchan:2009:1068} provided a theoretical framework for why \emph{primordial} disks around low-mass stars could persist on longer timescales than those around higher-mass stars, in spite of most observational evidence suggesting primordial disks are dispersed around low mass stars in less than 100 Myr. For a low-mass star (M0), the timescales for dust removal by Poynting-Robertson drag and grain-grain collisions are $\sim$10 times longer and 40\% longer than for a higher mass star (G0), respectively \citep{plavchan:2009:1068}. Primordial disks around low-mass stars have been observed to be longer lived than those around higher-mass stars \citep[e.g.,][]{ribas:2015:a52}, potentially due to the longer timescales for Poynting-Robertson drag to remove grains from these systems relative to higher-mass systems \citep[$\sim$10 times longer for an M0 star versus a G2 star;][]{plavchan:2009:1068}. However, there is currently little to no observational data to support primordial disks around low-mass stars surviving past 10s of Myr, hinting that the evolution of primordial disks around low-mass stars follow a similar evolution to primordial disks around Solar-mass stars.
	
	A search for low-mass stars exhibiting extreme MIR excesses was conducted by \citet[][hereafter TW14]{theissen:2014:146}. Their initial sample was pulled from the Sloan Digital Sky Survey \citep[SDSS;][]{york:2000:1579} Data Release 7 \citep[DR7;][]{abazajian:2009:543} spectroscopic sample of M dwarfs \citep[70,841 stars;][]{west:2011:97}. TW14 discovered 168 low-mass field stars exhibiting large amounts of excess MIR flux, and estimated a collision rate of $\sim$130 collisions per star over its main sequence lifetime. This rate is significantly higher than the rate estimated by \citet{weinberger:2011:72} for A--G type stars (0.2 impacts per star). The TW14 result suggests that collisions may be more common around low-mass stars, possibly due to a longer timescale over which collisions can act, coupled with the extremely long main sequence lifetimes of low-mass stars \citep[$\gg$10 Gyr;][]{laughlin:1997:420}, and/or the higher density of planets with small semi-major axes. One limitation of the TW14 study was the use of the SDSS DR7 spectroscopic sample, which was not produced in a systematic way, making estimates of completeness difficult. To further investigate the mechanism responsible for creating these observed extreme MIR excesses, a larger sample must be gathered, and methods to estimate the completeness of the sample must be developed.
	
	Although many large spectroscopic samples exist for low-mass stars, such as the SDSS spectroscopic M dwarf sample \citep{west:2011:97}, the Large Sky Area Multi-Object Fibre Spectroscopic Telescope \citep[LAMOST;][]{cui:2012:1197} Data Release 1 \citep[DR1;][]{luo:2015:1095} M dwarf catalog \citep[93,619 stars;][]{guo:2015:1182}, and the Palomar/Michigan State University (PMSU) Nearby Star Spectroscopic Survey \citep[$\sim$2400 stars;][]{hawley:1996:2799, reid:1995:1838}, these samples are dwarfed by the millions of photometric data products for low-mass stars that are currently available. Unfortunately, many photometric objects share similar colors and point-source-like morphologies with low-mass stars (e.g., giants, quasars, distant luminous galaxies). One way of distinguishing dwarf stars from other similarly colored objects is through the use of proper motions. Distant objects will show little to no tangential motion on the sky, while nearby stars will show significant, measurable motion in reference to background stars. 
	
	The largest catalog of low-mass stars with proper motions to date is the Motion Verified Red Stars catalog \citep[MoVeRS, containing $\sim$8.7 million stars;][]{theissen:2016:41}. MoVeRS was created using data from SDSS, the Two Micron All-Sky Survey \citep[2MASS;][]{skrutskie:2006:1163}, and the \emph{Wide-field Infrared Survey Explorer} \citep[\emph{WISE};][]{wright:2010:1868}. The Late-Type Extension to MoVeRS was recently released with additional very-low-mass objects later than M5 \citep[LaTE-MoVeRS;][]{ theissen:2017:92}. The MoVeRS catalog enables the search for extreme MIR excesses in a larger capacity than was previously available. 
	
	This paper performs a thorough investigation of the mass-, spatial-, and age-dependence of extreme MIR excesses around low-mass field stars. In Section~\ref{data}, we describe the sample from which the stars are drawn. Section~\ref{methods} briefly discusses the methods used in estimating stellar parameters (Section~\ref{params}) and distances (Section~\ref{photoparallax}), describes how we curate the sample of stars (Section~\ref{sample}), account for interstellar extinction (Section~\ref{extinction}), distinguish extreme MIR excess candidates (Section~\ref{irprops}), investigate the fidelity of the \emph{WISE} measurements (Section~\ref{spitzer}), obtain spectroscopic observations for youth (Section~\ref{spectroscopic}), and the inherent biases in the sample (Section~\ref{biases}). Section~\ref{model} provides details about the Galactic model, which we use to estimate the completeness of the sample, and discuss the completeness corrected results. In Section~\ref{nonexcess2} we investigate the non-significant MIR excess sample for trends with stellar age. In Section~\ref{discussion}, we summarize the conclusions and provide a discussion of our results. Details regarding the methods for estimating stellar parameters, including the Markov Chain Monte Carlo (MCMC) method for estimating $T_\mathrm{eff}$ and log $g$, and the methods for estimating stellar size are found in Appendix~\ref{stellarparams}. Details for building and using the Low-mass Kinematics (\emph{LoKi}) galactic model to estimate the level of completeness are discussed in Appendix~\ref{loki}.

\section{Data: The MoVeRS Catalog}\label{data}

	The occurrence rate for low-mass stars exhibiting extreme IR excesses was shown to be extremely low by TW14 ($\sim$0.4\%). To build a larger sample of candidate stars with extreme IR excesses, a massive input catalog of \emph{bona fide} low-mass stars is required. Although photometric catalogs exist for large numbers of low-mass stars \citep[e.g.,][]{bochanski:2010:2679}, proper motions are a way to definitively separate dwarf stars from giants and extragalactic objects of similar photometric colors. \citet{theissen:2016:41} created the MoVeRS catalog, a photometric catalog of low-mass stars extracted from the SDSS, 2MASS, and \emph{WISE} datasets, and selected based on their significant proper motions. The MoVeRS catalog contains 8,735,004 stars, 8,534,902 of which have cross-matches in the \emph{WISE} AllWISE catalog. Along with proper motions computed in \citet{theissen:2016:41}, the current version of the MoVeRS catalog contains photometry from SDSS, 2MASS, and \emph{WISE}, where available, for each star.
	
	To build the MoVeRS catalog, \citet{theissen:2016:41} initially selected stars based on their SDSS, 2MASS, and \emph{WISE} colors, tracing the stellar locus for stars with $16 < r < 22$ and $r-z \geqslant 0.5$. Stars were then selected based on a number of quality flags and proximity to neighboring objects. Proper motions for the remaining objects were computed using astrometric information from SDSS, 2MASS, and \emph{WISE}, which spans a $\sim$12-year time baseline. The precision of the catalog is estimated to be $\sim$10 mas yr$^{-1}$. Only stars with significant proper motions ($\mu_\mathrm{tot} \geqslant 2\sigma_{\mu_\mathrm{tot}}$) were included in the final catalog, increasing the likelihood that the catalog contains nearby stars as opposed to other astrophysical objects. 

	To illustrate the effectiveness of removing giants using proper motions, we consider a giant star at the edge of the photometric selection criteria used for MoVeRS ($r = 16$). A giant star would be approximately 1000 times more luminous than its dwarf counterpart, putting a giant approximately 30 times farther than a dwarf for a given magnitude. The median photometric distance for stars in the MoVeRS sample is 200 pc, putting a giant star at 6 kpc. The minimum required proper motion within MoVeRS is approximately 20 mas yr$^{-1}$. For a giant at a distance of 6 kpc, this translates to a tangential velocity of 570 km s$^{-1}$. \emph{Gaia} Data Release 1 \citep[DR1;][]{gaia-collaboration:2016:a2} Figure 6 shows that red giants with such high tangential velocities (hypervelocity stars) are a negligible fraction of the entire population, and are likely to be unbound from the Galaxy.
	
	If we assume a similar proper motion distribution between giant stars and QSOs (both essentially non-moving on the sky) for motions measured with \emph{WISE}+SDSS+2MASS, we can use Figure 3 from \citealt{theissen:2016:41} to make a statical estimate of the contamination rate of giants. The average time-baseline of 12 years translates to a combined proper motion uncertainty of 10 mas yr$^{-1}$ for a non-moving population. This gives a point-source with a proper motion of 20 mas yr$^{-1}$ a 4.5\% chance of being a giant. Combined with the relative fraction of all point sources that are giants (versus dwarfs) at the blue limit of the MoVeRS samples \citep[$\sim$2\%;][]{covey:2008:1778}, gives the likelihood of having an interloping giant with a proper motion of 20 mas yr$^{-1}$ less than 0.1\%. The vast majority of MoVeRS stars have proper motions that exceed 20 mas yr$^{-1}$, making the likelihood for contamination by giants significantly smaller than this. More information about the construction and properties of the MoVeRS catalog can be found in \citet{theissen:2016:41}. The Late-Type Extension to MoVeRS was recently released and contains stars with spectral types later than M5 \citep[LaTE-MoVeRS;][]{ theissen:2017:92}.
	
	Photometry from \emph{WISE}, taken in four MIR bands ($W1$, $W2$, $W3$, and $W4$ with effective wavelengths at 3.4, 4.6, 12, and 22 \um, respectively), is particularly crucial for finding extreme MIR excesses around K and M dwarfs due to the fact that dust orbiting within the snow-line, where terrestrial planets form, is warm ($\sim$300 K), with its thermal emission peaking in the MIR. The $W3$ band also samples the 10 \um\ silicate feature prominent in the types of disks expected to produce these extreme MIR excesses. The sensitivity of \emph{WISE}, particularly the $W3$ band \citep[$\sim$730 $\mu$Jy at 12 \um;][]{wright:2010:1868}, allows these extreme MIR excesses to be detected at much higher precision than previous all-sky MIR observatories (e.g., the \emph{Infrared Astronomical Satellite}, \citealt{neugebauer:1984:l1}, and \emph{Akari}, \citealt{murakami:2007:369}).

\section{Methods}\label{methods}

\subsection{Estimating Stellar Parameters}\label{params}

	An important step in identifying and quantifying the significance of a MIR excess is measuring the deviation between the expected photospheric MIR values and the measured photometric values, which requires an estimate of the fundamental stellar parameters (e.g., $T_\mathrm{eff}$). Additionally, estimates for stellar temperature ($T_\mathrm{eff}$) and size ($R_\ast$) put constraints on dust temperature and orbital distance \citep{jura:1998:897, chen:2005:493}. Photospheric models for low-mass stars are limited in replicating the myriad of complex molecules found in low-mass stellar atmospheres due to the low temperature environments \citep{schmidt:2016:2611}. Furthermore, the onset of potential clouds forming in the coolest stars provides further complications for modeling \citep{allard:2013:128}. However, these models are good at producing the overall expected stellar energy distributions (SEDs), and are effective for constraining many of the fundamental stellar parameters. TW14 estimated stellar parameters using a grid of BT-Settl models based on the PHOENIX code \citep{allard:2012:2765, allard:2012:3}, which have taken into account many molecular opacities and cloud models. 
	
	TW14 compared synthetic photometry and spectra from models to data from SDSS, 2MASS and \emph{WISE} to estimate goodness-of-fit. Due to the lack of spectra for the MoVeRS sample, we only considered synthetic photometry in deriving the goodness-of-fit. This process involved fitting synthetic photometry, derived using relative spectral response curves for SDSS \citep{doi:2010:1628}, 2MASS \citep{cohen:2003:1090}, and \emph{WISE} \citep{wright:2010:1868}, to actual measurements from each photometric survey.
	
	TW14 derived stellar parameters by computing reduced-$\rchi^2$ values over the entire range of models, a method which is intractable computationally for the large number of stars in the MoVeRS catalog. To reduce the parameter space, we employed a Markov chain Monte Carlo (MCMC) technique to sample and build posterior probability distributions for each of the stars, used to estimate best-fit parameters and uncertainties (using the 50$^\mathrm{th}$ percentile value, and the 16$^\mathrm{th}$ and 84$^\mathrm{th}$ percentile values, respectively). Details of the MCMC method are described in Appendix~\ref{mcmc}. Using this process, we estimated $T_\mathrm{eff}$ and log $g$ values for \emph{all} 8.7 million sources in the MoVeRS catalog. We used the $T_\mathrm{eff}$ values to derive a color-$T_\mathrm{eff}$ relationship, also found in Appendix~\ref{mcmc}. With distance estimates, the scaling values derived from this fitting procedure were used to estimate stellar size ($R_\ast$) and a radius-color relation (also found in Appendix~\ref{mcmc}). The new MoVeRS catalog (MoVeRS 2.0), with the estimated stellar parameters, is available through SDSS CasJobs\footnote{\url{http://skyserver.sdss.org/casjobs/}} and VizieR\footnote{\url{http://vizier.u-strasbg.fr/}}.

\subsection{Estimating Distances: Photometric Parallax}\label{photoparallax}

	Distances to stars are important for estimating luminosities, radii, and many other stellar and kinematic parameters (see TW14 for details). For stars with resolved disks, distances can be used to convert angular sizes into absolute sizes. For unresolved disks, stellar sizes can give approximate orbital distances for circumstellar dust, and approximate dust masses. Few parallax measurements have been made for M dwarfs, relative to higher mass stars, due to their intrinsic faintness. The two largest astrometry databases, the General Catalog of Trigonometric Stellar Parallaxes, Fourth Edition \citep[the Yale Parallax Catalog;][]{van-altena:1995:} and the \emph{Hipparcos} catalog \citep{perryman:1997:l49, van-leeuwen:2007:653} are both severely incomplete for M dwarfs and brown dwarfs \citep{dittmann:2014:156}. Although large parallax databases are incomplete for low-mass stars, two nearby stellar samples now have many parallax measurements, the REsearch Consortium On Nearby Stars \citep[RECONS;][]{riedel:2014:85, winters:2015:5} and MEarth \citep{nutzman:2008:317}. The RECONS sample includes parallaxes for over 1400 M dwarfs within 25 pc \citep{winters:2015:5}, and the MEarth sample includes over 1500 M dwarfs within 33 pc \citep{dittmann:2014:156}. There is very little overlap between the two samples since the RECONS survey began operating in the southern hemisphere, while MEarth started as a survey in the northern hemisphere, only recently adding telescopes to the southern hemisphere \citep{irwin:2015:767}. Additionally, a few studies have measured trigonometric parallaxes for sub-stellar objects \citep[e.g.,][]{faherty:2012:56, manjavacas:2013:a52, marocco:2013:161, marsh:2013:119, smart:2013:2054, zapatero-osorio:2014:a6, weinberger:2016:24}, but these studies are limited by small numbers.
	
	Unfortunately, none of these trigonometric parallax surveys have data in SDSS passbands, which makes deriving a photometric relationship impossible without adding in additional errors from color transformations. The most commonly used photometric parallax relationship for low-mass stars with SDSS colors comes from \citet[][hereafter B10]{bochanski:2010:2679}. These relationships are derived from 86 low-mass stars with trigonometric parallax measurements from various sources (B10). The average uncertainty in these relationships is $\sim$0.4 mags in absolute $r$-band magnitude ($M_r$), due in part to luminosity differences between stars of different metallicities \citep[see][]{savcheva:2014:145} and magnetic activity \citep[see][]{bochanski:2011:98}. This uncertainty in absolute magnitude corresponds to distance uncertainties of $\sim$20\%. Efforts are underway to obtain SDSS magnitudes for many of the low-mass stars with parallax measurements in the samples listed above (C. Theissen et al., 2017, in preparation), however, to date, such measurements do not exist. For this purpose, we chose to use the B10 $r-z$ relationship to estimate distances for the entire MoVeRS sample. Using these distances, we also estimated stellar radii for the MoVeRS sample (see Appendix~\ref{radii}). The new MoVeRS 2.0 catalog also includes our distances estimates.

\subsection{Sample Selection for Stars with MIR Excesses}\label{sample}

	To compile a clean set of stars for our analysis, we used a number of selection criteria, most of which have been adapted from TW14. We applied the following selection criteria to the MoVeRS sample:

\begin{enumerate}

\item We selected stars that did not have a \emph{WISE} extended source flag (\textsc{ext\_flg} = 0). This requirement ensured a point-source morphology through all \emph{WISE} bands. This cut left 8,483,499 stars.

\item We selected stars that did not have a contamination or confusion flag in either $W1$, $W2$, or $W3$ (\textsc{cc\_flg}$_{W1,W2,W3}$ = 0). This ensured clean photometry for those bands. This cut left 7,899,559 stars.

\item We selected stars with at least a signal-to-noise ratio (S/N) of 3 in $W1$, $W2$, and $W3$ (\textsc{WxSNR}$_{\mathrm{x}=W1,W2,W3} \geqslant 3$). This cut left 185,121 stars.

\item We kept only the highest fidelity stars, retaining relatively bright stars satisfying Equation (12) of \citet{theissen:2016:41}. This cut ensures stars have high-precision proper motion measurements and fall within the regime confirmed with independent checks to other proper motion catalogs. This cut left 145,526 stars.

\item Lastly, to minimize source confusion, and reduce contamination due to dust extinction, we removed stars close to the Galactic plane ($|b| < 20^\circ$) and in the Orion region ($-30^\circ < b < 0^\circ$ and $190^\circ < l < 215^\circ$). This cut left 126,976 stars.

\end{enumerate}

\subsubsection{WISE Sensitivity Limits}\label{wiselimits}

	To directly address one of the limitations of the TW14 study, we constructed a uniform sample of stars. We broadly categorized the stars into three groups: 1) stars which are close enough that \emph{WISE} can significantly detect their photospheres at $12$ \um; 2) stars that are far enough away that their photospheres are undetectable at 12 \um, but for which an extreme MIR excess (on the order of those found in TW14) is significantly detectable by \emph{WISE}; and 3) stars which are too far away to be detectable by \emph{WISE}, even if they have an extreme MIR excess. We were only interested in stars that have measurable detections in $W3$. Below, we discuss the methods for building the ``full" sample, stars that meet criterion (2), and the ``clean" sample, stars that meet criterion (1), which is a subset of the ``full" sample. We first discuss selecting stars exhibiting excess MIR flux (Section~\ref{irprops}), and will apply further criteria to select stars with extreme MIR excesses ($L_\mathrm{IR} / L_\ast \geqslant 0.01$) in Section~\ref{EME}.

\begin{figure}
\centering
 \includegraphics[width=\linewidth]{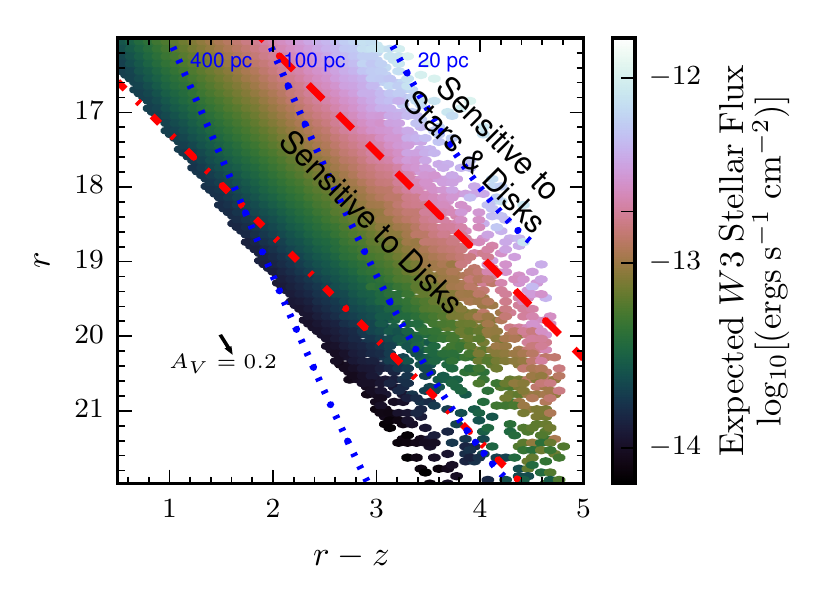}
\caption{Expected stellar photospheric $W3$ flux for a given $r$-band magnitude and $r-z$ color. Each bin is (0.1 mags)$^2$. The $W3$ flux limit of $1.89 \times 10^{-13}$ ergs s$^{-1}$ cm$^{-2}$ is shown as the red dashed line. We indicate where a 12 \um\ excess 10 times the expected photospheric value would reach the $W3$ flux limit (red dash-dotted line), a limit where we are sensitive to large IR excesses (large disks), but not necessarily to stellar photosphere flux levels. We also show approximate magnitude and color ranges expected at 20, 100, and 400 pc using the $r-z$ photometric parallax relationship from B10 (blue dotted lines). We also plot the extinction vector corresponding to the 90\% percentile extinction value in our sample (Section~\ref{extinction}), showing that extincted stars tend to move parallel to our selection criteria.
\label{fig:sensitivity}}
\end{figure}

	The $W3$ 5$\sigma$ point-source sensitivity limit is estimated to be 730 $\mu$Jy (also the approximate 95\% completeness limit; hereafter referred to as the $W3$ flux limit), based on external checks with \emph{Spitzer} COSMOS data\footnote{\url{http://wise2.ipac.caltech.edu/docs/release/allwise/expsup/sec2\_3a.html}}, which translates to a flux density of $\sim$$1.89 \times 10^{-13}$ ergs s$^{-1}$ cm$^{-2}$. Using the sample of 126,976 stars, we computed the expected photospheric $W3$ flux for each star by scaling the best-fit stellar model to the measured $z$-band flux. This gave us a measure of the expected $W3$ flux from the stellar photosphere for each star. The map of expected $W3$ stellar flux for a given $r-z$ color and $r$-band magnitude is shown in Figure~\ref{fig:sensitivity}. Figure~\ref{fig:sensitivity} shows that a constant expected $W3$ flux is approximately linear in this color-magnitude space. 
	
	To quantify the relationship between $r$, $r-z$, and expected $W3$ flux, we started at $r=16$ and binned each 0.1 mag along the $r$-band axis, and binned each slice in 0.1 mag $r-z$ bins. We identified the $r$, $r-z$ value where the expected $W3$ flux dropped below $1.89 \times 10^{-13}$ ergs s$^{-1}$ cm$^{-2}$ (the $W3$ flux limit). We repeated this process between $16 \leqslant r \leqslant 22$, and then fit a line to the $r$, $r-z$ values. Our linear fit is shown as a red dashed line in Figure~\ref{fig:sensitivity}, and given by,
\begin{equation}\label{eqn:rlimit}
r = 13.40+1.38(r-z).
\end{equation}
\noindent Every star brighter than this limit should fall within the $W3$ flux limit, regardless of if the star has a 12 \um\ excess or not. This gives us a very uniform sample, free from a $W3$ sensitivity bias. Stars equal to or brighter than Equation~(\ref{eqn:rlimit}) will be referred to as the ``clean" sample, which consists of 6,129 stars. 

	Many of the stars in the TW14 sample had extremely large $W3$ excesses above the expected photospheric values, with the majority of observed 12 \um\ fluxes being 10 times greater than the expected photospheric values. Considering that we were looking for similarly large excesses, the volume of space over which we might get a true $W3$ detection can be increased. To illustrate this point, Figure~\ref{fig:sensitivity} shows the expected $r$, $r-z$ limit at which stars with 12 \um\ excesses 10 times their photospheric values would equal the $W3$ detection limit (dash-dotted line). However, to increase the detections (source counts) of stars with MIR excesses, we must also consider the larger sample of stars that reside outside the $W3$ bias-free limit, where a MIR excess could be detected (at larger distances, and hence larger volumes). This is illustrated in Figure~\ref{fig:sensitivity}, where we plot the estimated distance limits corresponding to different $r$, $r-z$ values.

	The \emph{WISE} sensitivity limits are highly dependent on the source position on the sky, due to different depths of coverage and zodiacal foreground emission. Therefore, many of the stars fainter than the imposed limit can yield true detections, but stricter criteria must be implemented in their selection. Sensitivity maps for the \emph{WISE} bands have been created using a profile-fit photometry noise model\footnote{\url{http://wise2.ipac.caltech.edu/docs/release/allwise/expsup/sec2\_3a.html}}. These sensitivity maps have been checked using 2MASS stars with spectral types earlier than F7 to estimate the sensitivity of the $W3$ band at different positions over the entire sky. The external comparison against 2MASS has shown that the $W3$ sensitivity map may slightly underestimate the sensitivity of the AllWISE catalog\footnote{\url{http://wise2.ipac.caltech.edu/docs/release/allwise/expsup/sec2\_3a.html}}, but provides a consistent model against which we can examine the measured $W3$ fluxes for significance as a function of stellar position on the sky.
	
	To select the highest-fidelity stars outside the limits of the clean sample, we required that each source have a $W3 \geqslant$ the $W3$ flux limit for its position on the sky according to the noise model sensitivity map. This sample, termed the ``full" sample, consists of the clean sample and an additional 19,354 stars, for a total count of 25,483 stars.

\subsubsection{Visual Inspection}\label{visual}

	To retain the highest quality detections, we performed visual inspection for each of the stars. The $W3$ band is especially susceptible to background and nearby contaminants due to its large point-spread-function (PSF; $\sim$6.5\arcsec). Visual inspection removed stars superimposed on top of galaxies or blended with other nearby stars, which could cause the elevated MIR fluxes. Visual inspection also removed stars close to nearby bright objects that could produce additional MIR flux, or stars in areas of high IR cirrus. During visual inspection, we viewed SDSS and \emph{WISE} archival images to ensure that the candidate objects were real MIR detections, a process similar to the procedure in TW14. Stars were assigned a \textsc{quality} flag, with \textsc{quality} $= 1$ indicating a star free from any contaminants, and of the highest visual quality, and \textsc{quality} $= 2$ indicating that the 12 \um\ source is good but may be affected by nearby or background contamination, slightly offset {between other \emph{WISE} bands}, or low contrast in $W3$. After visual inspection, we were left with 20,502 stars in the full sample, and 5,786 stars in the clean sample. The breakdown of the samples and quality flags is shown in Table~\ref{tbl:quality}. This provides a clean sample from which to select stars with excess MIR flux (Section~\ref{irprops}) and account for interstellar extinction (Section~\ref{extinction}).
	
\begin{deluxetable}{cc}
\tabletypesize{\footnotesize}
\tablecolumns{2}
\tablecaption{Visual Inspection Quality\label{tbl:quality}}
\tablehead{
\colhead{Quality} & \colhead{Number of} \\
\colhead{Flag} & \colhead{Stars} 
}
\startdata
\multicolumn{2}{c}{Full Sample}\\ 
\hline
2 	& 18281	\\
1 	& 2221	\\
\hline
\multicolumn{2}{c}{Clean Sample}\\ 
\hline
2	& 4849	\\
1	& 937	
\enddata
\end{deluxetable}

\subsubsection{Accounting for Interstellar Extinction}\label{extinction}

	Due to the distances to the stars in the sample ($\gtrsim$ 100 pc), interstellar extinction may affect the photometry. Since dust grains along a line-of-sight in the interstellar medium both extinct and redden an object's SED, interstellar extinction increases the likelihood of a false MIR excess detection. For wavelengths longer than $\sim$5 \um, extinction effects should be negligible, with the exception of the 10 \um\ silicate feature \citep{gao:2013:7}. Although we expect extinction to minimally affect the SED fits for the sources in our sample, due to the requirement that stars reside at relatively high Galactic latitudes ($|b| > 20^\circ$), extinction must still be evaluated, especially since the $W3$-band samples the 10 \um\ silicate feature.
	
	Directly measuring extinction for a star is most accurately done with an optical spectrum that samples the ``knee" of the extinction curve, and a comparison to an un-extincted template of the same spectral-type \citep{jones:2011:44}. However, because optical spectra are unavailable for the vast majority of the MoVeRS sample, we employed a more broad approach. SDSS provides estimates for the relative extinction, $A_\lambda/A_V$ (the ratio of extinction in a given bandpass to extinction in the $V$ band), for each star and each band in the photometric catalog. These extinction values were estimated along the line-of-sight using the \citet{schlegel:1998:525} dust maps, created using galactic extinction measurements from the \emph{Cosmic Microwave Background Explorer} \citep[\emph{COBE};][]{boggess:1992:420} and the \emph{Infrared Astronomical Satellite} \citep[\emph{IRAS};][]{neugebauer:1984:l1}. These maps estimate the \emph{total} extinction along a line-of-sight out of the Galaxy, and may therefore overestimate the actual extinction values for stars closer than 1--2 kpc. Extinction effects may also occur due to circumstellar material, expected of the MIR excess candidates. However, the probability that an optically thick disk is seen directly edge-on is small assuming inclinations are random \citep{beatty:2010:1433}, although edge-on has the highest probability ($\sim$3.5\% chance to view within $\pm 2^\circ$ of edge-on). Therefore, we may assume the disk to be optically thin at visible wavelengths (similar to \citealt{weinberger:2011:72}).
	
\begin{figure}
\centering
\includegraphics[width=\linewidth]{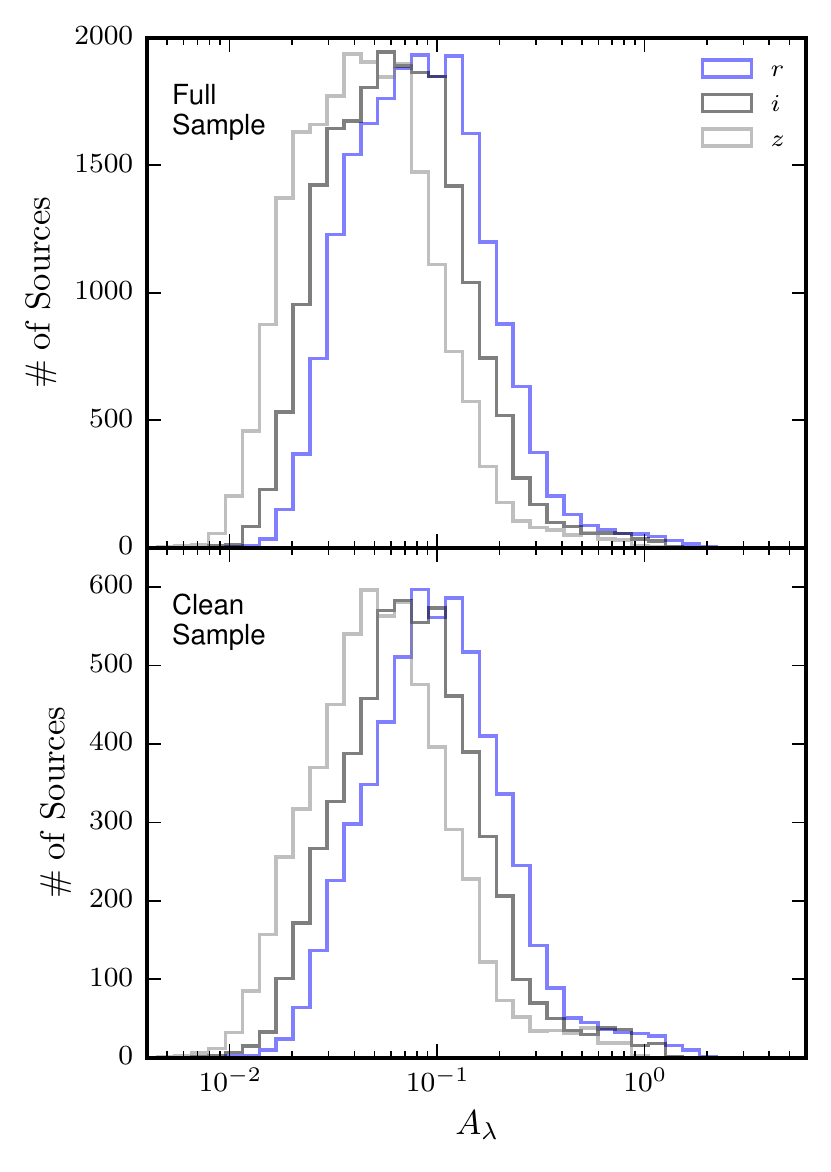}
\caption{Histograms for the $r$-, $i$-, and $z$-band extinctions from SDSS. The median extinctions for all bands in both samples are $<$ 0.1 mags, making extinction negligible for most of the stars in our samples.
\label{fig:extinction}}
\end{figure}

	To estimate the extinction in the sample, we used the SDSS extinction estimates for the $riz$-bands ($A_r$, $A_i$, and $A_z$). The extinction values for the clean and full samples are shown in Figure~\ref{fig:extinction}. The vast majority of the samples have small extinction values ($<$ 0.1 mags), with median values for $A_r$, $A_i$, and $A_z$ of 0.08, 0.06, and 0.04 for the full sample, and 0.09, 0.07, and 0.05 for the clean sample, respectively. Therefore, we do not expect extinction to affect the majority of our model fits from Appendix~\ref{mcmc}. Furthermore, extinction tends to move stars parallel to our initial selection criteria (see Figure~\ref{fig:sensitivity}), and should minimally bias our selected sample (Section~\ref{wiselimits}). For our full and clean samples, we corrected for extinction using the the SDSS estimates for $A_r$, $A_i$, and $A_z$, and the relative extinction values ($A_\lambda/A_V$) for SDSS bandpasses from \citet{schlegel:1998:525} Table 6 to compute $A_V$ values. We then applied corrections to the $rizJHK_s$ bandpasses using relative extinction measurements from the Asiago Database \citep{moro:2000:361, fiorucci:2003:781}, and an $R_V = 3.1$. Further details of this method can be found in \citet{theissen:2014:146}. 
	
	\citet{rieke:1985:618} found that the relative extinction at 10 \um\ due to the Galactic ISM extinction curve can be as large as the relative extinction in the $K$-band. \citet{davenport:2014:3430} used 1,052,793 main sequence stars from SDSS DR8 \citep{aihara:2011:29} with $|b| > 10^\circ$ to measure the dust extinction curve relative to the $r$-band for the first three \emph{WISE} bands. \citet{davenport:2014:3430} derived $A_\lambda/A_{K_s} = 0.60$, 0.33, and 0.87 for $W1$, $W2$, and $W3$, respectively. Another study by \citet{xue:2016:23} using GK-type giants from the SDSS Apache Point Observatory Galaxy Evolution Experiment \citep[APOGEE;][]{eisenstein:2011:72} spectroscopic survey found that the MIR relative extinction values were extremely sensitive to the NIR extinction, commonly expressed as a power-law $A_\lambda \propto \lambda^{-\alpha}$. This power-law also corresponds to the relative extinction between the $J$- and $K_s$-bands, i.e., $A_J / A_{K_s} = (\lambda_J / \lambda_{K_s})^{-\alpha}$. \citet{rieke:1985:618} measured $\alpha = 1.65$ using a small number of stars, however, \citet{xue:2016:23} measured a slightly larger value of $\alpha = 1.79$. The value of $\alpha$ corresponding to the measurements from \citet{davenport:2014:3430} is 1.25, significantly less steep than other studies. \citet{wang:2014:l12} studied the universality of the NIR extinction law using color excess ratios of APOGEE M and K giants, and found that the extinction law shows very little variation across different environments. We chose to adopt the relative extinction values from \citet{xue:2016:23}, whose measurement of $\alpha$ is consistent with other measurements from the diffuse ISM \citep{martin:1990:113}, to correct for extinction in each \emph{WISE} passband. Using the extinction corrected photometry, we reran the full and clean samples through the stellar parameters pipeline (Section~\ref{params}) to obtain new estimates for $T_\mathrm{eff}$ and $R_\ast$. For the remainder of this study we use the unreddened photometry.

\subsection{Determining Infrared Excesses}\label{irprops}

\begin{figure}
\centering
\includegraphics[width=\linewidth]{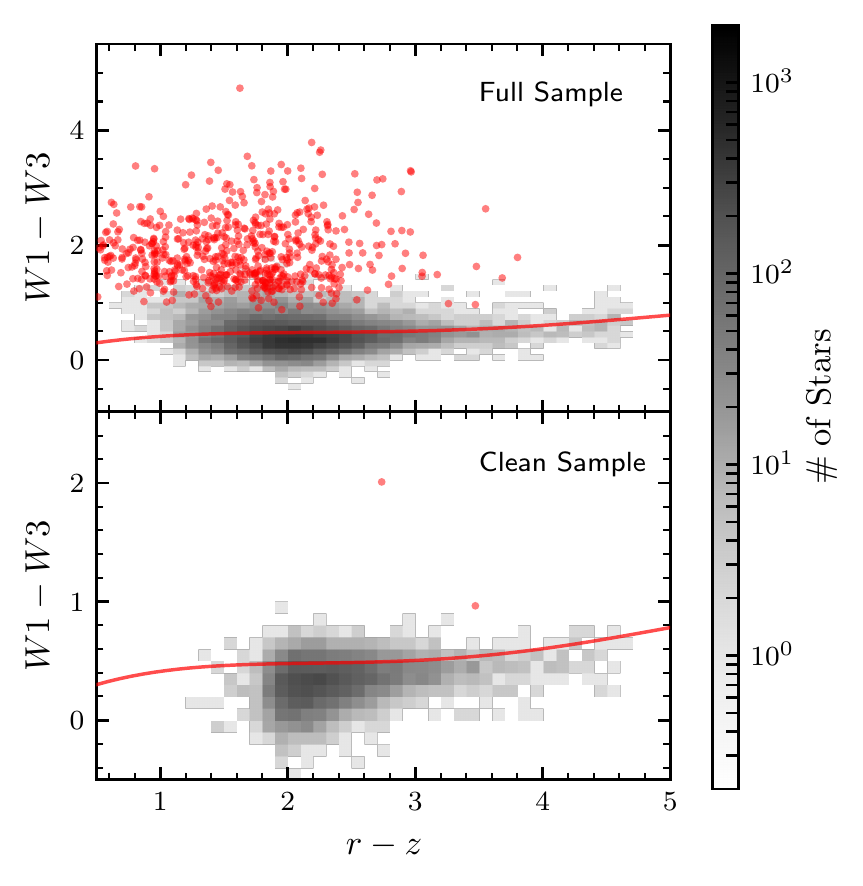}
\caption{SDSS and \emph{WISE} color-color 2D histogram for the full sample (top) and the clean sample (bottom). Each bin is (0.1 mag)$^2$. The red line is the empirical calibration of expected colors for main sequence dwarfs from TW14. Stars with $\sigma^\prime$ greater than the significance threshold defining true MIR excesses ($\sigma^\prime > 3.48$ for the full sample and $\sigma^\prime > 2.53$ for the clean sample) are marked as red points.
\label{fig:rzw1w3}}
\end{figure}

\begin{figure} 
\centering
 \includegraphics[width=\linewidth]{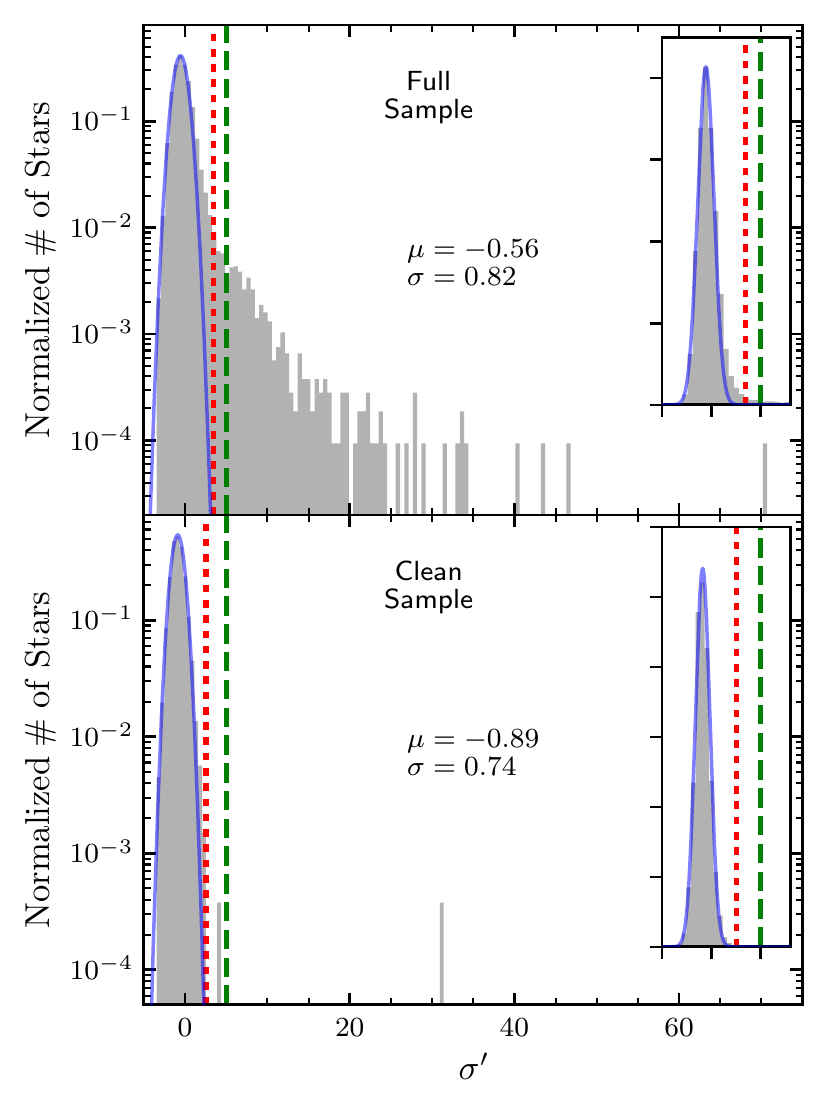}
\caption{Normalized distributions of $\sigma^\prime$ for the full sample (top) and the clean sample (bottom). Also plotted is the best-fit normal distribution (blue line). The red dotted line signifies the criteria for stars with MIR excess at the 99\% confidence level ($\sigma^\prime = 3.48$ for the full sample, and $\sigma^\prime = 2.53$ for the clean sample). The cutoff significance value used by TW14 ($\sigma^\prime = 5$) is denoted by the green dashed line for comparison. The inset plot shows the linear distributions. The clean sample (bottom) is well represented by a normal distribution, with a long tail out to high-significance MIR excesses. Both distributions are shifted slightly negative of zero, suggesting either the TW14 calibration needs to be recalibrated, or that some effect, such as the metal content of the stellar ensemble, has shifted these values.
\label{fig:sigprimes}}
\end{figure}

	TW14 explored two different methods to determine which stars showed high levels of excess IR fluxes over the expected photospheric values (``extreme" MIR excesses will be evaluated in Section~\ref{EME}). The first method, and the method ultimately used by TW14, is a modified version of the empirical calibrations from \citet{avenhaus:2012:a105}, using main sequence stars to determine the expected \emph{WISE} colors as a function of $r-z$ color (denoted as $\sigma^\prime$). Figure~\ref{fig:rzw1w3} shows the $r-z$ versus $W1-W3$ distribution for the full and clean samples, along with the empirical calibration of TW14. Figure~\ref{fig:sigprimes} shows the residual distribution with the TW14 empirical calibration (red line; Figure~\ref{fig:rzw1w3}) subtracted. Although it is common to define stars with disks to be only those with highly-significant deviations from the expected photospheric values in a binary fashion, we acknowledge that the distribution is continuous, and many of the stars with non-significant deviations may have true detections but smaller disk masses or dust that is becoming optically thin. Although we used the more classical binary description of stars with an excess versus stars without an excess, we will address this continuous distribution in Section~\ref{nonexcess}.
	
	Rather than making a blanket cut on stars with $\sigma^\prime \geqslant 5$, as was done in TW14, we used the distributions from Figure~\ref{fig:sigprimes} to evaluate the false-positive probabilities of the candidates. To obtain stars with a 99\% probability of hosting a true MIR excess, we define the probability threshold (assuming normal distributions),
\begin{equation}
P_{FP}(\mathrm{MIR\; Excess}) \times N_\mathrm{sample} < 0.01,
\end{equation}
where $P_{FP}(\mathrm{MIR\; Excess})$ is the probability that the MIR excess is a false-positive, and $N_\mathrm{sample}$ is the number of sources within the given sample. For the full sample, $P_{FP}(\mathrm{MIR\; Excess}) < 4.88\times10^{-7}$, and for the clean sample $P_{FP}(\mathrm{MIR\; Excess}) < 1.73\times10^{-6}$. Converting these false-positive probabilities into $\sigma^\prime$ values for each sample, we define stars with true MIR excesses to have $\sigma^\prime > 3.48$ for the full sample (4.90$\sigma$), and $\sigma^\prime > 2.53$ for the clean sample (4.64$\sigma$), both limits are shown in Figure~\ref{fig:sigprimes} (red dotted line), and candidates that meet these thresholds are marked as red points in Figure~\ref{fig:rzw1w3}. 
	
	Figure~\ref{fig:sigprimes} indicates that the TW14 calibration appears to be shifted to slightly redder \emph{WISE} colors than the bulk of the stellar population. The peak of the distribution is shifted negative of zero, which suggests that either the TW14 relationship needs to be recalibrated, or that some other effect is shifting the distribution, such as metallicity. Recently, \emph{WISE} bands have been shown to be sensitive to the metal content of stars, with metal poor stars showing redder $W1-W2$ color \citep{schmidt:2016:2611}. Although this analysis was only completed for late-K and early-M dwarfs, it is reasonable that a similar metallicity trend will hold for lower-mass stars. No metallicity relationship has been shown to exist for the $W1-W3$ color, however, if the primary metallicity sensitive band is $W1$, then we might expect metallicity to have a small effect on the $W1-W3$ color. 

	The second method takes the difference between the measured flux, and the expected flux (estimated from a stellar photospheric model), weighted by the measurement uncertainty. This value is commonly abbreviated as
\begin{equation}\label{eqn:excessrelationship}
\rchi_{12} = \frac{F_{12 \mu\mathrm{m,\; measured}} - F_{12 \mu\mathrm{m,\; model}}} {\sigma_{F_{12 \mu\mathrm{m,\; measured}}}}.
\end{equation}
	
\begin{figure} 
\centering
\includegraphics[width=\linewidth]{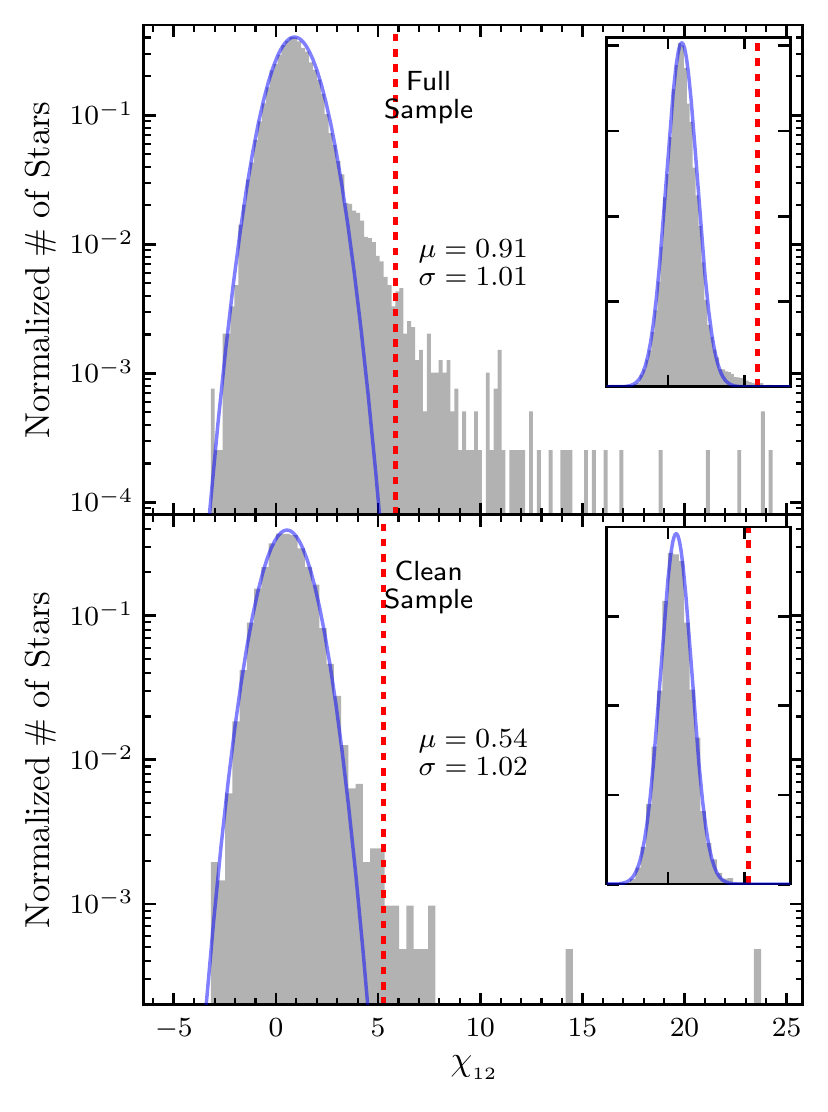}
\caption{Normalized distributions of $\rchi_{12}$ for the full sample (top) and the clean sample (bottom). Also plotted is the best-fit normal distribution (blue line). The inset plot shows the linear distributions. Both distributions are qualitatively similar in shape to those for the $\sigma^\prime$ values (Figure~\ref{fig:sigprimes}). Again, the red dotted line signifies the criteria for stars with MIR excess at the 99\% confidence level ($\rchi_{12} = 5.85$ for the full sample, and $\rchi_{12} = 5.25$ for the clean sample).
\label{fig:chis}}
\end{figure}

	Using stellar parameters and scaling values from the MCMC method (Section~\ref{params}), we computed the expected 12 \um\ flux densities for stars in both the full and clean samples. Next, we converted $W3$ magnitudes to flux densities using the \emph{WISE} all-sky explanatory supplement\footnote{\url{http://wise2.ipac.caltech.edu/docs/release/allsky/expsup/sec4\_4h.html}} (further details can be found in TW14). Figure~\ref{fig:chis} shows the distribution of $\rchi_{12}$ values for the full and clean samples. The majority of both samples are well represented by normal distributions with similar widths, although the full sample is shifted to slightly higher $\rchi_{12}$ values due to a distance bias which will be discussed in Section~\ref{biases}. 
			
	\citet{avenhaus:2012:a105} showed that the empirical method outlined above was able to detect the disk around AU Mic at 22 \um, while methods involving SED fitting were unable to significantly detect the same disk using observational data at similar wavelengths \citep{liu:2004:526, plavchan:2009:1068, simon:2012:114}. Presumably this indicates that $\sigma^\prime$ is a stronger discriminator of MIR excess significance. Although the SED fitting is important for estimating parameters that will allow us to then constrain disk parameters, we chose to select excess sources based solely on their $\sigma^\prime$ significance, similar to TW14. 

	Selecting stars with MIR excesses using the aforementioned criteria produced 609 stars in the full sample, and two stars in the clean sample. The cumulative false-positive probabilities for our selected stars are 0.0386\% ($\sim$0.24 stars) for the full sample, and $8.699\times10^{-6}$\% ($\ll 1$ star) for the clean sample. We used more stringent criteria in the selection of stars exhibiting MIR excesses than those used in TW14. Additionally, the parent population of stars for this sample (MoVeRS) is different than the parent population of TW14 (W11). To quantify this, the MoVeRS sample contains 15,262 of the W11 catalog ($\sim$22\%). Of the 15,262 matches in MoVeRS, 57 (of 168) are from the TW14 study of stars with MIR excesses ($\sim$34\%). Based on the selection criteria above, only 9 (of the 57) stars with MIR excesses would meet the new criteria ($\sim$16\%). These values will be considered when comparing our results to those from TW14 in Section~\ref{discussion}. Additionally, 181 of the MIR excess candidates in the full sample, and one of the MIR excess candidates in the clean sample, have $W4$ detections with S/N $> 2$. We will consider these $W4$ detections when we fit for fractional IR luminosities (Section~\ref{EME}).

\subsubsection{Extreme MIR Excesses}\label{EME}

\begin{figure} 
\centering
\includegraphics[width=\linewidth]{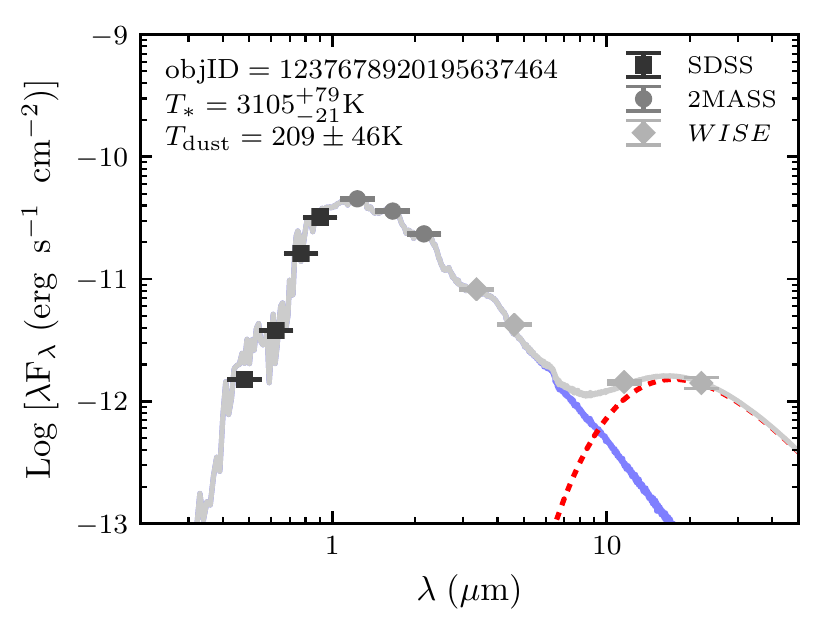}
\caption{SEDs for an object with a $W4$ detection. Plotted are the best-fit photosphere model (blue line), the best-fit dust blackbody (red dotted line), and the combined model (gray line). Model parameters are listed in the top left corner. The excess MIR flux is well fit by a single blackbody function.
\label{fig:dustfit}}
\end{figure}

	Extreme MIR excesses arising from planetary collisions are expected to produce large amounts of dust, and hence large fractional IR luminosities ($L_\mathrm{IR} / L_\ast \gtrsim 10^{-2}$). The primary focus of this study are these extreme MIR excesses, however, this requires knowledge about the total IR flux of the dust grains. For stars that have both $W3$ and $W4$ detections, we can fit a simple blackbody to the excess MIR flux, similar to what was done in TW14. We acknowledge that the disks we are interested in observing should emit a strong silicate features \citep[e.g.,][]{meng:2014:1032}, which would make $W3$ a poor indicator of the underlying blackbody continuum of the dust. However, with no ability to discern the blackbody continuum from the silicate emission (e.g., a MIR spectrum), we use the approximation that $W3$ is dominated by the continuum radiation. Using the extreme MIR excess candidates that had a $W4$ detection with a S/N $> 2$, we fit a combined model comprised of the best-fit photospheric model found in Section~\ref{extinction}, and a simple blackbody function. To determine the best fit blackbody function, we used a least-squares minimization, fitting for $T_\mathrm{dust}$ and the multiplicative scaling factor for the blackbody. For the least-squares fit, we used the best-fit photosphere model, and fit the dust blackbody function to the $W3$ and $W4$ measurements, weighted by the measurement uncertainty. An example fit from this process is shown in Figure~\ref{fig:dustfit}. For stars without a $W4$ detection, we assume the peak SED flux is at $W3$, giving an estimate for $T_\mathrm{dust} \approx 317.4$ K (TW14).
	
	To compute $L_\mathrm{IR} / L_\ast$, we integrated the best-fit photospheric model to estimate $L_\ast$, and for $L_\mathrm{IR}$, we subtracted the stellar model from the combined fit (stellar model plus best-fit blackbody), and integrated the residual flux to estimate $L_\mathrm{IR}$, taking the ratio of the two values \citep[similar to][]{patel:2014:10, patel:2017:54}. Keeping only the stars with $L_\mathrm{IR} / L_\ast \geqslant 10^{-2}$, we were left with 584 stars in the full sample and two stars in the clean sample, removing none of our stars. This is likely due to the fact that our initial selection criteria required significant MIR excesses. We will address ``non-significant" MIR excesses in the following section, and again in the discussion (Section~\ref{discussion}).

\subsubsection{Non-significant MIR Excesses}\label{nonexcess}

	In studies of disks that are inferred from their MIR excesses, it is common to only select stars with significant excesses, which deviate from the expected photospheric value. However, the distribution of stars with or without excesses is continuous, with a very subtle area between what is considered to have an excess and what is not considered to have an excess. Many of the stars that are not included in the bona fide sample of stars with MIR excesses are \emph{indeed} stars with excess MIR emission above their photospheric values. For example, the region between the 2$\sigma$ value and our cutoff limit ($1.09 < \sigma^\prime < 3.48$; Figure~\ref{fig:sigprimes}) contains many stars with real excesses and may trace the end of a collisional cascade where the dust is becoming optically thin. The problem is that we cannot confidently identify individual stars that have excesses in this range, since some of the stars in the $1.09 < \sigma^\prime < 3.48$ range are interlopers from the stellar distribution of $\sigma^\prime$. Instead, we can statistically examine this population.
	
\begin{figure} 
\centering
\includegraphics[width=\linewidth]{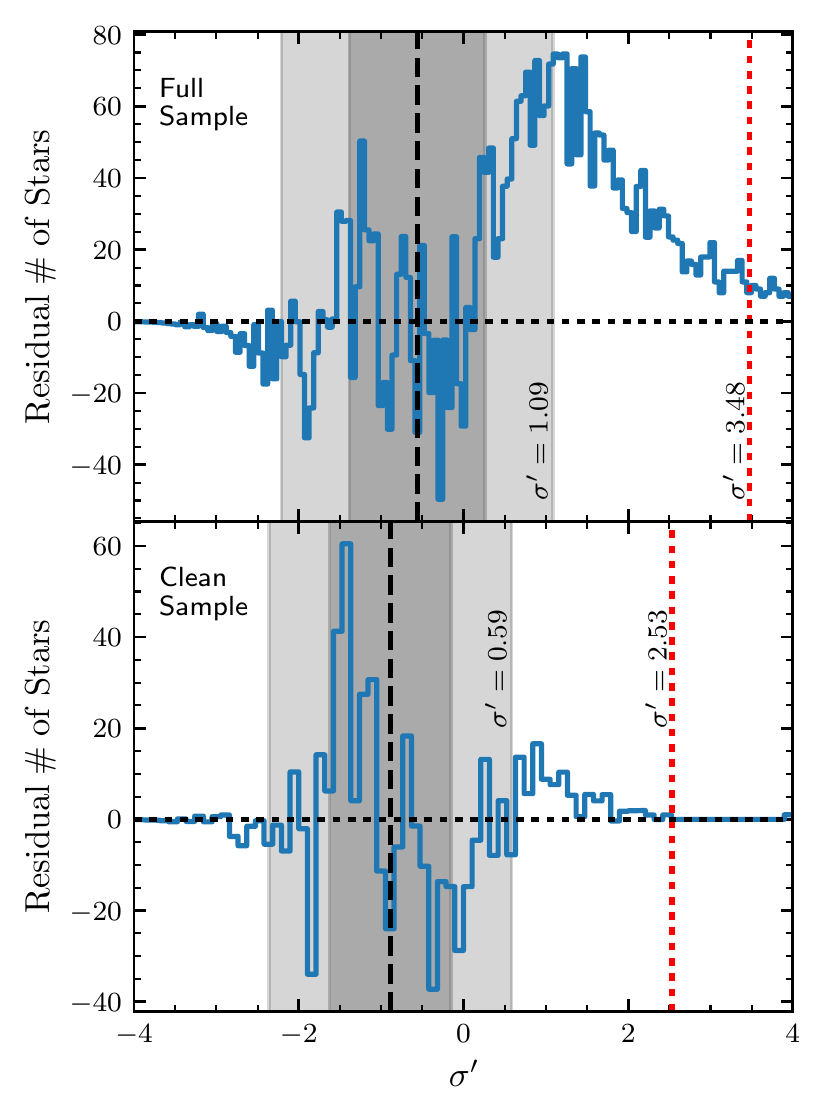}
\caption{Residual histograms subtracting the best-fit normal distribution from the distribution of $\sigma^\prime$ values (see Figure~\ref{fig:sigprimes}). The red dotted line represents the cutoff for significance used in identifying MIR excess candidates (Section~\ref{irprops}). The black dashed line shows the mean of the best-fit normal distribution (the approximate center of the distribution), and the black dashed line denotes a residual value of 0. The gray shaded areas show the 1- and 2-$\sigma$ regions about the mean. Within the 2$\sigma$ regions, the positive/minus scatter is approximately equal, and can be thought of as noise. In the positive region, at values of $\sigma^\prime$ larger than 2$\sigma$, there are significant bumps out to the imposed cutoff limit (red dotted line), indicating a large portion of true MIR excesses within this significance region.
\label{fig:sigprimeresiduals}}
\end{figure}

	Using the $\sigma^\prime$ distributions (Figure~\ref{fig:sigprimes}), we explored the number of excesses that exist within the non-significant excess region. We fit normal distributions to the core of the $\sigma^\prime$ distributions to minimize effects from the long tail of excess sources (blue line; Figure~\ref{fig:sigprimes}). Next, we subtracted the best-fit normal distribution (scaled from the normalized distribution to the true distribution) interpolated at the mid-point of each bin from the distribution of $\sigma^\prime$ values. The residual histograms are shown in Figure~\ref{fig:sigprimeresiduals}. The scatter within the 1$\sigma$ range (and to a lesser extent the 2$\sigma$ range) can be considered noise since the distribution is not perfectly normally distributed. However, the bumps at $\sigma^\prime$ values greater than 2$\sigma$ can be considered real since there is no corresponding scatter at similar negative $\sigma^\prime$ values about the mean. These bumps represent real sources harboring MIR excesses
	
	To quantify the number of potentially missing stars with MIR excesses, we integrated the region between the 2$\sigma$ limit (light gray region, $\sigma^\prime = 1.09$ for the full sample and $\sigma^\prime = 0.58$ for the clean sample; Figure~\ref{fig:sigprimeresiduals}) and the significant cutoff we imposed (red dotted line, $\sigma^\prime = 3.48$ for the full sample and $\sigma^\prime = 2.53$ for the clean sample; Figure~\ref{fig:sigprimeresiduals}). We estimate that $\sim$1400 stars are excluded from the full sample and $\sim$90 stars from the clean sample. However, this assumes that all missing stars are hosts to ``extreme MIR excesses." We computed fractional IR luminosities using the same method from the preceding section, finding that 5.6\% of the non-excess stars in the full sample and 0.5\% of the non-excess stars in the clean sample hosted extreme MIR excesses. This translates into $\sim$80 and $\sim$1 star(s) missing from the full and the clean samples, respectively. Although we cannot definitively say which stars within this region actually harbor a true MIR excess, it is important to consider this missing population in the context of the frequency of low-mass field stars exhibiting MIR excesses. If we consider the clean sample (as the full sample has a number of inherent biases that we will account for in Section~\ref{model}), then accounting for the missing sources, we estimate the fraction of stars exhibiting a MIR excess is $\sim$0.05\%. We will discuss this statistic further in Section~\ref{discussion}.

\subsection{Fidelity of Excesses: Cross-match to \emph{Spitzer}}\label{spitzer}

	To examine the validity of the extreme MIR excess detections, we cross-matched the candidates with the \emph{Spitzer} Enhanced Imaging Products catalog (this includes both IRAC and MIPS observations). We found ten candidates with \emph{Spitzer} photometry matched to within 6\arcsec. A search through the literature indicated that none of the Spitzer data for these sources have been published previously. Figure~\ref{fig:spitSED} shows the SEDs for these ten matching stars, demonstrating that the \emph{Spitzer} photometry is consistent with the \emph{WISE} photometry (for both $W3$ and $W4$ detections). All of these stars appear to have true MIR excesses. We are confident that the detected MIR excesses are true excesses originating from their host stars. However, younger populations of stars are expected to exhibit MIR excesses, therefore, we must test for youth where available in the samples.

\begin{figure*}[!htb]
\centering
\includegraphics{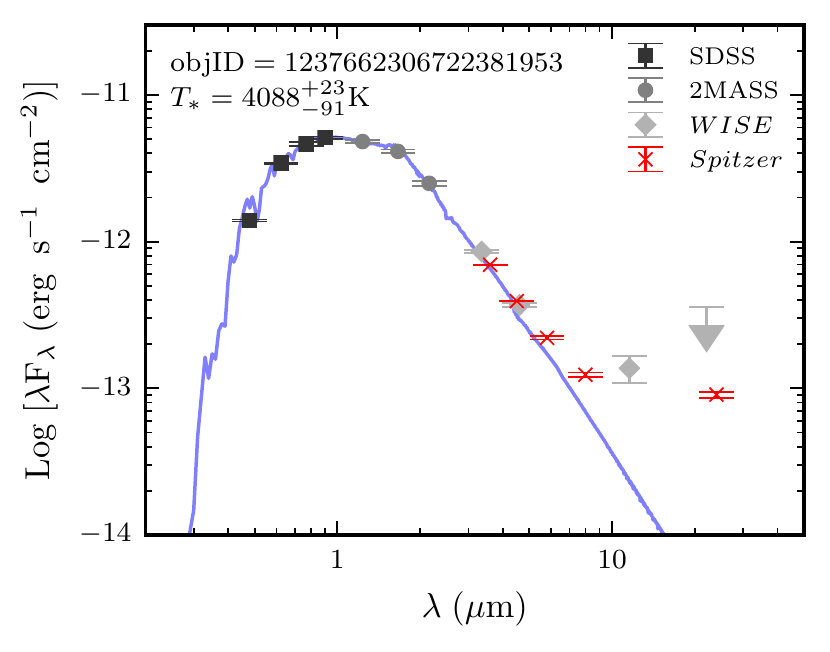}
\includegraphics{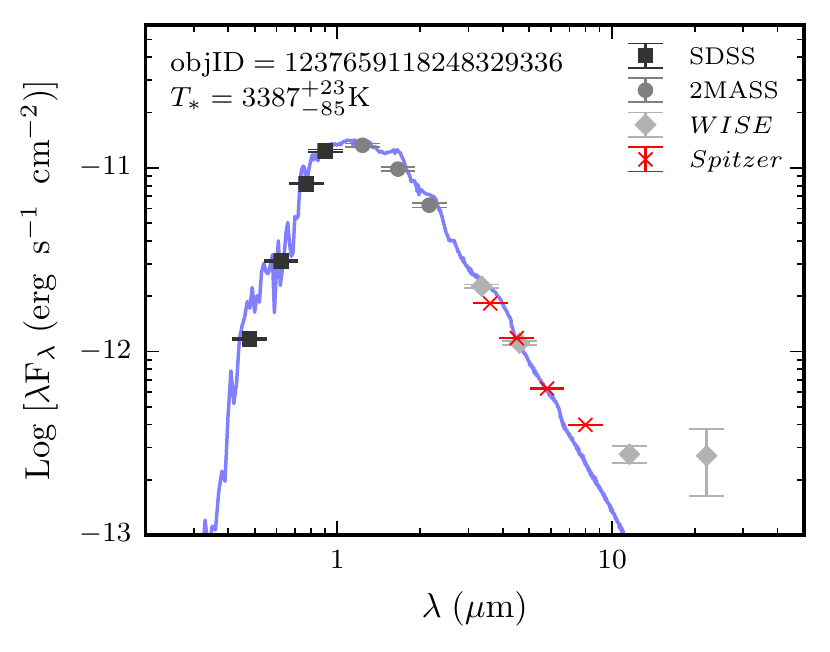}
\includegraphics{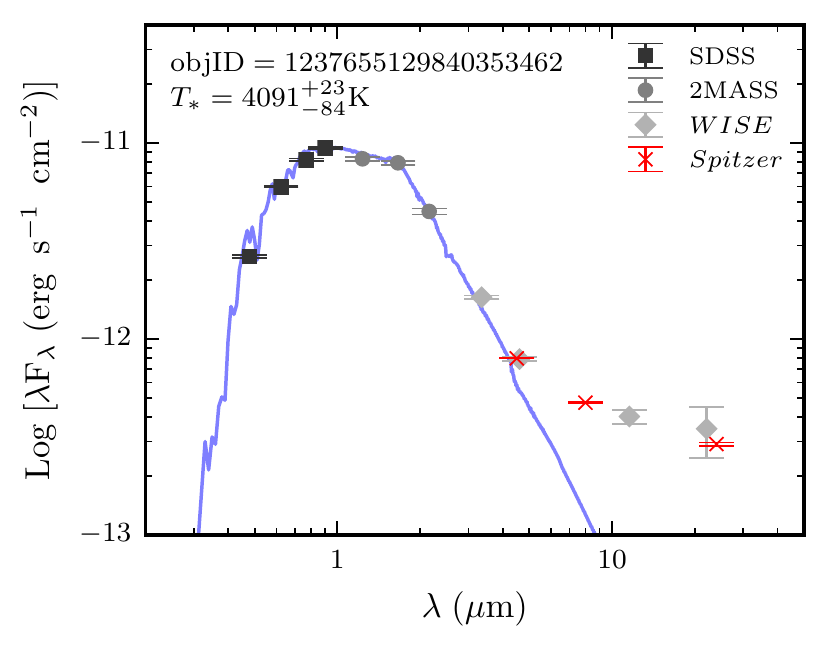}
\includegraphics{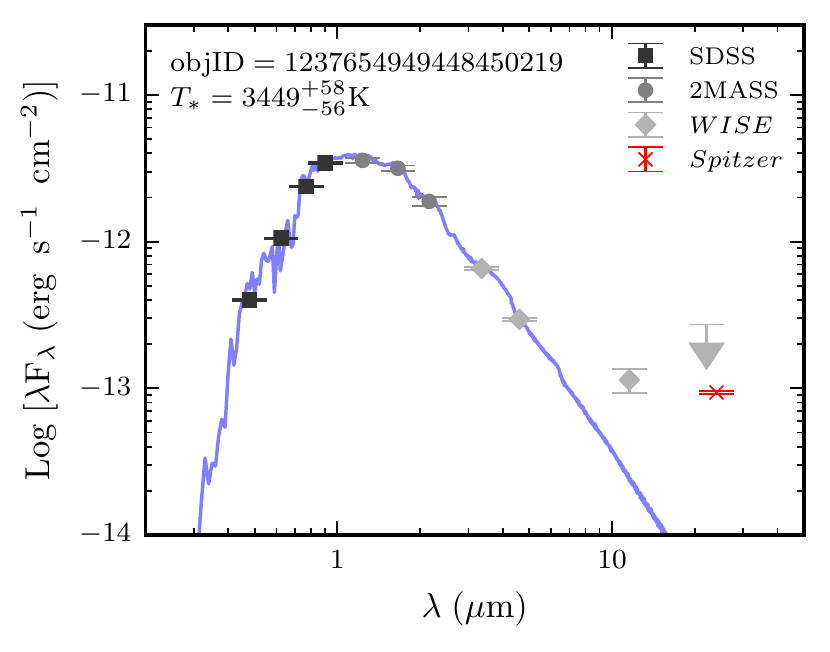}
\includegraphics{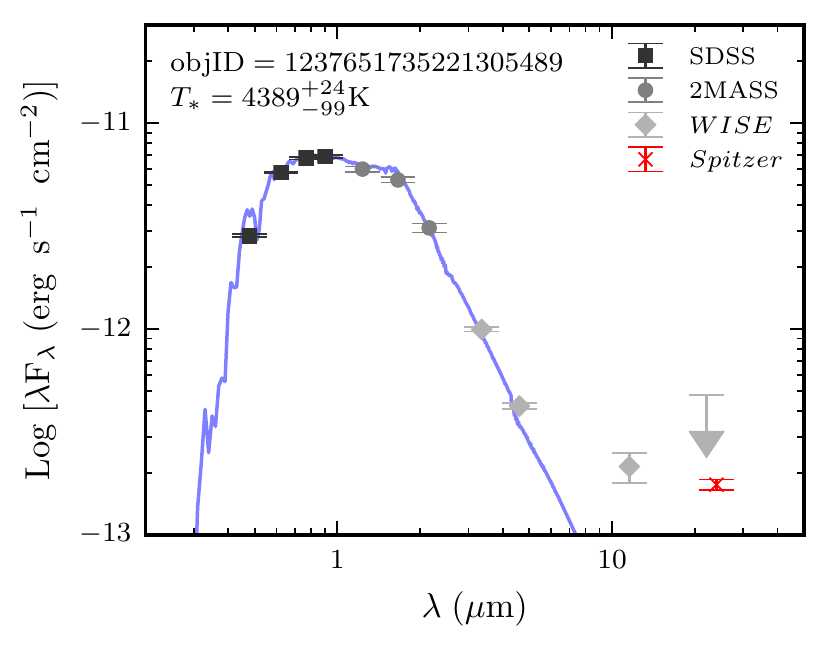}
\includegraphics{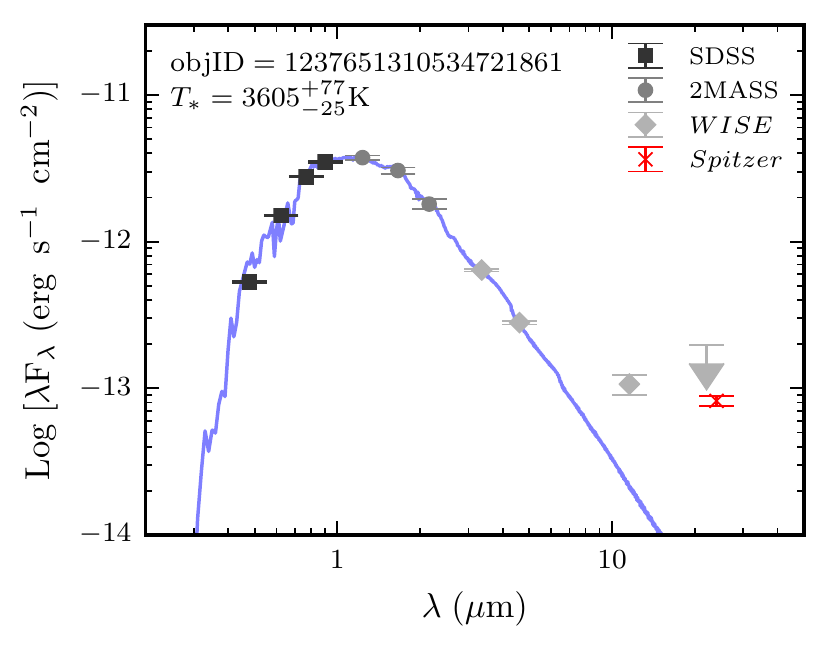}
\captcont{SEDs for all objects with \emph{Spitzer} detections. For all sources, there is good agreement between the \emph{WISE} and \emph{Spitzer} photometry, with all stars appearing to have true MIR excesses.
\label{fig:spitSED}}
\end{figure*}

\begin{figure*}[!htb]
\centering
\includegraphics{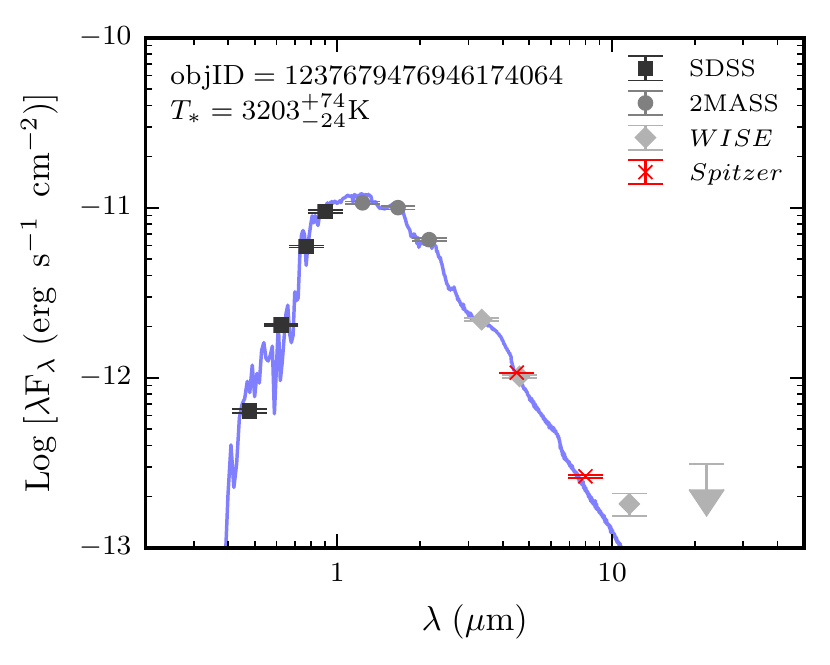}
\includegraphics{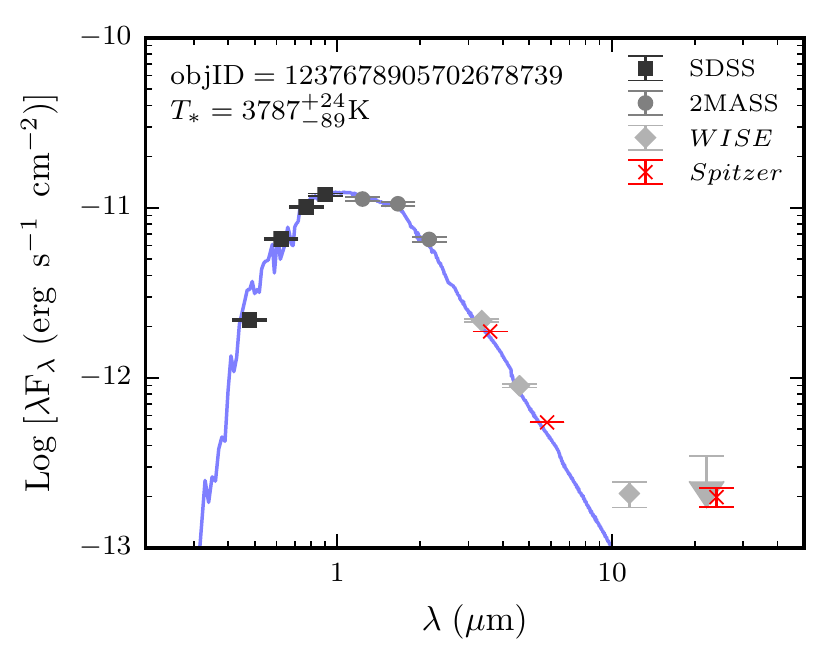}
\includegraphics{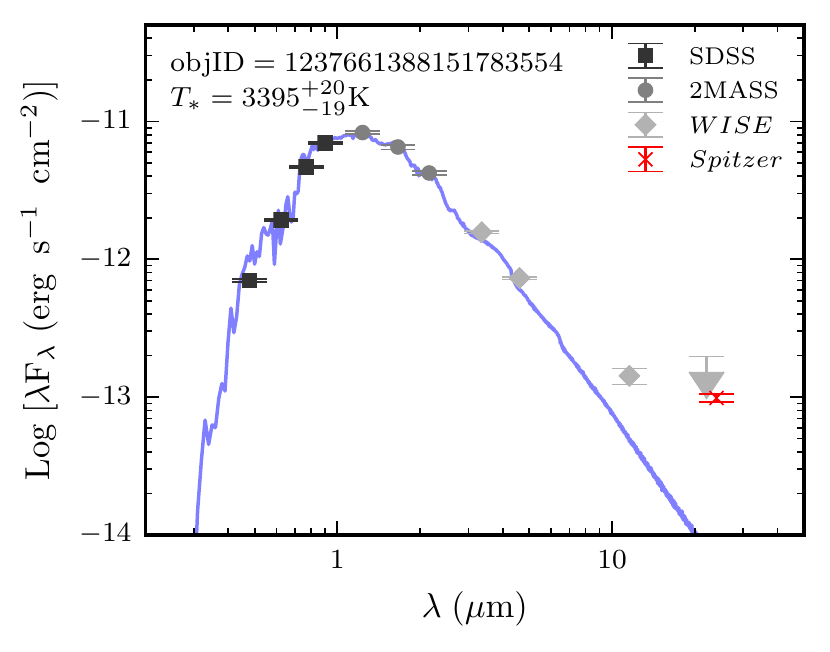}
\includegraphics{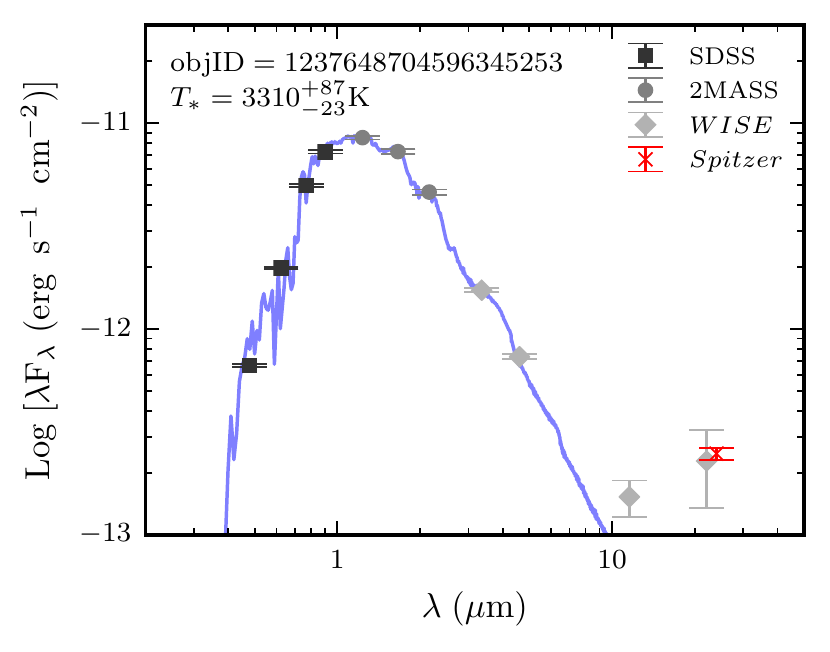}
\caption{Continued.}
\end{figure*}

\subsection{Spectroscopic Tracers of Youth}\label{spectroscopic}

	One strength of the TW14 sample over the MoVeRS sample is the availability of optical SDSS spectra for each star. This ensured that all objects were low-mass stars and made possible an investigation for youth. TW14 used age diagnostics such as H$\alpha$ emission to determine that the stars in their sample were older fields stars and not young, pre-main sequence stars, the latter of which we expect to host circumstellar disks (and therefore MIR excesses). To examine possible age diagnostics and confirm our selection of low-mass dwarfs for the sample, we identified ten SDSS spectroscopic targets within the extreme MIR excess sample, and received time on the Discovery Channel Telescope (DCT) to obtain optical spectra for 15 additional extreme MIR excess candidates. Unfortunately, none of the spectroscopic subsample overlapped with the stars with \emph{Spitzer} data (Section~\ref{spitzer}).
	
	TW14 used two age-dependent spectroscopic diagnostics: H$\alpha$ \citep[e.g.,][]{west:2006:2507, west:2008:785} and Li \textsc{i} \cite[e.g.,][]{cargile:2010:l111}. H$\alpha$ emission (in addition to other Balmer transitions) is a strong indicator of accretion, resulting in large equivalent width (EW) measurements\footnote{As is convention in studies of small stars, positive EW measurements indicate emission.} \citep[EW $\gtrsim 4$ \AA;][]{barrado-y-navascues:2003:2997} and broad lines \citep[10\% widths $> 270$ km s$^{-1}$;][]{white:2003:1109}. Stars exhibiting H$\alpha$ due to accretion are also young ($< 10$ Myr), and typically found in young associations rather than the field.
	
	For older populations of stars ($\gg$ 100 Myr), H$\alpha$ emission (and other Balmer transitions) is also tied to ``magnetic activity," as strong magnetic fields lead to chromospheric heating \citep{west:2015:3}. \citet{west:2008:785} demonstrated that the lifetime for magnetic activity (as traced through H$\alpha$ emission) is mass-dependent in the M dwarf regime. For the highest mass M dwarfs, the lifetime for magnetic activity is 500 Myr--1 Gyr, increasing to $>8$ Gyr for the lowest-mass M dwarfs. This makes H$\alpha$ emission a moderate age diagnostic for field stars, when coupled with stellar mass or spectral type. A lack of detectable H$\alpha$ emission in the earliest-type stars in our sample would indicate a relatively old ($>$ 1--2 Gyr) field population. We used the same regions as TW14 to measure the EW of H$\alpha$, and determine stars for which an EW measurement could or could not be made.
	
	Lithium absorption is more strongly correlated with youth than H$\alpha$ emission, but it is also mass dependent. Modeling results by \citet{chabrier:1997:1039} demonstrated that the initial lithium abundance will deplete by a factor of 10 in 10 Myr for a 0.7$M_\odot$ star ($\sim$M0), while a star with a mass of 0.08$M_\odot$ ($\sim$M8) will take $\sim$100 Myr to deplete by the same factor. This makes Li \textsc{i} absorption a strong discriminator of youth.
	
	Due to the difficulty in measuring the EW of Li \textsc{i} (primarily caused by the strong TiO features around Li \textsc{i} and typically low S/N), we applied a comparative technique, using SDSS template spectra \citep{bochanski:2007:531}, similar to what was done by \citet{cargile:2010:l111}. The template spectra from \citet{bochanski:2007:531} were built from a composite of SDSS field stars spectra. Therefore, they should indicate the baseline shape of the spectrum near the Li \textsc{i} feature for low-mass field stars devoid of Li \textsc{i} absorption. A comparison between the spectra and the \citet{bochanski:2007:531} template spectra provides a means to detect Li \textsc{i} absorption without making a direct measurement of the EW. Further details of the method are described in TW14.
	
\begin{figure} 
\centering
 \includegraphics[width=\linewidth]{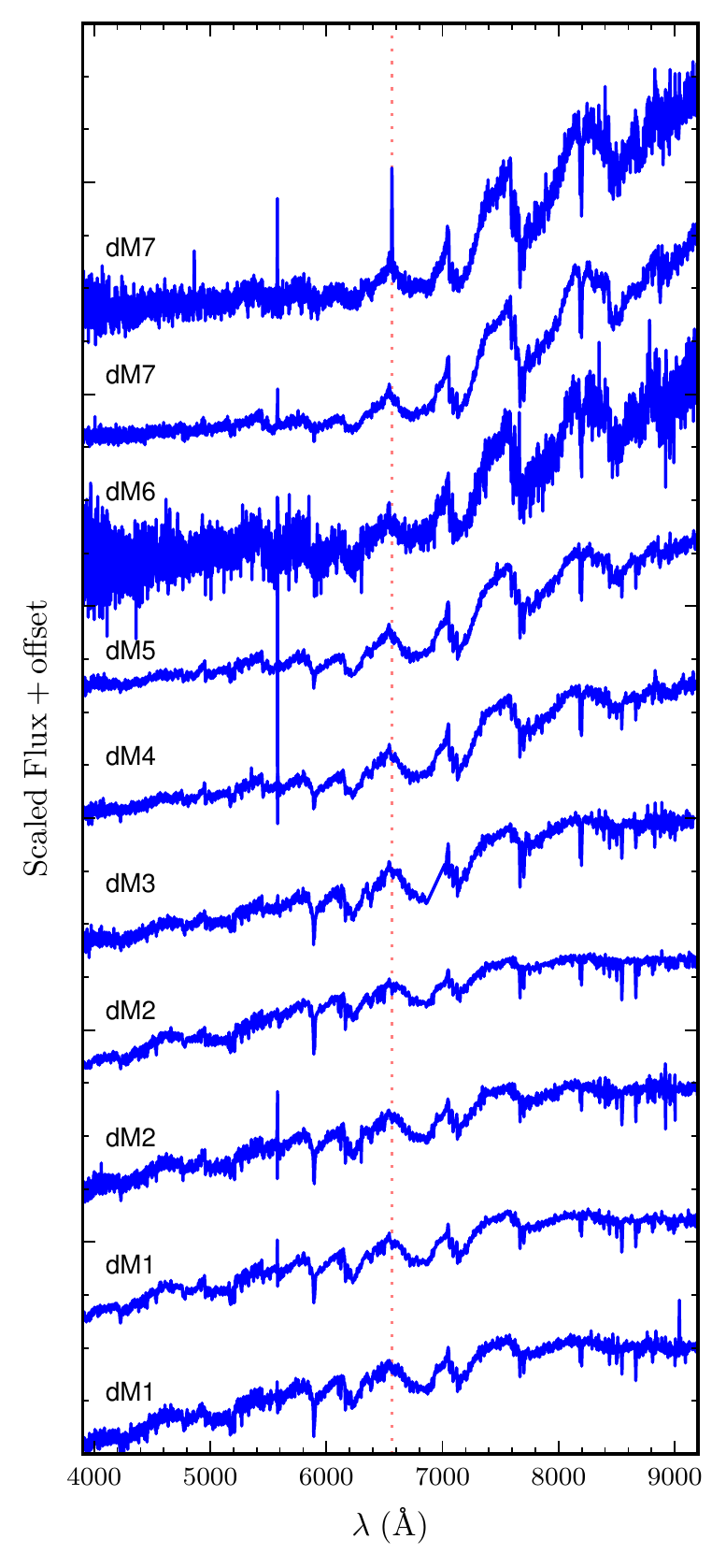}
\caption{Scaled and RV corrected SDSS spectra. All SDSS spectra appear to be low-mass stars (M dwarfs), confirming our sample selection. The dotted red line indicates the wavelength corresponding to H$\alpha$. Only one of the objects has detectable H$\alpha$ emission and none of the objects show detectable amounts of Li \textsc{i} absorption. Spectral types using the \emph{PyHammer} python package \citep{kesseli:2017:} are listed above each spectrum. The large feature commonly found at 5600 \AA\ is an artifact caused by the SDSS spectrograph and is not a real feature \citep{silvestri:2006:1674, morgan:2012:93}.
\label{fig:sdssspectra}}
\end{figure}

	We discovered that ten of the extreme MIR excess candidates had been previously observed through one of the SDSS spectroscopic programs and had spectra available. Nine of these stars were included in TW14 because they were part of the SDSS DR7 spectroscopic sample of M dwarfs \citep{west:2011:97}, and one of the stars was observed after the \citet{west:2011:97} sample was compiled. All ten of these stars are classified as M dwarfs, confirming our selection of low-temperature dwarfs. The radial velocity (RV) corrected SDSS spectra are shown in Figure~\ref{fig:sdssspectra}. Only one of these stars (an M7) showed significant H$\alpha$ emission. The average activity lifetime of an M7 star is $\sim$8 Gyr \citep{west:2008:785}. None of these stars had detectable amounts of lithium. Our Li \textsc{i} analysis sets a lower age limit of $> 100$ Myr. The lack of H$\alpha$ emission for stars earlier than M7 indicates a typical minimum age of $\sim$ 1 Gyr for the sample \citep{west:2008:785}, indicative of an older field population.
	
\begin{figure} 
\centering
\includegraphics[width=\linewidth]{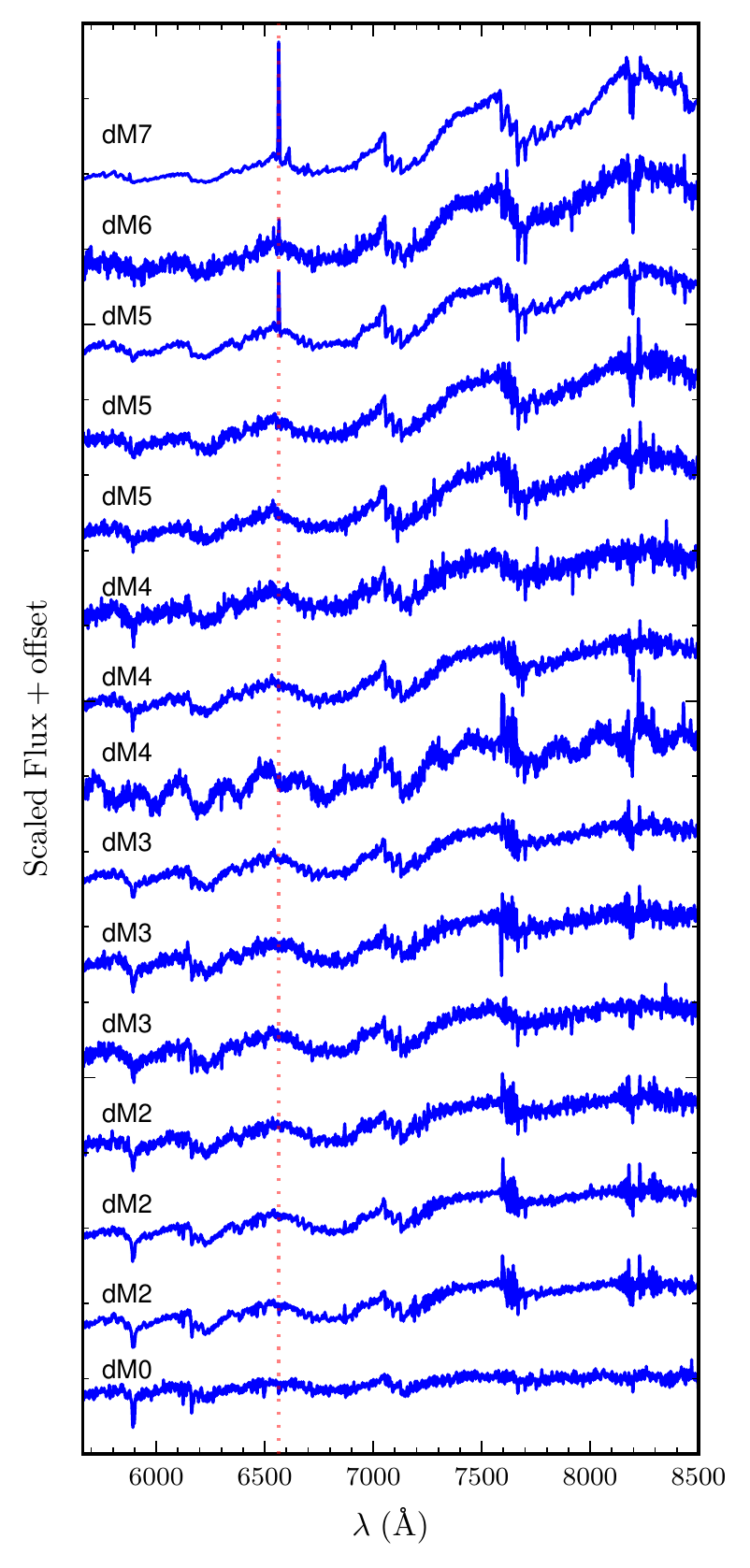}
\caption{Scaled and RV corrected spectra from the DCT. The dotted red line indicates the wavelength corresponding to H$\alpha$. The top three stars have detectable amounts of H$\alpha$ emission. Spectral types using the \emph{PyHammer} python package \citep{kesseli:2017:} are listed above each spectrum. All spectra appear to be M-type stars. The eighth spectrum from the bottom has peculiar characteristics, partially consistent with a cool star and a carbon star (discussed in the text). 
\label{fig:spectra}}
\end{figure}

	To further assess the age for the sample of extreme MIR excess candidates, we obtained optical spectra with the DeVeny Spectrograph on the 4.3-m DCT for an additional 15 candidates with high-significance MIR excesses ($\sigma^\prime > 10$), shown in Figure~\ref{fig:spectra}. The spectra cover the range $\sim$ 5600\AA--9000\AA\ at a resolution of $\lambda/\Delta\lambda \approx 2850$ (2.5 pixel). The candidates were selected based on location in the sky, and should represent a relatively unbiased subsample of the full sample.
	
	Spectra were reduced using a modified version of the \emph{pyDIS} Python package \citep{davenport:2016:d}, originally designed for use with the APO 3.5-m Dual Imaging Spectrograph (DIS). Stars were spectral typed using the \emph{PyHammer}\footnote{\url{https://github.com/BU-hammerTeam/PyHammer}} Python package \citep{kesseli:2017:}. Although this is a small portion of the total sample, we expect a similar age distribution for the parent population.
	
	The spectroscopic observations collected indicate that the DCT sample is also made up of low-temperature stars, further confirming our sample selection. One of the stars (SDSS objID 1237668734684955989; 2MASS J18351414+4026520) has peculiar features. The TiO bands found at 7053\AA\ are consistent with a cool star, but other features are consistent with a carbon dwarf \citep[dC;][]{green:2013:12}, while some of the features are not. This object motivates further investigation to determine its true nature. From the full spectroscopic sample of 25 stars, we estimate a contamination rate of 4\% for our entire sample due to objects that are not typical low-mass stars.
	
	We observed that only three of the stars for which we have DCT spectra, all within the fully convective regime ($\gtrsim$ M4), showed signs of H$\alpha$ emission. Additionally, none of the stars had detectable amounts of Li \textsc{i}. This lack of Li \text{i} absorption is consistent with the stars having ages $\gg$ 100 Myr estimated from the SDSS spectra. Considering the stars without H$\alpha$ emission, this indicates the average of the population is $\gtrsim$ 1 Gyr \citep{west:2008:785}, again consistent with the findings from the SDSS spectra. Based on the age limits from the two spectroscopic subsamples, we concluded (as did TW14) that the orbiting dust (inferred from the MIR excesses) was not primordial in nature, since the primordial disk is expected to be dispersed on timescales much shorter than the presumed ages 
	
\begin{deluxetable*}{lccccccc} 
\tabletypesize{\footnotesize}
\tablecolumns{8}
\tablecaption{Spectroscopic Parameters\label{tbl:observations}}
\tablehead{
\colhead{SDSS DR8$+$ objID} & \colhead{R.A.} & \colhead{Decl.} & \colhead{Radial Velocity} & \colhead{Spectral} & \colhead{H$\alpha$ EW\tablenotemark{a}} & \colhead{Telescope} & \colhead{$\langle L\rangle$} \\
 & \colhead{(H:M:S)} & \colhead{(D:M:S)} & \colhead{$\pm 7$ (km s$^{-1}$)} & \colhead{Type} & \colhead{(\AA)} & &
}
\startdata
1237665369038782628 &	10:17:40.54	& $+$28:58:51.62	& $+39.5$		& M1		& ...				& SDSS	& 21.34\\
1237651250974556408 &	15:47:54.70	& $+$52:48:57.52	& $-32.5$		& M1		& ...				& SDSS	& 13.77\\
1237657071156723794 &	01:27:51.44	& $+$00:16:33.17	& $+6.2$		& M2		& ...				& SDSS	& 21.98\\
1237655692480151822 & 15:16:10.43	& $-$01:42:37.24	& $-48.4$		& M2		& ...				& SDSS	& 16.95\\
1237671125374861409 &	09:32:04.26	& $+$14:08:26.51	& $+39.0$		& M3		& ...				& SDSS	& 92.45\\
1237662619722449089 & 15:38:25.49	& $+$32:28:44.59	& $-10.0$		& M4		& ...				& SDSS	& 36.11\\
1237667254011101278 & 11:30:25.02	& $+$29:14:16.37	& $+25.6$		& M5		& ...				& SDSS	& 59.30\\
1237659161736315205 &	15:48:31.45	& $+$42:53:07.14	& $-21.1$		& M6		& ...				& SDSS	& 179.04\\
1237665128545911020 &	12:42:03.86	& $+$34:55:37.74	& $-45.7$		& M7	 	& ...				& SDSS	& 240.58\\
1237661068171346281 &	09:31:07.08	& $+$10:06:07.25	& $+16.2$		& M7		& $10.3 \pm 0.9$	& SDSS	& 327.29\\
1237668331488084142 &	14:12:46.44	& $+$15:01:52.55	& $-42.1$		& M0		& ...				& DCT	& -1.97\\
1237651250974556408 &	15:47:54.70	& $+$52:48:57.52	& $-8.4$		& M2		& ...				& DCT	& 17.59\\
1237655749395022353 &	18:04:45.57	& $+$46:36:57.79	& $-51.4$		& M2		& ...				& DCT	& 41.55\\
1237672026249167591 &	22:41:17.31	& $+$33:40:21.14	& $-43.6$		& M2		& ...				& DCT	& 22.31\\
1237664852033142893 &	14:15:55.43	& $+$32:54:33.84	& $+25.1$		& M3		& ...				& DCT	& 11.19\\
1237662500006461639 &	16:01:09.94	& $+$36:35:30.07	& $+5.2$		& M3		& ...				& DCT	& 38.02\\
1237655747779363146 &	17:45:18.61	& $+$57:53:59.65	& $+4.3$		& M3		& ...				& DCT	& 28.02\\
1237668734684955989 &	18:35:14.13	& $+$40:26:51.95	& $+93.0$		& Pec\tablenotemark{b}	& ...	& DCT	& ...\\
1237671941420483289 &	19:06:24.80	& $+$64:36:19.88	& $-56.3$		& M4		& ...				& DCT	& 40.04\\
1237656241159012941 &	21:58:10.54	& $+$11:42:01.70	& $-122.0$	& M4		& ...				& DCT	& 30.99\\
1237659330309456141 &	15:35:00.41	& $+$48:53:42.51	& $-111.1$	& M5		& ...				& DCT	& 51.76\\
1237655465932292383 &	16:17:07.09	& $+$45:52:14.97	& $-70.0$		& M5		& ...				& DCT	& 86.02\\
1237652943699509565 &	22:00:46.74	& $+$12:44:01.96	& $-32.4$		& M5		& $6.3 \pm 0.5$	& DCT	& 76.04\\
1237652937790915940 &	20:53:41.55	& $+$08:35:14.57	& $-26.7$		& M6		& $3.5 \pm 0.9$	& DCT	& 241.09\\
1237678920195637464 &	22:35:47.06	& $+$11:42:15.67	& $-43.5$		& M7		& $15.8 \pm 1.8$	& DCT	& 103.27
\enddata
\tablenotetext{a}{Positive EW measurements indicate emission. Inconclusive measurements are not listed.}
\tablenotetext{b}{This object shows peculiar spectral features. The TiO bands at $\sim$7050 are indicative of a low-mass star. However, the numerous bumps in the spectrum may indicate a carbon dwarf.}
\end{deluxetable*}

\subsubsection{Spectroscopic Estimates of Luminosity Classes}

	We also make an estimate on the contamination rate of giants in our subsample of the MoVeRS catalog using the collected spectra. A thorough investigation into separating M-type stars based on luminosity class was undertaken by \citet{mann:2012:90}, using a modified method similar to \citet{gilbert:2006:1188} for \emph{Kepler} target stars. The spectroscopic features \citet{mann:2012:90} used for determining luminosity classes included: 1) the CaH2 (6814--6846 \AA) and CaH3 (6960--6990 \AA) indices \citep[][]{reid:1995:1838}; 2) the Na \textsc{i} doublet \citep[8172--8197 \AA;][]{schiavon:1997:902}; 3) the Ca \textsc{ii} triplet \citep[8484--8662 \AA;][]{cenarro:2001:959}; 4) the mix of atomic lines (Ba \textsc{ii}, Fe \textsc{i}, Mn \textsc{i}, and Ti \textsc{i}) at 6470--6530 \AA \citep{torres-dodgen:1993:693}; and 5) the K \textsc{i} (7669--7705 \AA) and Na \textsc{i} lines identified in \citet{mann:2012:90}. The Ca \textsc{ii} triplet falls within a region prone to fringing at the red-end of the DCT spectra, therefore, we omitted measuring this feature. Most of the spectroscopic features above change with surface gravity and temperature, therefore, we compare the above spectroscopic indices against the TiO5 index \citep[][]{reid:1995:1838}, which is sensitive to both metallicity and temperature \citep{woolf:2006:218, lepine:2007:1235}, but relatively insensitive to surface gravity \citep[e.g.,][]{jao:2008:840}. All other aforementioned features were measured using the available SDSS and DCT spectra following the same prescription outlined in \citet{mann:2012:90}. Table~\ref{tbl:specmeasurements} contains the information for the continuum region(s) and band region used to measure EWs and spectral indices.

\begin{deluxetable}{lccc} 
\tabletypesize{\footnotesize}
\tablecolumns{4}
\tablecaption{Spectroscopic Indices\label{tbl:specmeasurements}}
\tablehead{
\colhead{Index Name} & \colhead{Band} & \colhead{Continuum} \\
& \colhead{(\AA)} & \colhead{(\AA)} &
}
\startdata
Na \textsc{i} (a)\tablenotemark{a}							&	5868--5918	& 6345--6355	\\
Ba \textsc{ii}/Fe \textsc{i}/Mn \textsc{i}/Ti \textsc{i}\tablenotemark{a} 	&	6470--6530	& 6410--6420	\\
CaH2\tablenotemark{b} 									&	6814--6846	& 7042--7046	\\
CaH3\tablenotemark{b} 									&	6960--6990	& 7042--7046	\\
TiO5\tablenotemark{b}									&	7126--7135	& 7042--7046	\\
K \textsc{i}\tablenotemark{a} 								&	7669--7705	& 7677--7691, 7802--7825\\
Na \textsc{i} (b)\tablenotemark{a} 							&	8172--8197	& 8170--8173, 8232--8235
\enddata
\tablenotetext{a}{Measured as an EW. Linear interpolation is done through the continuum ranges to estimate the continuum.}
\tablenotetext{b}{Measured as a band index by calculating the mean flux within each wavelength range, and taking the ratio between the band mean flux to the continuum mean flux.}
\end{deluxetable}
	
	To determine the expected EWs and spectral indices for low-mass dwarfs, we measured the same features for 38,722 stars from the \citet{west:2011:97} spectroscopic sample of M dwarfs with good photometry (\textsc{goodphot} $=1$) and good proper motions (\textsc{goodpm} $=1$). Although there is expected to be some small amount of giant contamination within this sample, it is estimated to be less than 2\%, and the use of good proper motions should further minimize giant contamination. We also obtained optical spectra for 154 giant stars from \citet{fluks:1994:}, \citet{danks:1994:382}, \citet{serote-roos:1996:93} and SDSS. All giant spectra were sampled to the same resolution as our sample spectra prior to measuring spectroscopic indices to remove any potential bias.

\begin{figure*} 
\centering
\includegraphics[width=\linewidth]{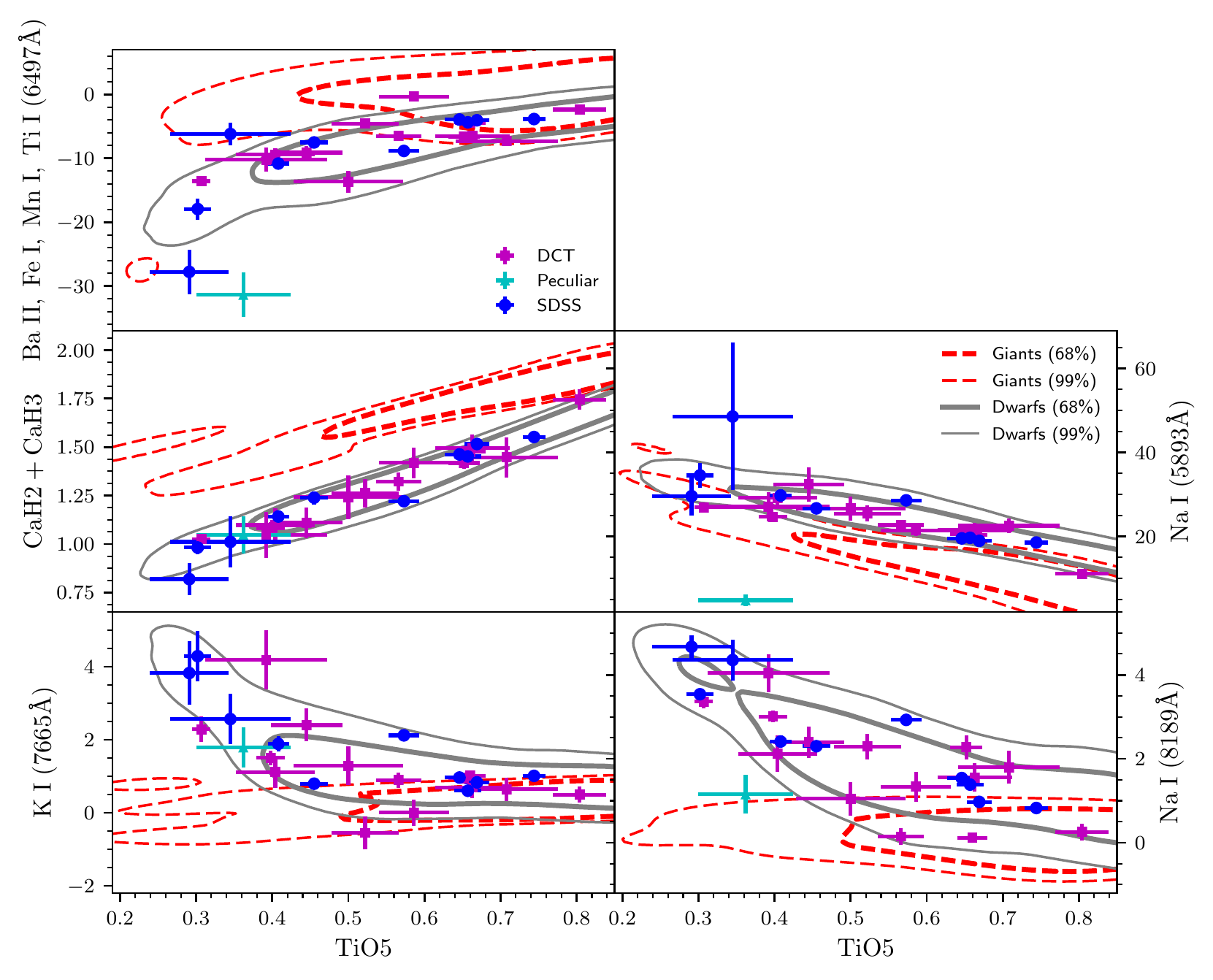}
\caption{Comparison of spectroscopic indices for dwarfs and giants. Our dwarf training set (gray solid lines) and giant training set (red dashed lines) show the 68\% and 90\% confidence intervals. Also plotted are our sample from the DCT (purple squares), SDSS (blue circles), and our peculiar source (cyan triangle). The likelihood of each source being a dwarf versus a giant is shown in Table~\ref{tbl:observations}
\label{fig:giantdwarf}}
\end{figure*}

	To estimate the likelihood that each star in our sample is either a dwarf or a giant, we built 2-D probability distributions for both the dwarfs and giant comparison samples for each spectroscopic tracer using a Gaussian Kernel Density Estimation using Silverman's Rule \citep{silverman:1986:}, as is shown in Figure~\ref{fig:giantdwarf}. The likelihood that source $i$ is a dwarf given spectroscopic index $j$ is estimated by the log-likelihood,
\begin{equation}
L_{i,j} = \log_{10}\left(\frac{P_\mathrm{dwarf}}{P_\mathrm{giant}}\right).
\end{equation}
The likelihood given all indices that a source is a dwarf versus a giant is
\begin{equation}\label{eqn:prob}
\langle L_{i}\rangle = \frac{\sum_j w_j L_{i,j}}{\sum_j w_j },
\end{equation}
where $w_j$ is a weighting factor for spectroscopic index $j$. \citet{mann:2012:90} found that setting weights to unity (allowing all spectroscopic tracers to be equally weighted) did not significantly alter results. We chose to equally weight all the measured spectroscopic indices, simplifying Equation~(\ref{eqn:prob}) to $\langle L_{i}\rangle = \sum_j L_{i,j}$.

	Each source was then either assigned to the category of dwarf star ($L_i > 2$), giant star ($L_i < -2$), or undetermined ($-2 < L_i < 2$), based on the 99\% confidence that one training set was more likely to host the source. All but one of our sources has a high probability of being a dwarf versus a giant. The earliest type star in our sample has an inconclusive classification, primarily due to all spectroscopic indices for both training sets beginning to converge for the earliest type stars (largest values of TiO5). Given this object's measured proper motion in multiple catalogs, this is most likely a dwarf star. The inclusion of this object in \emph{Gaia} DR1 indicates that both a higher precision proper motion measurement and a trigonometric distance are forthcoming, which will definitively determine the luminosity class of this object. We did not attempt to ascribe a luminosity class to our peculiar object due to multiple non-similarities in its spectrum as compared to both our training sets. Based on our above analysis, we do not change our estimated contamination rate of $\sim$4\%.

\subsection{Disk Properties}\label{disks}

\begin{figure} 
\centering
\includegraphics[width=\linewidth]{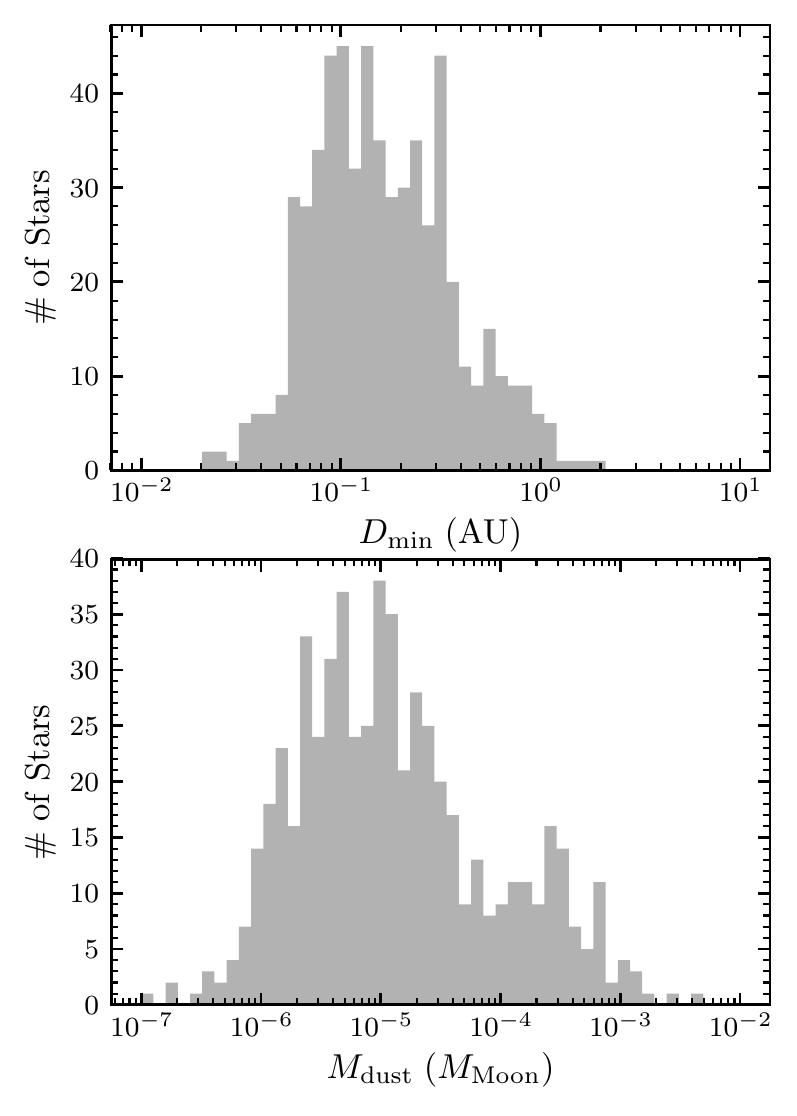}
\caption{
\emph{Top}: Distribution of minimum orbital distances of dust surrounding the stars with MIR excesses. The vast majority of dust populations are estimated to be within 1 AU of their host star, typically within the snow-line for low-mass stars \citep[$\sim$0.3 AU;][]{ogihara:2009:824}. 
\emph{Bottom}: Distribution of minimum dust masses. The median value of $10^{-5} M_\mathrm{Moon}$ is comparable to the TW14 study of low-mass stars with extreme MIR excesses.
\label{fig:diskparams}}
\end{figure}

	We can further explore the properties of our extreme MIR excess systems by making some basic assumptions about the disk properties. Dust temperatures allow us to estimate both the orbital distance of the dust, and the minimum dust mass. Using the dust grain temperature estimates (Section~\ref{EME}), we calculated the minimum orbital distance of the dust assuming the dust grains are in thermal equilibrium with the host star, given by,
\begin{equation}
D_\mathrm{min} = \frac{1}{2} \left(\frac{T_\ast}{T_\mathrm{gr}} \right)^2 R_\ast,
\end{equation}
where $T_\ast$ and $T_\mathrm{gr}$ are the stellar effective temperature and dust grain temperature, respectively, and $R_\ast$ is the stellar radius. Assuming a simple geometry for the orbiting dust a dust mass ($M_d$) can be estimated. Similar to TW14, we assumed the dust is in a thin shell, orbiting a distance $D_\mathrm{min}$ from the host star, with a particulate radius $a$ and density $\rho_s$, and a cross section equal to the physical cross section of a spherical grain. We take $\langle a \rangle = 0.5$ \um\ and $\rho_s = 2.5$ g cm$^{-3}$, similar to TW14. The dust mass is then defined as,
\begin{equation}
M_d \geqslant \frac{16}{3} \pi \frac{L_\mathrm{IR}}{L_\ast} \rho_s \langle a \rangle D_\mathrm{min}^2.
\end{equation}
Further details regarding this process can be found in TW14. The orbital distances and dust masses for the extreme MIR excess candidates are shown in Figure~\ref{fig:diskparams}. The majority of stars harbor dust within 1 AU, with the peak of the distribution at a few tenths of an AU, within the snow-line for low-mass stars \citep[$\sim$0.3 AU;][]{ogihara:2009:824}. For the majority of our sample, which only have $W3$ measurements, the dust temperature was assumed to be 317.4 K, which predetermined the estimated orbital distance of the dust to be within the snow-line. A colder disk ($<$ 317.4 K) would need to be even more massive to have a similar flux level at $W3$, making it more likely that we are observing a less massive, hotter disk. Our dust mass estimates are comparable to those found in TW14, with the median value of $10^{-5} M_\mathrm{Moon}$. Obtaining MIR spectra of these stars with the next generation of telescope will help to further characterize these dust populations (e.g., constrain mineralogy).
	
\subsection{The Extreme MIR Excess Sample}\label{EMEsample}

\begin{figure}
\centering
\includegraphics[width=\linewidth]{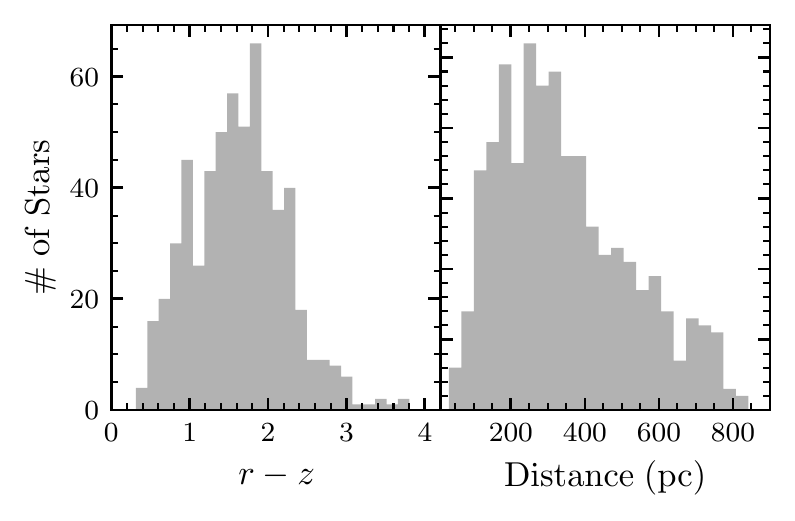}
\includegraphics[width=\linewidth]{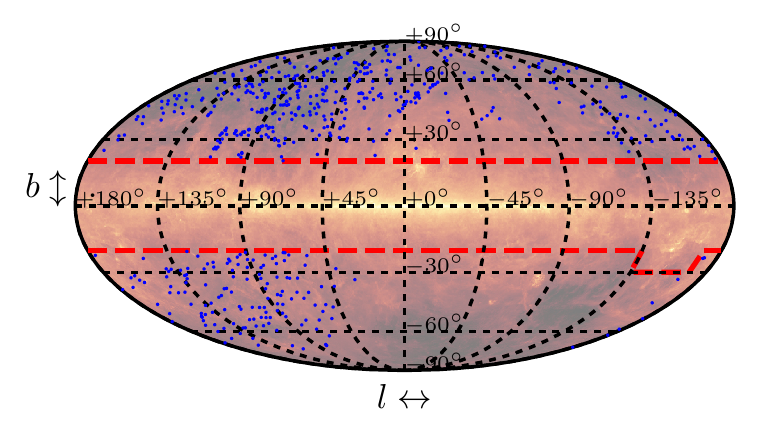}
\caption{
\emph{Top Left}: The distribution of $r-z$ colors for the sample. The peak of the distribution corresponds to a spectral type of dM4 \citep[$M_\ast \approx 0.125M_\odot$;][]{baraffe:1996:l51}, approximately where the initial mass function peaks \citep{chabrier:2003:763}.
\emph{Top Right}: The distribution of distances for the sample. The majority of stars are found within 500 pc, which is consistent with other samples of low-mass stars from SDSS (TW14).
\emph{Bottom}: The Galactic distribution of the stars with extreme MIR excesses. Candidate stars are blue points on top of the \emph{IRAS}/\emph{COBE} 100 \um\ dust map \citep{schlegel:1998:525}. Red dashed lines denote regions removed from our search (Section~\ref{sample}).
\label{fig:EMEsample}}
\end{figure}

	The general characteristics of our sample of stars with extreme MIR excesses are similar to those from TW14. We show the $r-z$ color distribution, distance distribution, and Galactic spatial distribution of sources in Figure~\ref{fig:EMEsample}. The $r-z$ color distribution peaks at $r-z \approx 2$, which is equivalent to a dM4, which corresponds to the peak of the initial mass distribution \citep[$M_\ast \approx 0.125M_\odot$;][]{baraffe:1996:l51,chabrier:2003:763}. The distance distribution peaks at approximately 200 pc, which is consistent with other low-mass stellar samples from SDSS \citep[e.g.,][]{west:2011:97}.
	
	The candidates are fairly spread out within the SDSS footprint. To test for clumping of objects, we ran a friends-of-friends algorithm to test for spatial groupings within 10 pc of one another (see TW14 for further details). We found 10 pairs of stars within 10 pc of each other, with no other groupings larger than two stars. We tested each pair for similar 2-D kinematics (are moving together through the Galaxy) using Equation (6) from \citet{dhital:2010:2566}, given by:
\begin{equation}
\left( \frac{\Delta\mu_\alpha}{\sigma_{\Delta\mu_\alpha}}\right) + \left( \frac{\Delta\mu_\delta}{\sigma_{\Delta\mu_\delta}}\right) \leqslant 2,
\end{equation}
where $\Delta\mu_\alpha$ and $\Delta\mu_\delta$ are the differences between the two proper motion components for each pair, and their uncertainties are the quadrature sum of each individual proper motion uncertainty. The smallest value for this metric among the pairs was 5, indicating that none of these pairs showed similar 2-D kinematics. This indicates that these distances are more likely chance alignments than actual physical groupings. The catalog of candidates is available through the online journal and the column descriptions are listed in Table~\ref{tbl:catalogschema}.

\LongTables
\begin{deluxetable}{cll} 
\tabletypesize{\footnotesize}
\tablecolumns{3}
\tablecaption{Extreme MIR Excess Candidates Catalog Schema\label{tbl:catalogschema}}
\tablehead{
\colhead{Column} & \colhead{Column} & \colhead{Units} \\
\colhead{Number} & \colhead{Description} & 
}
\startdata
1	& SDSS Object ID & ... \\
2	& SDSS R.A. & deg. \\
3	& SDSS Decl. & deg. \\
4	& SDSS $u$-band PSF mag. & mag \\
5	& SDSS $u$-band PSF mag. error & mag \\
6	& SDSS $u$-band extinction & mag \\
7	& SDSS $u$-band unreddened PSF mag. & mag \\
8	& SDSS $g$-band PSF mag. & mag \\
9	& SDSS $g$-band PSF mag. error & mag \\
10	& SDSS $g$-band extinction & mag \\
11	& SDSS $g$-band unreddened PSF mag. & mag \\
12	& SDSS $r$-band PSF mag. & mag \\
13	& SDSS $r$-band PSF mag. error & mag \\
14	& SDSS $r$-band extinction & mag \\
15	& SDSS $r$-band unreddened PSF mag. & mag \\
16	& SDSS $i$-band PSF mag. & mag \\
17	& SDSS $i$-band PSF mag. error & mag \\
18	& SDSS $i$-band extinction & mag \\
19	& SDSS $i$-band unreddened PSF mag. & mag \\
20	& SDSS $z$-band PSF mag. & mag \\
21	& SDSS $z$-band PSF mag. error & mag \\
22	& SDSS $z$-band extinction & mag \\
23	& SDSS $z$-band unreddened PSF mag. & mag \\
24	& 2MASS $J$-band PSF mag. & mag \\
25	& 2MASS $J$-band PSF corr. mag. unc. & mag \\
26	& 2MASS $J$-band PSF total mag. unc. & mag \\
27	& 2MASS $J$-band SNR & ... \\
28	& 2MASS $J$-band $\rchi^2_\nu$ goodness-of-fit & ... \\
29	& 2MASS $J$-band extinction & mag \\
30	& 2MASS $J$-band unreddened PSF mag. & mag \\
31	& 2MASS $H$-band PSF mag. & mag \\
32	& 2MASS $H$-band PSF corr. mag. unc. & mag \\
33	& 2MASS $H$-band PSF total mag. unc. & mag \\
34	& 2MASS $H$-band SNR & ... \\
35	& 2MASS $H$-band $\rchi^2_\nu$ goodness-of-fit & ... \\
36	& 2MASS $H$-band extinction & mag \\
37	& 2MASS $H$-band unreddened PSF mag. & mag \\
38	& 2MASS $K_s$-band PSF mag. & mag \\
39	& 2MASS $K_s$-band PSF corr. mag. unc. & mag \\
40	& 2MASS $K_s$-band PSF total mag. unc. & mag \\
41	& 2MASS $K_s$-band SNR & ... \\
42	& 2MASS $K_s$-band $\rchi^2_\nu$ goodness-of-fit & ... \\
43	& 2MASS $K_s$-band extinction & mag \\
44	& 2MASS $K_s$-band unreddened PSF mag. & mag \\
45	& 2MASS photometric quality flag & ... \\
46	& 2MASS read flag & ... \\
47	& 2MASS blend flag & ... \\
48	& 2MASS contamination \& confusion flag & ... \\
49	& 2MASS extended source flag & ... \\
50	& \emph{WISE} $W1$-band PSF mag. & mag \\
51	& \emph{WISE} $W1$-band PSF mag. unc. & mag \\
52	& \emph{WISE} $W1$-band SNR & ... \\
53	& \emph{WISE} $W1$-band $\rchi^2_\nu$ goodness-of-fit & ... \\
54	& \emph{WISE} $W1$-band extinction & mag \\
55	& \emph{WISE} $W1$-band unreddened PSF mag. & mag \\
56	& \emph{WISE} $W2$-band PSF mag. & mag \\
57	& \emph{WISE} $W2$-band PSF mag. unc. & mag \\
58	& \emph{WISE} $W2$-band SNR & ... \\
59	& \emph{WISE} $W2$-band $\rchi^2_\nu$ goodness-of-fit & ... \\
60	& \emph{WISE} $W2$-band extinction & mag \\
61	& \emph{WISE} $W2$-band unreddened PSF mag. & mag \\
62	& \emph{WISE} $W3$-band PSF mag. & mag \\
63	& \emph{WISE} $W3$-band PSF mag. unc. & mag \\
64	& \emph{WISE} $W3$-band SNR & ... \\
65	& \emph{WISE} $W3$-band $\rchi^2_\nu$ goodness-of-fit & ... \\
66	& \emph{WISE} $W3$-band extinction & mag \\
67	& \emph{WISE} $W3$-band unreddened PSF mag. & mag \\
68	& \emph{WISE} $W4$-band PSF mag. & mag \\
69	& \emph{WISE} $W4$-band PSF mag. unc. & mag \\
70	& \emph{WISE} $W4$-band SNR & ... \\
71	& \emph{WISE} $W4$-band $\rchi^2_\nu$ goodness-of-fit & ... \\
72	& \emph{WISE} $W4$-band extinction & mag \\
73	& \emph{WISE} $W4$-band unreddened PSF mag. & mag \\
74	& \emph{WISE} contamination \& confusion flag & ... \\
75	& \emph{WISE} extended source flag & ... \\
76	& \emph{WISE} variability flag & ... \\
77	& \emph{WISE} photometric quality flag & ... \\
78	& \emph{Spitzer} IRAC Ch1 PSF flux density & $\mu$Jy \\
79	& \emph{Spitzer} IRAC Ch1 PSF flux density unc. & $\mu$Jy \\
80	& \emph{Spitzer} IRAC Ch2 PSF flux density & $\mu$Jy \\
81	& \emph{Spitzer} IRAC Ch2 PSF flux density unc. & $\mu$Jy \\
82	& \emph{Spitzer} IRAC Ch3 PSF flux density & $\mu$Jy \\
83	& \emph{Spitzer} IRAC Ch3 PSF flux density unc. & $\mu$Jy \\
84	& \emph{Spitzer} IRAC Ch4 PSF flux density & $\mu$Jy \\
85	& \emph{Spitzer} IRAC Ch4 PSF flux density unc. & $\mu$Jy \\
86	& \emph{Spitzer} MIPS Ch1 PSF flux density & $\mu$Jy \\
87	& \emph{Spitzer} MIPS Ch1 PSF flux density unc. & $\mu$Jy \\
88	& Proper motion in R.A. ($\mu_\alpha \cos \delta$) & mas yr$^{-1}$ \\
89	& Proper motion in Decl. & mas yr$^{-1}$ \\
90	& Total error in R.A. proper motion & mas yr$^{-1}$ \\
91	& Total error in Decl. proper motion & mas yr$^{-1}$ \\
92	& Full Sample Flag & ... \\
93	& Clean Sample Flag & ... \\
94	& Visual Quality Flag & ... \\
95	& Photometric distance & pc \\
96	& Distance from the Galactic plane & pc \\
97	& $\sigma^\prime$\tablenotemark{a} & ... \\
98	& $T_\mathrm{eff}$ estimate & K \\
99	& Upper $T_\mathrm{eff}$ limit & K \\
100	& Lower $T_\mathrm{eff}$ limit & K \\
101	& Log $g$ estimate & dex \\
102	& Upper Log $g$ limit & dex \\
103	& Lower Log $g$ limit & dex \\
104	& $\rchi_{12}$\tablenotemark{a} & ... \\
105	& $\rchi_{22}$\tablenotemark{a} & ... \\
106	& $L_\mathrm{IR} / L_\ast$ & ... \\
107	& $D_\mathrm{min}$ & AU \\
108	& $M_\mathrm{d}$ & $M_\mathrm{moon}$ \\
109	& $T_\mathrm{gr}$ & K \\
110	& $\sigma_{T_\mathrm{gr}}$ & K 
\enddata
\tablenotetext{a}{Defined in Section~\ref{irprops}.}
\end{deluxetable}

\subsection{Distance and Color (Temperature) Bias}\label{biases}

\begin{figure} 
\centering
\includegraphics[width=\linewidth]{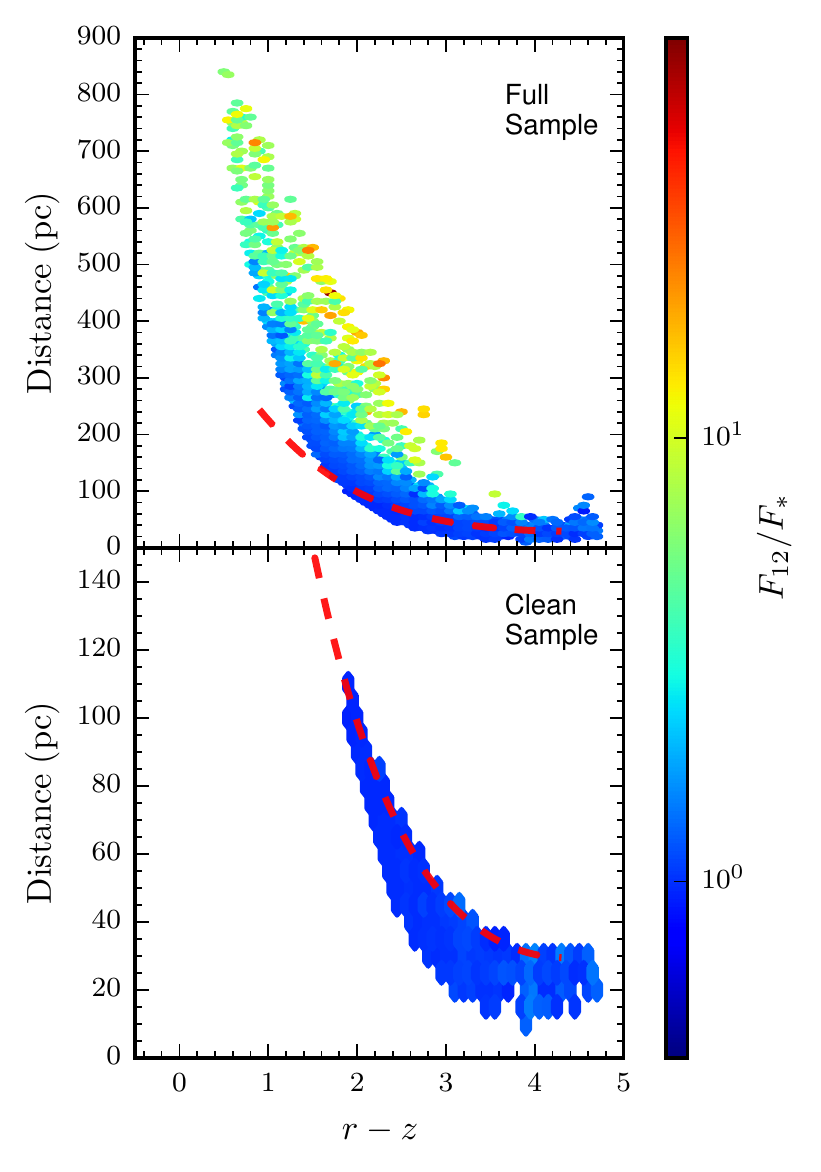}
\caption{Distance as a function of $r-z$ color for the full (top) and clean (bottom) samples. Each bin is 0.1 mag $\times$ 10 pc, and the color is the mean flux ratio (measurement/model) in the $W3$ band. 
The distances are compared to the estimated maximum distance corresponding to the $W3$ flux limit (see Section~\ref{wiselimits}) used for the clean sample (red dashed line). 
For the full sample, there is an inherent bias due to the distances for the bluest stars in the sample, requiring stars to exhibit large MIR excesses to be detected in $W3$. 
The clean sample is located much closer (within the bias distance limit), and should not have any significant bias.
\label{fig:rz_ratio}}
\end{figure}

	Due to SDSS being a magnitude limited survey, our selection of stars suffers a distance bias that is dependent on stellar effective temperature. For each stellar temperature range, there will be a minimum and maximum distance over which a dwarf star can be observed due to the saturation and faintness limits of SDSS, respectively. To explore where this bias occurs, we examined the flux ratios ($F_{12 \mu\mathrm{m,\; measured}} / F_{12 \mu\mathrm{m,\; model}}$) as a function of $r-z$ color and distance (Figure~\ref{fig:rz_ratio}). Figure~\ref{fig:rz_ratio} also shows the distance corresponding to the $W3$ flux limit (730 $\mu$Jy; see Section~\ref{wiselimits}). 

	For the full sample, the spread in distances are typically larger than the limit corresponding to the distance at which the photospheric flux level would be detectable at the $W3$ flux limit (dashed line). This makes many of the stars in the full sample undetectable (at this flux limit) unless they have a MIR excess (assuming no line-of-sight dependence on sensitivity). Figure~\ref{fig:rz_ratio} further illustrates that we can only detect the bluest stars in $W3$ if they have an extreme MIR excess, since their distances are too large to detect their photospheres at the $W3$ flux limit. This is true for some of the redder sources as well, but we have the ability to observe many of their photospheres at 12 \um. Due to the distance spread above the $W3$ flux limit distance in the full sample, there is a bias for which we must account. 
	
	The case is different for the clean sample, where the distance spread for all $r-z$ colors is closer than the distance corresponding to the $W3$ flux limit. Therefore, the clean sample should be free from a higher limit distance bias, unlike the full sample, but may suffer from a lower distance limit bias due to saturation. The clean sample also does not cover the same $r-z$ color range (a proxy for stellar temperature and mass) as the full sample, restricting its use for only mid- to late-spectral type low-mass stars. The distance bias will be accounted for using a Galactic model.

\section{LoKi Galactic Model: Estimating Stellar Counts and Proper Motions for Completeness}\label{model}

	A major limitation of the extreme MIR excess study completed by TW14 was a non-uniform sample, and no method to estimate completeness. To estimate the completeness of the current sample, we used a Galactic model to estimate how many stars were missing from the sample (e.g., within a local volume or along a line-of-sight). Galactic models have been used to simulate stellar densities \citep[e.g.,][]{juric:2008:864, van-vledder:2016:425}, kinematics \citep[e.g.,][hereafter D10]{ivezic:2008:287, dhital:2010:2566, dhital:2015:57}, or both \citep{robin:2003:523, sharma:2011:3}. Galactic models are typically comprised of three main components, the thin disk (cold component), the thick disk (warm component), and the halo. Each component is individually modeled in terms of its mixing fractions and kinematics. We created a model, dubbed the Low-mass Kinematics (\emph{LoKi}) galactic model\footnote{\url{https://github.com/ctheissen/LoKi}}, to estimate the total number of stars we would expect to observe within a given volume, and their respective kinematics. The model incorporates a luminosity function \citep[LF;][]{bochanski:2010:2679} to select stars in proportion to their abundance in the Galaxy, in addition to simulating their positions and kinematics. We ran 100 realizations of the model over the entire simulated volume, and kept only stars with significant proper motions (dependent on stellar color and line-of-sight; see Appendix~\ref{loki}) that would have been included in the MoVeRS sample. The methods involved in building and using \emph{LoKi} are described in detail in Appendix~\ref{loki}.

\subsection{Extreme MIR Excess Fractions}\label{mirfractionscolor}

	Using the larger photometric sample from MoVeRS and the \emph{LoKi} galactic model, we were able to extend the findings of TW14. Using \emph{LoKi}, we were able to explore the occurrence of extreme MIR excesses as a function of color (a proxy for stellar mass), and Galactic height (a proxy for stellar age). This was done by simulating the total number of stars expected to be observed within the given volume observed by SDSS. These simulations provide stellar counts and Galactic height distributions, which we used to investigate the occurrence of extreme MIR excesses in low-mass stars. 
		
	TW14 compared the stars with MIR excesses to the entire W11 catalog to calculate the fraction of stars exhibiting an extreme MIR excesses ($\sim$0.4\% of field M dwarfs exhibit an extreme MIR excess), or the ``extreme MIR excess fraction" (i.e., the ratio of the number of stars exhibiting an extreme MIR excess to the total number of stars). Using the same parent population selection criteria as TW14 (i.e., using all 390,006 stars with $J \leqslant 17$), we calculated a global extreme MIR excess fraction from the MoVeRS sample of $\sim$0.1\%. However, because MoVeRS is not a volume complete catalog, these fractions are likely overestimates and need to be corrected using a Galactic model. In addition, as described in Section~\ref{nonexcess}, we exclude a number of potentially real extreme MIR excesses. Without the ability to determine which of these stars harbor true excesses, as they fall within the statistical scatter of the parent population, the results in this section should be taken as lower limits.
	
	We used the \emph{LoKi} galactic model to simulate the number of stars expected in the observed footprint (see Appendix~\ref{loki} for details), and their distribution in the Galaxy. Using the model, we computed volume complete fractions, i.e., estimated the denominator value for the number of stars for which we should have been able to detect an extreme MIR excess. We computed the global extreme MIR excess fraction from the model stellar counts using the mean value of the stellar counts across all 100 simulations, estimating an extreme MIR excess fraction of $\sim$0.02\%. The model complete MIR excess fraction is an order of magnitude smaller than that found by TW14, but still orders of magnitude larger than the extreme MIR excess fraction estimated for A--G type stars by \citet[$\sim$0.0007\%;][]{weinberger:2011:72}. We will discuss this further in Section ~\ref{discussion}.
	
	Galactic height is strongly correlated with stellar age for ensembles of stars. This is due to the fact that stars are born close to the Galactic plane, and, over time, are dynamically heated away from the plane \citep[e.g.,][]{west:2006:2507,west:2008:785}. This method of assigning ages to ensembles of stars based on absolute distance from the Galactic plane is commonly referred to as ``Galactic stratigraphy" \citep{west:2015:3}.
	
	TW14 identified a weak trend of decreasing MIR excess fractions as a function of increasing stellar age. However, their sample was small and incomplete. To further investigate the findings of TW14, we computed MIR excess fractions using stars with extreme MIR excesses (\fullsampleE\ stars in the full sample and \cleansampleE\ stars in the clean sample, Section~\ref{EME}; numerator value), and model stellar counts (denominator value) over the same volume as the SDSS observations, and with proper motions detectable by MoVeRS (dependent on stellar color and line-of-sight; see Appendix~\ref{loki}). Figure~\ref{fig:zfractions} shows the model corrected extreme MIR excess fractions as a function of absolute distance from the Galactic plane ($Z$). Each bin has two points corresponding to the 1st and 99th percentile values across all model runs, with error bars representing the greatest and smallest binomial errors between the two percentiles. The fact that much of the sample is not at low Galactic latitudes should result in very few young stars. The estimated ages from Section~\ref{spectroscopic}, and the results from TW14, suggest that the vast majority of stars within SDSS at high Galactic latitudes are members of the field population ($\gg$100 Myr). Figure~\ref{fig:zfractions} shows a declining trend with Galactic height, with the majority of stars with extreme MIR excesses found within 100 pc of the Galactic plane. To assess the statistical significance of this trend, we performed a least-squares linear fit (of the form $y = mx + b$) to the average fraction for each bin, weighted by the average binomial uncertainty, finding a slope of $m = (-6.836 \pm 1.468)\times10^{-7}$ pc$^{-1}$. This indicates that younger field populations are more likely to have extreme MIR excesses, and that stars are less likely to host extreme MIR excesses as they age \citep[using ``Galactic stratigraphy";][]{west:2006:2507, west:2008:785}. This also indicates that there is some typical age after which the mechanism responsible for creating an extreme MIR excess ceases to act.

\begin{figure}
\centering
\includegraphics[width=\linewidth]{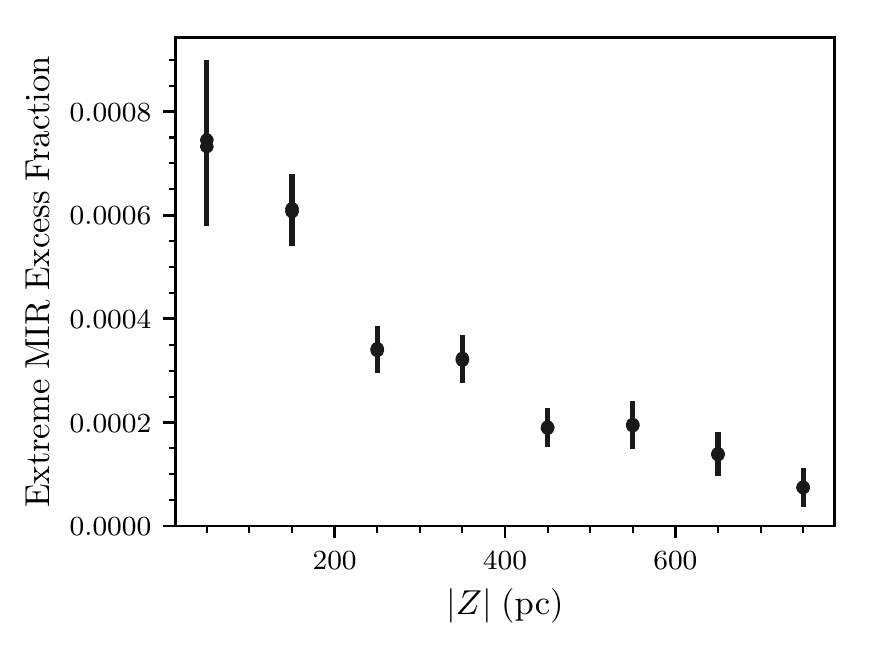}
\caption{The fraction of stars exhibiting an extreme MIR excess as a function of absolute distance from the Galactic plane (a proxy for stellar age). Points with error bars represent the model corrected completeness values for the 1$^\mathrm{st}$ and 99$^\mathrm{th}$ percentile values for each bin (from the 100 model realizations) with binomial errors. We see a steady decline in MIR excess fraction away from the Galactic plane, which has been shown to strongly correlate with age \citep[e.g.,][]{west:2008:785}. This trend indicates that younger field populations are more likely to have extreme MIR excesses, with the likelihood of hosting an extreme MIR excess decreasing as a function of increasing stellar age. This also indicates that the mechanism responsible for creating extreme MIR excesses ceases after some typical stellar age.
\label{fig:zfractions}}
\end{figure}
	
\begin{figure}
\centering
\includegraphics[width=\linewidth]{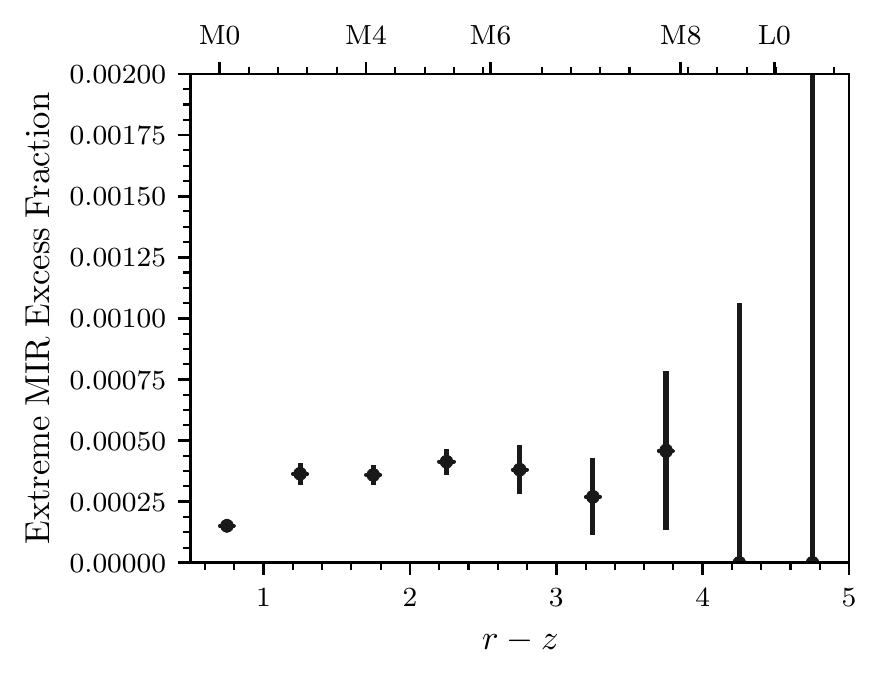}
\caption{The fraction of stars exhibiting an extreme MIR excess as a function of color (a proxy for stellar mass). Points and error bars are the same as Figure~\ref{fig:zfractions}. We see a relatively flat MIR excess fraction for all stellar colors (stellar masses), possibly indicating that the mechanism for creating MIR excesses is independent of stellar mass. Each of these bins samples different volumes, which accounts for the lack of MIR excesses in the reddest bins due to smaller volumes, and hence fewer stellar counts. This also implies an age bias since the bluest bins tend to be at farther distances (older populations) than the redder bins. Approximate spectral types taken from \citet{hawley:2002:3409} and \citet{bochanski:2007:531}.
\label{fig:colorfractions}}
\end{figure}

	TW14 did not attempt to examine a stellar mass dependence with MIR excess fractions. However, with the larger sample of extreme MIR excess candidates and the Galactic model, we were able to examine the MIR excess fractions as a function of $r-z$ color (a proxy for stellar mass). Figure~\ref{fig:colorfractions} shows the fraction of stars exhibiting an extreme MIR excess as a function of $r-z$ color. Again we fit a linear function to the trend and found a slope of $m = (1.486 \pm 0.424) \times 10^{-4}$ pc$^{-1}$, indicating an upward trend. There is a slight distance (and hence age) bias in Figure~\ref{fig:colorfractions}, as bluer stars tend to be at greater distances (older) than redder stars. This effect is due to SDSS observing primarily out of the plane of the Galaxy, which makes distance strongly correlated with vertical distance from the Galactic plane \citep[e.g., see][]{bochanski:2010:2679}. Furthermore, the vertical distribution of stars from the Galactic plane is strongly correlated with stellar age \citep{ma:2016:}, with older stellar populations found farther from the Galactic plane on average. Considering the upward trend with redder colors, this is consistent with Figure~\ref{fig:zfractions}, as younger stellar populations tend to have larger extreme MIR excess fractions.
	
\begin{figure*}
\centering
\includegraphics[width=\linewidth]{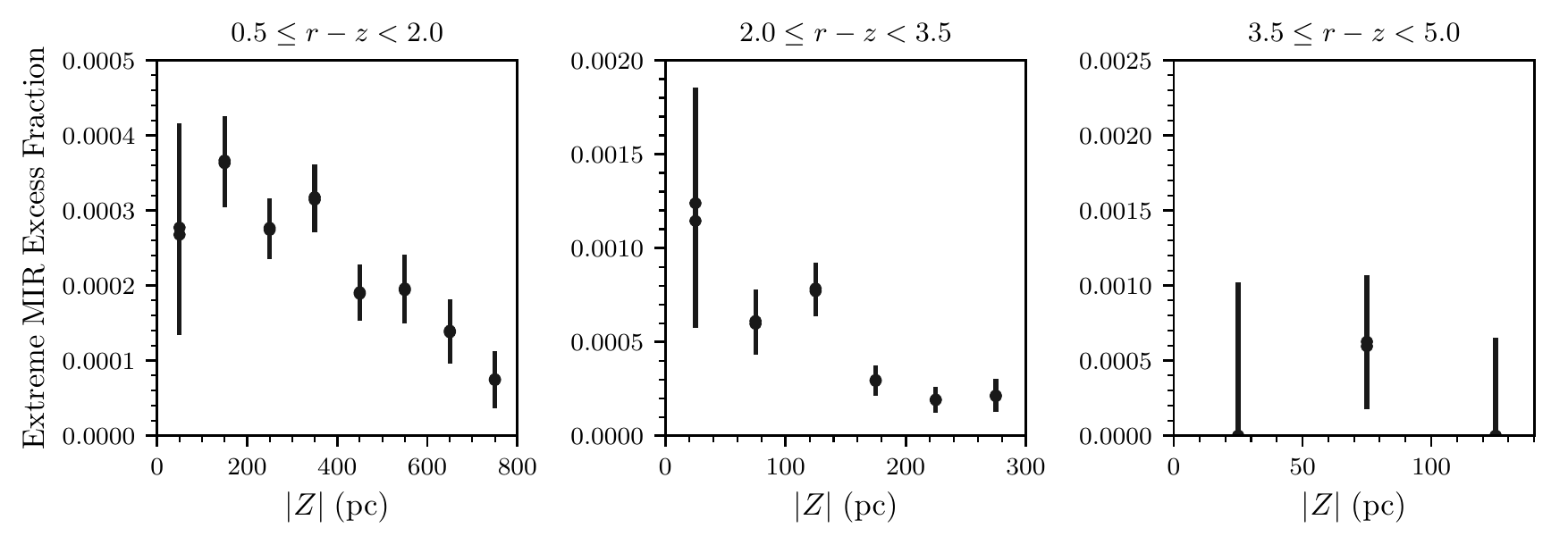}
\caption{The fraction of stars exhibiting an extreme MIR excess as a function of absolute distance from the Galactic plane in color bins. We see a declining trend in MIR excess fractions with Galactic height. The bluest bin suffers from a selection effect due to the majority of these stars being at relatively large distances (which is strongly correlated with distance from the Galactic plane), we are missing many of the stars that actually reside close to the Galactic plane due to the saturation limits of SDSS. The reddest bin does not sample a large enough volume to detect a larger number of stars with MIR excesses if they occur at similar rates across the stellar mass regime.
\label{fig:zcolorfractions}}
\end{figure*}

	To minimize selection effects and explore the interplay among extreme MIR excess fractions, stellar age, and stellar mass, we examined extreme MIR excess fractions as a function of absolute distance from the Galactic plane binned in three $r-z$ color regimes (Figure~\ref{fig:zcolorfractions}). The first bin ($0.5 \leqslant r - z < 2$) potentially suffers from selection effects due to the inherently large distances to these objects, dictated by the saturation limit of SDSS (see Figure~\ref{fig:rz_ratio}), placing the majority of observed stars farther away from the Galactic plane (76\% with $|Z| > 200$ pc). Although the model attempts to recover some fraction of these stars, we implemented the same magnitude and proper motion cuts on the model sample, therefore both the model and our sample will suffer from a similar selection effect. The intermediate mass stars within the sample ($2 \leqslant r-z < 3.5$) show a slight trend with $|Z|$, and these bins are likely to be relatively free from the selection effects affecting the other mass bins. The lowest mass bin ($3.5 \leqslant r-z < 5$) has very few sources and likely does not sample a large enough volume to detect MIR excesses if excesses occur at similar rates across all stellar masses. The measured best-fit slopes for all three color bins from bluest to reddest are $m = (-4.254 \pm 0.788) \times 10^{-7}$ pc$^{-1}$, $m = (-2.683 \pm 1.389) \times 10^{-6}$ pc$^{-1}$, and $m = (-3.358 \pm 16.809) \times 10^{-6}$ pc$^{-1}$.

\section{Non-significant MIR Excesses Revisited: A Further Investigation into Timescales}\label{nonexcess2}

\begin{figure}
\centering
\includegraphics[width=\linewidth]{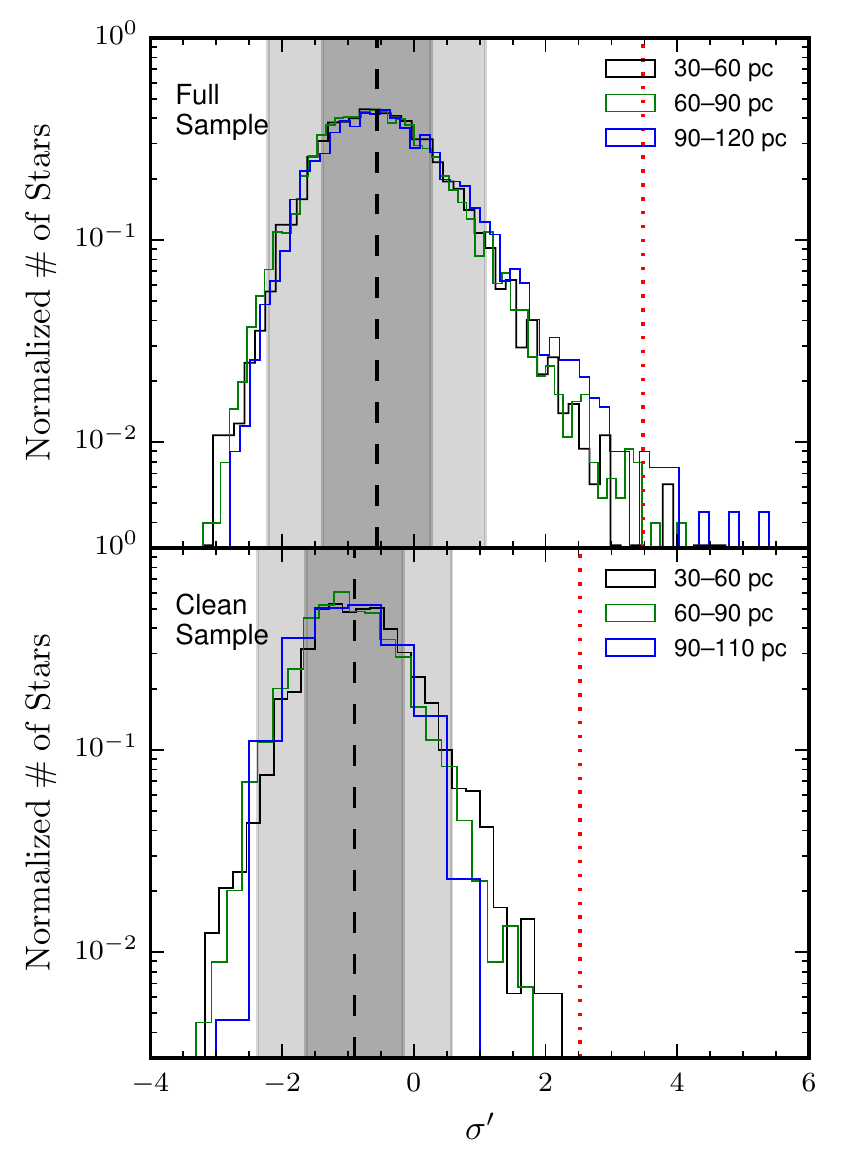}
\caption{Normalized distributions of $\sigma^\prime$ values as a function of $|Z|$ for stars with $2 \leqslant r-z < 3.5$. Dashed lines, dotted lines, and shaded regions are the same as Figure~\ref{fig:sigprimeresiduals}. The nearest bin (0--30 pc) has been omitted due to a bias from the SDSS saturation limit. The full sample shows a slight shift to higher $\sigma^\prime$ values at larger Galactic heights. This is likely due to a bias as fewer stars without MIR excesses are visible at distances greater than 100 pc. The clean sample shows the longest tail for the 30--60 pc bin, indicating a possible dependence on age for the stars with smaller MIR excesses.
\label{fig:sigprimedist}}
\end{figure}

	The strong trend of decreasing extreme MIR excess fraction with Galactic height indicates a trend with stellar age, and motivates further investigation. To explore if the overall distribution of non-significant excess sources changes as a function of age, we examined the $\sigma^\prime$ distribution as a function of $|Z|$ for the full and clean samples, using stars with $2 \leqslant r-z < 3.5$ to minimize selection effects due to distance. Figure~\ref{fig:sigprimedist} shows how the distribution of $\sigma^\prime$ changes as a function of $|Z|$. 
	
	To assess if there is a significant difference between the distributions in both the full and clean samples, we investigated the skew of each sample distribution. The underlying hypothesis is that all samples come from a nearly Gaussian parent distribution, with the stars with excess skewing that parent population to more positive $\sigma^\prime$ values. To statistically assess the skew of each distribution, we took 100,000 bootstrap samples of each distribution and measured the skew of the resulting distribution. We report the mean values along with the 68\% (16th and 84th percentiles) and 95\% (2.5th and 97.5th percentiles) confidence intervals in Table~\ref{tbl:skew}. The full sample shows a trend towards more excess sources (larger skewness) at farther distances away from the Galactic plane. This is most probably due to the fact that at larger distances, we are more sensitive to stars with excesses. 	
		
	The clean sample should be devoid of selection effects associated with distance at the expense of a smaller spread in Galactic height. In Figure~\ref{fig:sigprimedist} we see a decrease in the number of high $\sigma^\prime$ sources (MIR excess sources) at higher Galactic heights, which is also illustrated by the decreasing skew in Table~\ref{tbl:skew}, although the observed decrease is a tentative result. The decrease in skewness would be consistent with there being age evolution in all of the stars with MIR excesses, not only stars exhibiting extreme MIR excesses.

\begin{deluxetable}{lcc}
\tabletypesize{\footnotesize}
\tablecolumns{3}
\tablecaption{Sample Skewness\label{tbl:skew}}
\tablehead{
\colhead{Sample} & \colhead{Distance Range} & \colhead{Skewness\tablenotemark{a}} 
}
\startdata
Full &	30--60 pc		& $0.72^{+0.08 (0.15)}_{-0.08 (0.16)}$	\\
Full &	60--90 pc		& $0.83^{+0.07 (0.13)}_{-0.07 (0.14)}$	\\
Full &	90--120 pc	& $1.07^{+0.07 (0.14)}_{-0.07 (0.14)}$	\\
Clean &	30--60 pc		& $0.39^{+0.09 (0.17)}_{-0.09 (0.20)}$	\\
Clean &	60--90 pc		& $0.34^{+0.07 (0.13)}_{-0.07 (0.14)}$	\\
Clean &	90--110 pc	& $0.18^{+0.08 (0.16)}_{-0.08 (0.17)}$	
\enddata
\tablenotetext{a}{Confidence intervals correspond to the 68\% confidence and the 95\% confidence (inside parenthesis).}
\end{deluxetable}

\section{Conclusions and Discussion}\label{discussion}

	The large sample of low-mass stars contained within the MoVeRS catalog has allowed us to compile the largest sample of low-mass field stars exhibiting large MIR excesses to date (\fullsampleE\ stars). We examined the dependence of MIR excess occurrence with stellar mass (using $r-z$ color as a proxy), and stellar age (using Galactic height as a proxy). The sample is divided into a ``full" sample (\fullsampleE\ stars), consisting of stars with high-fidelity, high-significance MIR excess detections, and a ``clean" sample (\cleansampleE\ stars), which also contains high-fidelity, high-significance stars with excesses, but is magnitude (volume) limited. 
	
	To build the samples, we implemented cuts to ensure relatively bright sources, with high S/N \emph{WISE} observations. These stars were then visually inspected to reduce contaminants (e.g., crowded fields). The final samples, including both stars with and without excesses, were made up of \fullsample\ stars (full sample; \fullsampleE\ stars with extreme MIR excesses) and \cleansample\ stars (clean sample; \cleansampleE\ stars with extreme MIR excesses). Stars with extreme MIR excesses were selected using modified empirical criteria from TW14. A cross-match to the \emph{Spitzer} Enhanced Imaging Products catalog identified 10 stars and verified the \emph{WISE} MIR excesses. The full sample covers the range $0.5 \leqslant r-z < 5$, covering all spectral-sub types within the M dwarf regime ($0.1M_\odot \lesssim M_\ast \lesssim 0.7M_\odot$). The clean sample is biased towards later-spectral type stars ($2 \leqslant r-z < 5$; $0.1M_\odot \lesssim M_\ast \lesssim 0.35M_\odot$), and was chosen to minimize biases due to distance/magnitude and \emph{WISE} sensitivity. 
	
	Spectroscopic observations of 25 stars in the sample taken by SDSS and using the DCT support the hypothesis that the sample is made up of field stars and confirms the selection of M dwarfs, although one star has characteristics similar to a carbon dwarf, indicating a contamination rate of $\sim$4\%. Many carbon stars are known to show evidence for circumstellar material \citep{green:2013:12}, potentially making us more likely to select for them in this study, and indicating that the contamination rate for the MoVeRS catalog is likely much less than 4\%. For the remainder of the stars with spectra, the vast majority lack H$\alpha$ emission, consistent with an inactive, older ($\gg$100 Myr), field population. Furthermore, none of the stars have measurable Li \textsc{i} absorption, expected for stars with ages $< 100$ Myr. Since the magnetic activity lifetimes of lower-mass stars are one to several Gyrs, and none of the stars had detectable Li \textsc{i} absorption, the parent population likely has an average age $> 1$ Gyr. The samples and their derived quantities are available in the electronic format of this manuscript.
	
	Our primary finding is that there is a strong correlation with the fraction of field stars exhibiting an extreme MIR excess as a function of absolute distance from the Galactic plane. Although the bins with higher-mass stars suffer selection effects and are biased towards stars farther away from the Galactic plane (due to the brightness of these stars and the saturation limits of SDSS), and the lowest-mass stars are biased towards extremely close distances, and therefore small volumes, we find a significant decreasing trend for stars with MIR excesses at larger Galactic heights, specifically in the intermediate-mass stars, which are largely unbiased. These data strongly support an age dependency on the presence of extreme MIR excesses. We also find that MIR excesses have a correlation with $r-z$ color, indicating a possible dependence with stellar mass.
	
	Giant collisions between large planetismals or terrestrial planets are expected to create a collisional cascade that may last for $\sim$100,000 years \citep{weinberger:2011:72}. If we assume a typical stellar age for the sample of 1 Gyr, and a timescale over which a MIR excess can be detected of 0.1 Myr, then only 0.01\% of the sample should show a detectable excess, which reduces to $\sim$0.5 stars for the clean sample, roughly consistent with our findings. This is assuming a volume complete sample, and the ability for the mechanism creating MIR excesses to act at anytime during the lifetime of the star. Limiting the timescale over which the mechanism can act (to less than 1 Gyr), or increasing the lifetime of the collisional products would increase the number of predicted stars observed to have an extreme MIR excess. Although we are unable to link a distinct timescale over which a collision may occur, our findings are consistent with a short lifetime for the collisional cascade to create enough dust for a significant MIR detection. Additionally, multiple collisions can extend the lifetime of the collisional products past 100,000 years.
	
	Using the clean sample, which is relatively unbiased and complete, we reinvestigated the collision rate found in TW14. The estimated fraction of stars undergoing collisions is $(3.5 \pm 1.7) \times 10^{-4}$, an order of magnitude smaller than the TW14 value. However, when we consider the different selection criteria for the parent population (34\%, from Section~\ref{irprops}), and the more stringent criteria applied for a star to be included in the extreme MIR excess sample (16\%, from Section~\ref{irprops}), we find the TW14 fraction of 0.4\% is reduced to 0.02\%, consistent with this study. This fraction is still two orders of magnitude larger than the number estimated by \citet[$\sim 7 \times 10^{-6}$;][]{weinberger:2011:72} for A--G spectral type stars. Our updated fraction gives us a collision rate of $\sim$9 impacts per star up to its current age. This value is consistent with the findings of TW14 that planetary collisions occur more frequently around low-mass stars. 
	
	Investigating the continuous distribution of stars with excess MIR flux, versus simply the high-significance sample, we estimate there are potentially 80 stars with actual extreme MIR excesses excluded from our full sample, and one star excluded from the clean sample. Non-extreme MIR excesses may represent the more evolved state of the aforementioned collisional disks, at the end of the lifetime for a collisional cascade where the disk is becoming optically thin, or perhaps smaller collisions. The addition of these stars would imply the estimated fraction of stars undergoing collisions is underestimated by a factor of $\sim$4. Indicating that collisions may be even more frequent in low-mass stellar systems.
	
	Planetary collisions have also been put forth to explain a dichotomy found in the \emph{Kepler} data. \emph{Kepler} has found a wealth of planetary systems around low-mass stars, both singly-transiting systems and multi-transiting systems. Numerous studies have used ensemble statistics to reproduce \emph{Kepler} multi-planet observations with success \citep{lissauer:2011:8,fang:2012:92,tremaine:2012:94,fabrycky:2014:146}. However, as noted by \citet{lissauer:2011:8}, the best fitting models under-predict the number of observed singly-transiting systems by a factor of $\sim$2. \citet{lissauer:2011:8} postulate that a second population of systems with higher inclination dispersions and/or lower multiplicities may explain the dearth of singly-transiting systems. This proposed dual population has become known as the ``\emph{Kepler} dichotomy."
	
	 Recently, \citet{ballard:2016:66} simulated planetary systems with a range of mutual inclinations and multiplicities to replicate \emph{Kepler} results for the M dwarf population. \citet{ballard:2016:66} found that a high multiplicity ($N \approx 7$ planets per star) with a typical mutual inclination of 2$^\circ$ could produce a planetary population in good agreement with the \emph{Kepler} multi-planet yield, both with and without invoking a range of eccentricities. \citet{ballard:2016:66} accounted for the dearth of singly-transiting systems by invoking a second population of planetary systems, either with a single planet, or with 2--3 planets and a large scatter in mutual inclination (4$^\circ$--9$^\circ$). The best mixture between these two populations was found to be $\sim$50\%.
	 
	 \citet{ballard:2016:66} discuss two possible explanations for the \emph{Kepler} dichotomy, initial formation conditions and dynamical disruption. In the former of these scenarios, \citet{johansen:2012:39} posit that, for the case of Solar-mass stars, the formation, migration, or scattering of a giant planet could suppress planet formation. This is a scenario similar to the Grand Tack model \citep{walsh:2011:206}, which was put forward to explain the anomalously low mass of Mars in our own solar system. However, the lack of massive planets found orbiting most low-mass stars makes this an unlikely scenario. \citet{moriarty:2016:34} used $N$-body simulations of late stage planet formation to attempt to reproduce \emph{Kepler} observations, and found that two separate disk surface mass densities could reproduce the dichotomy. However, it is unclear if two distinct surface density profiles are observationally motivated.
	 
	 Dynamical disruption as an explanation for the \emph{Kepler} dichotomy has also been explored through the use of models. Simulations of tightly packed planetary systems \citep{pu:2015:44, volk:2015:l26} have shown that coplanar, high-multiple planetary systems are metastable, and are disrupted on Gyr timescales. Furthermore, in systems that experience dynamical instability, the most likely outcome is two planets colliding once they are excited to crossing orbits \citep{pu:2015:44}. Such collisions would likely result in massive amounts of orbiting dust, and potentially planets scattered to higher inclinations. Combined with the findings of \citet{quintana:2016:404}, that suppression of giant planets can extend the timescale over which collisions can occur to Gyrs, late-time occurring giant impacts are a plausible explanation for the \emph{Kepler} dichotomy.
	 
	 Our observed extreme MIR excesses support the hypothesis that the \emph{Kepler} dichotomy arises from late occurring ($>$ 1 Gyr) giant impacts due to dynamical disruption. Planetary collisions between orbiting planets with small semi-major axes would produce the massive dust populations inferred from these extreme MIR excesses. The high frequency of these impacts (relative to higher-mass stars) has strong implications on the habitability of terrestrial planets around low-mass stars. This analysis motivates the search for similar extreme MIR excesses in higher- and lower-mass stellar populations. 

	The upcoming \emph{Transting Exoplanet Survey Satellite} \citep[\emph{TESS};][]{ricker:2014:914320} will be instrumental in testing the evolution versus formation hypothesis for the \emph{Kepler} dichotomy through a larger sample of low-mass stars than \emph{Kepler} observed. \emph{TESS}, and to a lesser extent the \emph{Kepler} two-wheel mission (K2), will sample a larger distribution in Galactic height and rotation periods (both tracers of stellar age) to further estimate the timescale over which planetary collisions occur. Additionally, the upcoming \emph{James Webb Space Telescope} \citep[\emph{JWST};][]{gardner:2006:485} will allow us to constrain the mineralogy of the disks detected with \emph{WISE}, which can distinguish disks formed through violent collisions versus disks made of differentiated bodies, such as asteroids.

\acknowledgments

	The authors would like to thank the anonymous referee for extremely helpful comments and suggestions which greatly improved the manuscript.
	The authors would like to thank Adam Burgasser, Aurora Kesseli, Daniella Bardalez Gagliuffi, Julie Skinner, Saurav Dhital, Dylan Morgan, and Sebastian Pineda for their helpful discussions. 
	C.A.T. would like to acknowledge the Ford Foundation for his financial support. 
	A.A.W acknowledges funding from NSF grants AST-1109273 and AST-1255568. 
	A.A.W. and C.A.T. further acknowledge the support of the Research Corporation for Science Advancement's Cottrell Scholarship.
	This material is based upon work supported by the National Aeronautics and Space Administration under Grant No. NNX16AF47G issued through the Astrophysics Data Analysis Program.

	Funding for the Sloan Digital Sky Survey IV has been provided by
the Alfred P. Sloan Foundation, the U.S. Department of Energy Office of
Science, and the Participating Institutions. SDSS-IV acknowledges
support and resources from the Center for High-Performance Computing at
the University of Utah. The SDSS web site is www.sdss.org.

SDSS-IV is managed by the Astrophysical Research Consortium for the 
Participating Institutions of the SDSS Collaboration including the 
Brazilian Participation Group, the Carnegie Institution for Science, 
Carnegie Mellon University, the Chilean Participation Group, 
the French Participation Group, Harvard-Smithsonian Center for Astrophysics, 
Instituto de Astrof\'isica de Canarias, The Johns Hopkins University, 
Kavli Institute for the Physics and Mathematics of the Universe (IPMU) / 
University of Tokyo, Lawrence Berkeley National Laboratory, 
Leibniz Institut f\"ur Astrophysik Potsdam (AIP), 
Max-Planck-Institut f\"ur Astronomie (MPIA Heidelberg), 
Max-Planck-Institut f\"ur Astrophysik (MPA Garching), 
Max-Planck-Institut f\"ur Extraterrestrische Physik (MPE), 
National Astronomical Observatory of China, New Mexico State University, 
New York University, University of Notre Dame, 
Observat\'ario Nacional / MCTI, The Ohio State University, 
Pennsylvania State University, Shanghai Astronomical Observatory, 
United Kingdom Participation Group,
Universidad Nacional Aut\'onoma de M\'exico, University of Arizona, 
University of Colorado Boulder, University of Oxford, University of Portsmouth, 
University of Utah, University of Virginia, University of Washington, University of Wisconsin, 
Vanderbilt University, and Yale University.
 
	This publication makes use of data products from the Two Micron All Sky Survey, which is a joint project of the University of Massachusetts and the Infrared Processing and Analysis Center/California Institute of Technology, funded by the National Aeronautics and Space Administration and the National Science Foundation. This publication also makes use of data products from the \emph{Wide-field Infrared Survey Explorer}, which is a joint project of the University of California, Los Angeles, and the Jet Propulsion Laboratory/California Institute of Technology, funded by the National Aeronautics and Space Administration. 
	
	These results made use of Lowell Observatory's Discovery Channel Telescope. Lowell operates the DCT in partnership with Boston University, Northern Arizona University, the University of Maryland, and the University of Toledo. Partial support of the DCT was provided by Discovery Communications. 
	
	The authors are also pleased to acknowledge that much of the computational work reported on in this paper was performed on the Shared Computing Cluster which is administered by Boston University's Research Computing Services (\url{www.bu.edu/tech/support/research/}). This research made use of Astropy, a community-developed core Python package for Astronomy \citep{astropy-collaboration:2013:a33}. Plots in this publication were made using Matplotlib \citep{hunter:2007:90}.

\appendix

\section{Estimating Stellar Parameters}\label{stellarparams}

\subsection{Markov Chain Monte Carlo Method for Stellar Parameters}\label{mcmc}

	We calculated the parameters of the orbiting dust ($D_\mathrm{dust}$ and $M_\mathrm{dust}$) using our estimates of the fundamental stellar parameters ($T_\mathrm{eff}$ and $R_\ast$). We estimated stellar parameters using the BT-Settl models with solar abundances from \citet{caffau:2011:255}, and mixing lengths calibrated on 2-D/3-D radiative hydrodynamic simulations \citep[CIFIST2015;][]{freytag:2010:a19, freytag:2012:919, baraffe:2015:a42}. These models span temperatures ranging between 1200K--7000K in steps of 100K or 50K, dependent on surface gravity, and log $g$ values between 2.5--5.5 in steps of 0.5 dex, with metallicities and alpha abundances set to solar values. Using a previous version of the CIFIST models, \citet{mann:2013:188} found that the deviation between temperatures based on model comparisons to optical spectra and those derived empirically was 57 K.
	
	To produce the best model fits to stellar data requires probing parameter space to fit for $T_\mathrm{eff}$, [M/H], log $g$, $\alpha$-abundance, and the normalizing factor in the form of the square of the ratio of the stellar radius over the distance (i.e., $F_\lambda \propto L_\lambda/d^2$). To reduce the parameter space for fitting models to the millions of stars in the MoVeRS sample, a few basic assumptions were made that should not overly bias our results. Metallicity was set to solar abundances, removing this parameter from the search space. To further reduce the complexity of the algorithm, the normalization factor was removed from the parameter space by scaling the model fluxes to the measured $z$-band values (a similar process was used in TW14 using the $K_s$-band), leaving only two parameters for which to solve ($T_\mathrm{eff}$ and log $g$).
	
	We used the \emph{emcee} package \citep{foreman-mackey:2013:306}, a Python implementation of the \citet{goodman:2010:65} affine invariant sampler, to explore the remaining stellar parameter space. Since the BT-Settl models are not continuous across the parameter space, we interpolated between grid points using a nearest-neighbor method for model selection. For each step in the MCMC, the log-likelihood is given as,
\begin{equation}
\ln \mathcal{L}(\bm{\Theta}|\bm{X, \sigma}) = -\frac{1}{2}\sum_{n=1}^N \left [ \frac{(\Theta_n - X_n)^2}{\sigma_n^2} + \ln (2 \pi \sigma_n^2) \right ],
\end{equation}
where $\bm{\Theta}$ is a vector of length $N$ containing the model predicted, scaled fluxes for a given set of stellar parameters ($T_\mathrm{eff}$ and log $g$), $\bm{X}$ is a vector containing the observed fluxes, $\bm{\sigma}$ is a vector containing the measurement errors for the observed fluxes, and the length $N$ pertains to the number of bands in which data were available. Uniform priors were chosen across the parameter space, and assumed all the parameters were normally distributed.

\begin{figure*}
\centering
 \includegraphics[width=\linewidth]{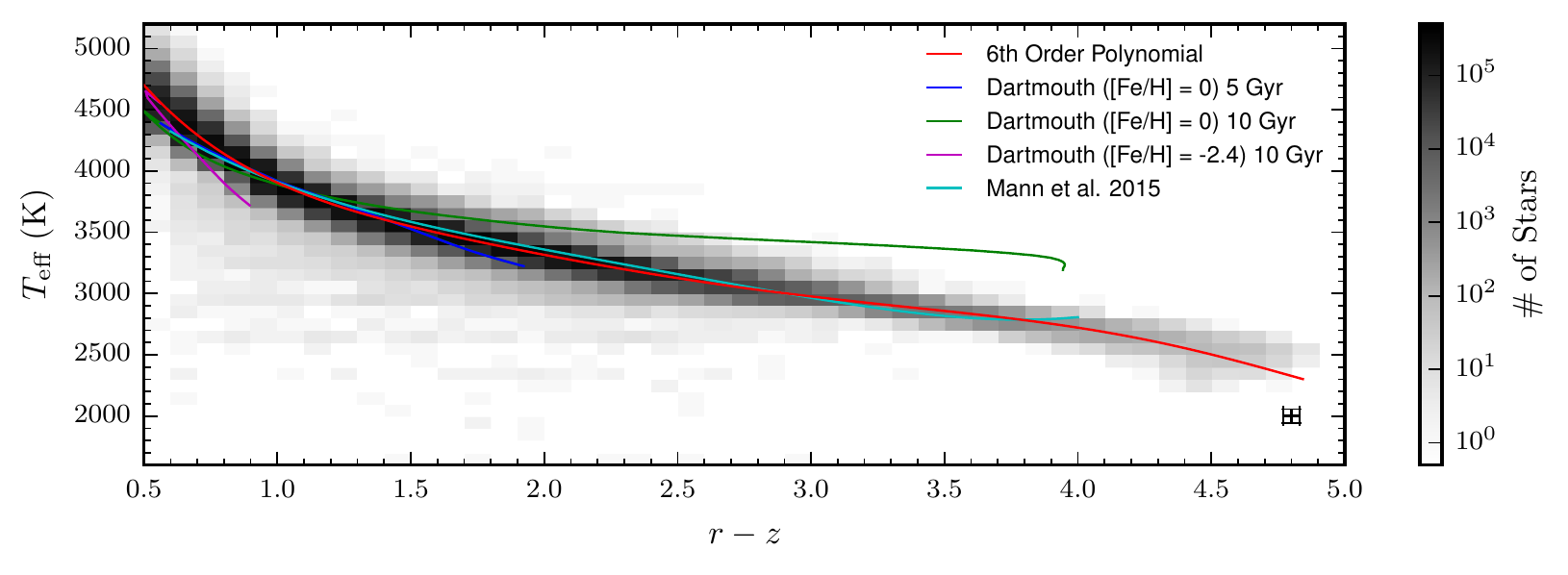}
\caption{Effective temperature as a function of $r-z$ color from the MCMC estimation. Each bin is 0.1 mag $\times$ 100 K. Typical errors are shown in the bottom right corner. We plot the best-fit 6$^\mathrm{th}$ order polynomial along with relationships from the Dartmouth Stellar Evolution Database \citep{dotter:2008:89, feiden:2013:183} and \citet{mann:2015:64}. Most relationships fail to replicate the reddest, or coolest, end of the main sequence.
\label{fig:modelparams}}
\end{figure*}

	Instead of collecting the entire posterior probability distributions for each of the stars, we calculated the 16th, 50th, and 84th percentiles of the distributions for both $T_\mathrm{eff}$ and log $g$. We plot the 50th percentile values as a function of $r-z$ color in Figure~\ref{fig:modelparams}. The $T_\mathrm{eff}$ estimates follow the expected trend with $r-z$ color. The width of the distribution is likely due to different metallicity classes \citep{mann:2015:64, schmidt:2016:2611}. Using an $F$ test, we compared different order polynomial relationships, and found the best-fit to the observed trend between $T_\mathrm{eff}$ and $r-z$ was a 6$^\mathrm{th}$ order polynomial,
\begin{equation}\label{eqn:poly}
T_\mathrm{eff} = a+bX+cX^2+dX^3+eX^4+fX^5+gX^6,
\end{equation}
where the coefficients are listed in Table~\ref{tbl:coeffs}. We find good agreement between our relationship and \citet{mann:2015:64}, except for the extremes where the \citet{mann:2015:64} fits are not well constrained.

\subsection{Estimating Stellar Radii}\label{radii}

\begin{figure}
\centering
 \includegraphics[width=\linewidth]{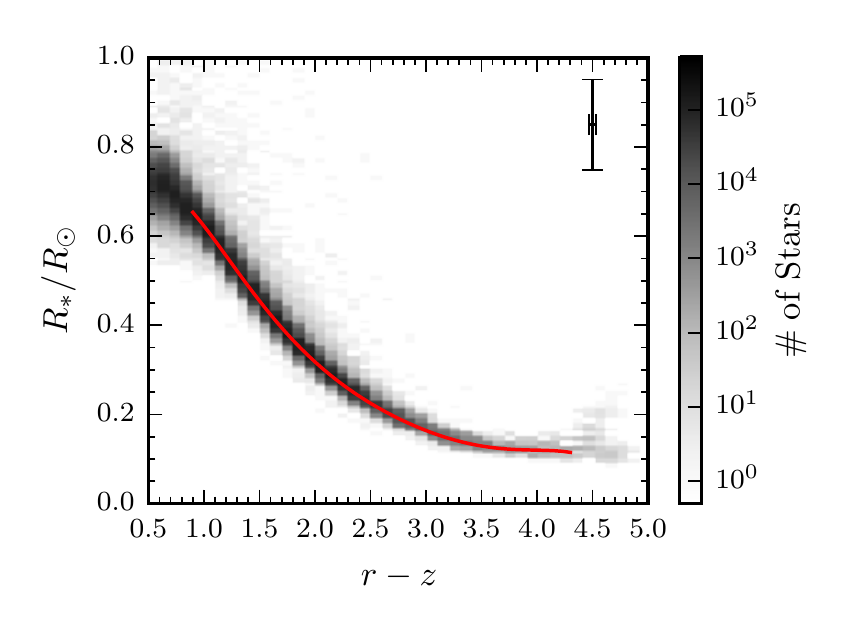}
\caption{$R_\ast$ as a function of $r-z$ color from the MCMC estimation. Each bin is 0.1 mag $\times$ 0.01 $R_\odot$. Typical errors are shown in the top right corner. We plot our best-fit 6$^\mathrm{th}$ order polynomial, only for the color range over which the B10 relationships are valid. The scatter at the red end is most likely an artifact of extrapolating the B10 photometric parallax relationships to redder colors.
\label{fig:radii}}
\end{figure}

	Stellar radii can be inferred using distances estimates (Section~\ref{photoparallax}), and the scaling factor of the best-fit model to the measured photometry (see Section~\ref{params} and \citealt{cushing:2008:1372}). Figure~\ref{fig:radii} shows the estimated stellar radii as a function of $r-z$ color. We again fit a polynomial relationship between $R_\ast$ and $r-z$ color and find a 6$^\mathrm{th}$ order polynomial provides the best-fit (using an $F$ test). Our polynomial relationship is shown in Figure~\ref{fig:radii} and described by an equation similar to Equation~(\ref{eqn:poly}), with coefficients listed in Table~\ref{tbl:coeffs}. The scatter we find for the reddest objects is likely an artifact of extrapolating the B10 relationships past their valid data range.

\begin{figure}
\centering
 \includegraphics[width=\linewidth]{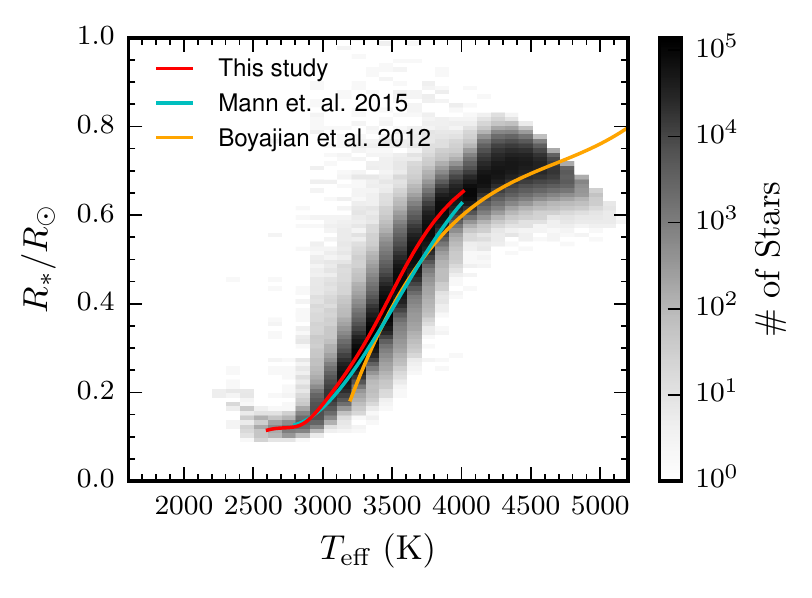}
\caption{$R_\ast$ as a function of $T_\mathrm{eff}$. The red line shows the relationship using the polynomial values from Table~\ref{tbl:coeffs}. Comparison with the \citet{mann:2015:64} relationship (cyan line) and \citet{boyajian:2012:112} relationship (yellow line) show an offset of $\sim0.05$ $R_\odot$ for hotter temperatures, but the relationship converges with \citet{mann:2015:64} at cooler temperatures. The \citet{mann:2015:64} and \citet{boyajian:2012:112} relationships were calibrated using nearby stars, and thus, the observed offset in the relationships for hotter stars could be due to SDSS sampling a less active and/or lower metallicity stellar population, or possible extinction effects. The upturn for the coolest stars is due to extrapolating the B10 polynomial to redder colors. 
\label{fig:radiiTemp}}
\end{figure}

	The relationship between effective temperature and stellar radii using our polynomial equations is shown in Figure~\ref{fig:radiiTemp}. The relationship follows similar trends to both the relationship by \citet{mann:2015:64} and \citet{boyajian:2012:112}. The upturn in radii at cooler temperatures is an artifact of the B10 photometric parallax relationship, which is not well-constrained for the reddest stars.

\begin{deluxetable*}{cccccccccccc}% 
\tabletypesize{\scriptsize} 
\tablecolumns{12} 
\tablewidth{0pt} 
\tablecaption{Polynomial Relationship Coefficients \label{tbl:coeffs}} 
\tablehead{ 
\colhead{Y} & \colhead{X} & \colhead{a} & \colhead{b} & \colhead{c} & \colhead{d} & \colhead{e} & \colhead{f} & \colhead{g} & \colhead{$\sigma$} & \colhead{$\rchi^2_\nu$} & \colhead{Range} }
\startdata
$T_\mathrm{eff}$ (K) & $r-z$\ & $6691.90$ & $-6000.26$ & $5135.52$ & $-2513.18$ & $679.434$ & $-94.2185$ & $5.18804$ & 47.41 & 1.39 & $0.5 \leqslant r-z \leqslant 4.84$\\
$R_\ast$ ($R_\odot$)& $r-z$ & $0.41895$ & $1.3345$ & $-1.9848$ & $1.1474$ & $-0.34214$ & $0.052184$ & $-0.0032136$ & 0.027 & 0.022 & $0.9 \leqslant r-z \leqslant 4.30$
\enddata 
\end{deluxetable*}

\section{Low-mass Kinematics (LoKi) Galactic Model}\label{loki}

\subsection{Stellar Density Profile}\label{density}

	We implemented a similar galactic model framework as that used in \citet{dhital:2010:2566}. In the model, the stellar density for each galactic component is given in terms of standard galactic coordinates. For the thin (cold component) and thick (warm component) disks, the stellar density profiles are given by,
\begin{equation}
\begin{aligned}
\rho_\mathrm{thin}(R,Z) = \rho(R_0, 0) &\exp\left(-\frac{|Z|}{H_\mathrm{thin}}\right) \\
							& \times \exp\left(-\frac{|R - R_0|}{L_\mathrm{thin}}\right),
\end{aligned}
\end{equation}
\begin{equation}
\begin{aligned}
\rho_\mathrm{thick}(R,Z) = \rho(R_0, 0) &\exp\left(-\frac{|Z|}{H_\mathrm{thick}}\right) \\
 							& \times \exp\left(-\frac{|R - R_0|}{L_\mathrm{thick}}\right),
\end{aligned}
\end{equation}
where $H$ is the scale heights above and below the plane, and $L$ is the scale length within the plane. The halo stellar density is expressed as a bi-axial power-law ellipsoid,
\begin{equation}
\rho_\mathrm{halo}(R,Z) = \rho(R_0, 0) \left(\frac{R_0}{\sqrt{R^2 + (Z/q)^2}}\right)^{r_\mathrm{halo}},
\end{equation}
where $q$ is the halo flattening parameter, and $r_\mathrm{halo}$ is the halo density gradient. In each of the above formulas, $R$ is the Galactic radius, $R_0$ is the Sun's distance from the Galactic center (8.5 kpc), and $Z$ is the Galactic height. To obtain the total stellar density at a specific radius and height in the Galaxy, all three density profiles weighted by the fraction of all stars in each component are summed,
\begin{equation}\label{eqn:density}
	\begin{aligned}
	\rho(R,Z) = f_\mathrm{thin} \cdot \rho_\mathrm{thin}(R, Z) &+ f_\mathrm{thick} \cdot \rho_\mathrm{thick}(R, Z)\\
	 &+ f_\mathrm{halo} \cdot \rho_\mathrm{halo}(R, Z),
	\end{aligned}
\end{equation}
with $f_\mathrm{thin} + f_\mathrm{thick} + f_\mathrm{halo} = 1$. The local stellar density scaled to the Galactic plane, $\rho$($R_0=8.5$ kpc, $Z=0$ pc), was obtained by integrating the bias-corrected, single star luminosity function (LF) from B10 for low-mass stars from SDSS. Table~\ref{tbl:params} contains the adopted disk parameters for the model.

\begin{deluxetable}{lccc} 
\tabletypesize{\footnotesize} 
\tablecolumns{4} 
\tablewidth{0pt} 
\tablecaption{Galactic Model Parameters \label{tbl:params}} 
\tablehead{ 
\colhead{Component} & \colhead{Parameter} & \colhead{Description} & \colhead{Value} }
\startdata
& $f_\mathrm{thin} $ & Fraction\tablenotemark{a} & $1-f_\mathrm{thick}-f_\mathrm{halo}$ \\ 
Thin disk & $H_\mathrm{thin} $ & Scale height & 300 pc \\ 
& $L_\mathrm{thin} $ & Scale length & 3100 pc \\ \hline
& $f_\mathrm{thick} $ & Fraction\tablenotemark{a} & 0.04 \\ 
Thick disk & $H_\mathrm{thick} $ & Scale height & 2100 pc \\ 
& $L_\mathrm{thick} $ & Scale length & 3700 pc \\ \hline
& $f_\mathrm{halo} $ & Fraction\tablenotemark{a} & 0.0025 \\ 
Halo & $r_\mathrm{halo} $ & Density gradient & 2.77 \\ 
& $q$ $(=c/a)\tablenotemark{b} $ & Flattening parameter & 0.64 
\enddata 
\tablecomments{The parameters were measured using M dwarfs for the disk (bias corrected values; B10) and MS turn-off stars for the halo \citep{ivezic:2008:287} in the SDSS footprint.}
\tablenotetext{a}{Evaluated in the solar neighborhood.}
\tablenotetext{b}{Assuming a bi-axial ellipsoid with axes $a$ and $c$.} 
\end{deluxetable}

\subsection{Stellar Densities and Distance Ranges}\label{densities}

	Perhaps the most fundamental parameter required in the model is the local stellar density. Many studies have measured the local stellar density $\rho(R_0,0)$, scaled to the Galactic plane \citep[][]{juric:2008:864, bochanski:2010:2679, van-vledder:2016:425}. Stellar number densities are commonly estimated through luminosity functions \citep[LFs; e.g.,][]{cruz:2007:439, bochanski:2010:2679}. We used the low-mass LF from B10 since the MoVeRS catalog (and hence, the sample) are built from the same photometric criteria used to create the B10 LF. However, as stated above, the B10 photometric parallax relationships extend to absolute magnitudes fainter than the B10 LF, therefore, care must be taken in obtaining stellar densities for the reddest stars. 
	
	The B10 LFs are given for both $M_r$ and $M_J$. $M_J$ is a commonly used metric for the LF function, however, the B10 photometric parallax relationships map SDSS colors to $M_r$. \citet{bochanski:2008:} gives a relationship between $M_r$ and $M_J$, which extends two magnitudes fainter in $M_r$ than the B10 $M_r$ LF. The \citet{bochanski:2008:} relationship also reaches to $M_J \approx 12$, which is also two magnitudes deeper than the B10 $M_J$ LF ($M_J \lesssim 10$). Using the $M_J$ LF from \citet{cruz:2007:439}, which begins where the B10 $M_J$ LF ends, fainter $M_r$ magnitudes were mapped to $M_J$ magnitudes (using the \citealt{bochanski:2008:} relationship), and estimated stellar densities past the limits of the B10 LFs. The stellar densities are shown in Table~\ref{tbl:modelinputs}. 
	
	The distance ranges are dictated by both the SDSS saturation limits and the maximum distance at which we would find an extreme MIR excess. For the lower distance limits, we binned the the MoVeRS sample in 0.5 magnitude bins in $r-z$ color and used the minimum distance value in each bin for the lower limit. The upper distance limit corresponds to the maximum MIR excess value above the photospheric value, since we can see an extremely large excess out to a farther distance than a smaller MIR excess. Figure~\ref{fig:rz_ratio} shows the distribution of MIR excess values above the photosphere, and we found that 95\% of the excesses had values up to 12 times the photospheric value. Using Equation~(\ref{eqn:rlimit}) scaled $\sim$2.7 magnitudes fainter (12 times greater than the expected photospheric flux), we derived new distance limits using the B10 photometric parallax relationships. The distance limits are shown in Table~\ref{tbl:modelinputs}. Since the B10 $M_r$ photometric parallax relationship did not go as red in $r-z$ as the sample, we used the \citet{baraffe:1998:403} 5 Gyr relationship between $4 < r-z \leqslant 5$. The \citet{baraffe:1998:403} model photometric parallax relationship is consistent with other photometric parallax relationships \citep{hawley:2002:3409, west:2005:706} to the reddest $r-z$ extent that it can be compared to empirical data (see B10 Figure 9).
	
\begin{deluxetable}{llcc} 
\tabletypesize{\footnotesize} 
\tablecolumns{4} 
\tablewidth{0pt} 
\tablecaption{Galactic Model Input Ranges\label{tbl:modelinputs}} 
\tablehead{ 
\colhead{$r-z$} & \colhead{$M_r$} & \colhead{$\rho(R_0, 0)$} & \colhead{Distance} \\
\colhead{} & \colhead{} & \colhead{(stars pc$^{-3}$)} & \colhead{(pc)} 
}
\startdata 
[0.5, 1.0)					& [6.52, 8.01)					& [0.00287, 0.00289] 	& [390, 1100] \\ \relax
[1.0, 1.5)					& [8.01, 9.59)					& [0.00257, 0.00259] 	& [215, 780] \\ \relax
[1.5, 2.0)					& [9.59, 11.18)					& [0.00677, 0.00680] 	& [90, 520] \\ \relax
[2.0, 2.5)					& [11.18, 12.74)				& [0.01005, 0.01010] 	& [60, 345] \\ \relax
[2.5, 3.0)					& [12.74, 14.19)				& [0.00657, 0.00660] 	& [35, 240] \\ \relax
[3.0, 3.5)					& [14.19, 15.46)				& [0.00489, 0.00493] 	& [15, 165] \\ \relax
[3.5, 4.0)					& [15.46, 16.50)				& [0.00461, 0.00464] 	& [10, 125] \\ \relax
[4.0, 4.5)\tablenotemark{a}	& [16.50, 17.50)\tablenotemark{b}	& [0.00143, 0.00146] 	& [10, 105] \\ \relax
[4.5, 5.0)\tablenotemark{a}	& [17.50, 18.50)\tablenotemark{b}	& [0.00086, 0.00089] 	& [10, 105] 
\enddata 
\tablenotetext{a}{This color range falls outside the limits of the B10 $M_r(r-z)$ relationship.}
\tablenotetext{b}{Values estimated from the 5 Gyr isochrone from \citet{baraffe:1998:403}.}
\end{deluxetable}

\subsection{Stellar Kinematics}\label{kinematics}

	Stellar kinematics are much more difficult to constrain than stellar densities, in part due to the difficulty in obtaining 3-dimensional kinematics of stars. Many studies have measured the mean velocities of stars as a function of Galactic height, and the velocity dispersions for the thin (cold component) and thick (warm component) disks, along with the halo \citep[e.g.,][]{west:2006:2507, bochanski:2007:2418, juric:2008:864, pineda:2017:}. An in-depth prescription of the kinematical model we used can be found in D10. Here we summarize the model, and explain some of the important differences in our specific model. 
	
	For a given stellar population, the \emph{average} stellar kinematics can be represented in Galactic cylindrical coordinates by the following equations:
\begin{equation}
\begin{aligned}
& \langle V_r(Z)\rangle = 0, \\
& \langle V_\theta(Z)\rangle = V_\mathrm{circ} - V_\mathrm{a} - f(Z), \\
& \langle V_z(Z)\rangle = 0,
\end{aligned}
\end{equation}
\noindent where $V_r$, $V_\theta$, and $V_z$ are the velocities in the radial, circular, and perpendicular directions, respectively. $V_\mathrm{circ}$ is the circular velocity, taken as 240 km s$^{-1}$ \citep{mcmillan:2011:2446,schonrich:2012:274}. The $V_\mathrm{a}$ term is due to interactions that stars undergo over their lifetimes, which cause circular orbits to become more eccentric and more inclined to the Galactic plane. These interactions cause the velocity component along the direction of Galactic rotation to lag the local standard of rest (LSR) for older stellar populations, a phenomenon known as asymmetric drift. $V_\mathrm{a}$ is approximately equal to 10 km s$^{-1}$ for low-mass stars in SDSS (D10). The last term for $V_\theta$ is a polynomial relationship between the average velocity and Galactic height, given by $f(Z) = a|Z| - b|Z|^2$ km s$^{-1}$, where $a = 0.013$ km s$^{-1}$ pc$^{-1}$ and $b = 1.56 \times 10^{-5}$ km s$^{-1}$ pc$^{-2}$ (taken from D10). This last term accounts for a mixture of thin and thick disk stars, with the ratio highly dependent on Galactic height. 
	
	For the velocity dispersions, we chose to explore different functional forms rather than a power law as was used in D10, which gives zero dispersion at the Galactic plane. Using results from the kinematic study of \citet{pineda:2017:}, we found that velocity dispersions grew approximately linearly with Galactic heights up to $\sim$1 kpc in all three velocity components for both thin and thick disk stars. The \citet{pineda:2017:} sample is an adequate representation of the candidate stars since they all fall within this Galactic height limit. The linear fits to the velocity dispersions take the form,
\begin{equation}\label{eqn:dispersion}
\sigma(Z)= k + n|Z|,
\end{equation} 
\noindent where the values of $k$ and $n$ are defined in Table~\ref{tbl:galparams}. For halo stars, we used velocity dispersion values from \citet{bond:2010:1}, using the dispersion relations taken at the Galactic plane ($Z=0$ pc). These velocity distributions can then be sampled to obtain expected galactic cylindrical $V_R$, $V_\theta$, and $V_Z$ velocity distributions for samples of stars at any location in the Galaxy. These $V_R$, $V_\theta$, and $V_Z$ velocities can be transformed into $UVW$ velocities, which can then be transformed into proper motions and radial velocities using the methods of \citet{johnson:1987:864}.

\begin{deluxetable}{lccc} 
\tabletypesize{\footnotesize} 
\tablecolumns{4} 
\tablewidth{0pt} 
\tablecaption{Galactic Kinematics \label{tbl:galparams}} 
\tablehead{ 
\colhead{Galactic} & \colhead{Velocity} & \colhead{$k$} & \colhead{$n$} \\
\colhead{Component} & \colhead{Component} & \colhead{(km s$^{-1}$)} & \colhead{(km s$^{-1}$ pc$^{-1}$)} }
\startdata 
						& $V_R$		& 22.43	& 0.04 \\ 
Thin disk\tablenotemark{a}	& $V_\theta$	& 13.92	& 0.03 \\ 
						& $V_Z$		& 10.85	& 0.03 \\ \hline
						& $V_R$		& 64.04	& 0.07 \\ 
Thick disk\tablenotemark{a}	& $V_\theta$	& 39.41	& 0.09 \\ 
						& $V_Z$		& 44.76	& 0.02 \\ \hline
						& $V_R$		& 135	& \\ 
Halo\tablenotemark{b}		& $V_\theta$	& 85		& \\ 
						& $V_Z$		& 85		& 
\enddata 
\tablenotetext{a}{The parameters were measured using M dwarfs from \citet{pineda:2017:} for the thin and thick disk components.}
\tablenotetext{b}{Halo components were taken from \citet{bond:2010:1}, using the values for the bins closest to the Galactic plane.}
\end{deluxetable}

\subsection{Model Comparisons: SDSS Source Counts}\label{comparison1}

\begin{figure}
\centering
 \includegraphics[width=\linewidth]{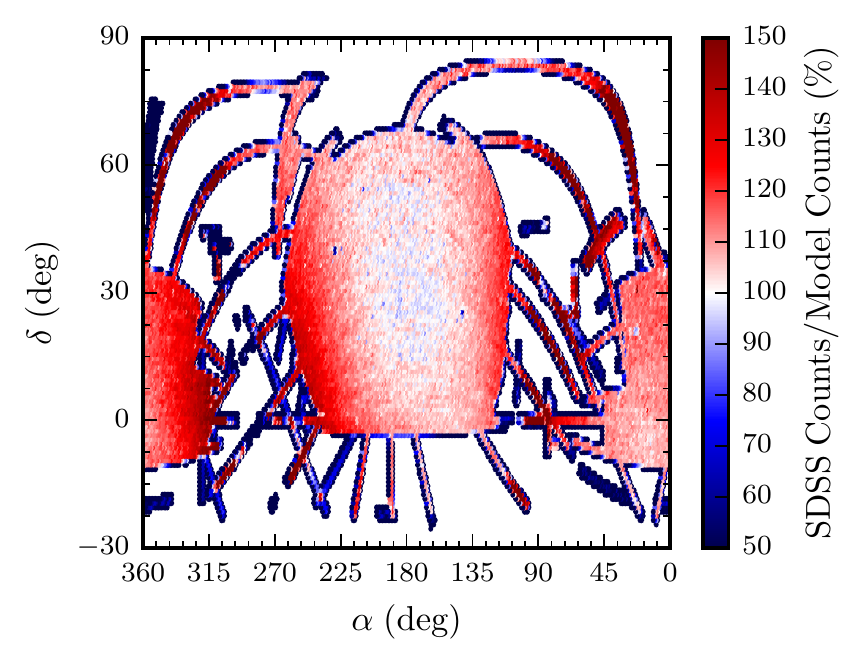}
\caption{Ratio of SDSS source counts to stellar counts from the model. Each bin is $1^\circ \times 1^\circ$. The model produces similar numbers to SDSS at high Galactic latitudes (typically less than 10\% difference). Close to the Galactic plane, SDSS source counts are much higher, likely due to reddened, higher mass stars that fall into the color selection. The model has higher source counts near edge regions of the SDSS stripes.
\label{fig:sdsscomparison}}
\end{figure}

	To assess the validity of the model, we compared stellar counts from the model against counts from SDSS for all objects with colors similar to those expected for low-mass stars. Specifically, we obtained source counts for $1^\circ \times 1^\circ$ size bins within the entire SDSS footprint, and required the following criteria (taken from \citealt{bochanski:2010:2679}):
\begin{enumerate}
\item Objects were {\sc primary} sources within the {\sc PhotoObjAll} table ({\sc mode} $= 1$),
\item Objects had point-source-like morphologies within the {\sc PhotoObjAll} table ({\sc type} $= 6$),
\item $i < 22$,
\item $z < 21.2$,
\item $r-i \geqslant 0.3$,
\item $i-z \geqslant 0.2$, and
\item $16 < r < 22$.
\end{enumerate}
	To compare SDSS source counts to the model, we integrated the B10 $M_r$ LF to get a total stellar density. Next, we integrated the model in $1^\circ \times 1^\circ$ size bins out to a distance of 2 kpc, the estimated depth of the B10 $M_r$ LF. A comparison between the stellar counts and SDSS source counts is shown in Figure~\ref{fig:sdsscomparison}. The model has better than 90\% agreement with SDSS at high Galactic latitudes. The model produces more stars in regions at the edges of the SDSS stripes, where we expect SDSS to be incomplete. Close to the Galactic plane, SDSS has a much higher number of sources. This is most likely due to bluer sources that are reddened and pulled into the color selection criteria from the higher extinction environment. Considering the input parameters for the model are based on SDSS data, it is not surprising that the model and SDSS source counts agree to such a high degree. Further comparisons must be made with independent observations to verify the model.

\subsection{Model Comparison: RECONS Sample}\label{comparison2}

	The REsearch Consortium on Nearby Stars \citep[RECONS;][]{henry:2006:2360, jao:2005:1954} has been compiling a sample of the low-mass stars within $\sim$25 pc in the southern hemisphere. The current realization of the RECONS samples was published by \citet{winters:2015:5}, and contains 1748 systems with an M dwarf primary and with parallax measurements (trigonometric or photometric). These stars also all have significant proper motions ($\mu \geqslant 180$ mas yr$^{-1}$), to remove possible giant stars. The completeness of this sample is unknown, but extrapolating results from the 5 pc sample, \citet{winters:2015:5} estimate their 25 pc sample to be between 48\%--77\% volume complete.

\begin{figure}
\centering
 \includegraphics[width=\linewidth]{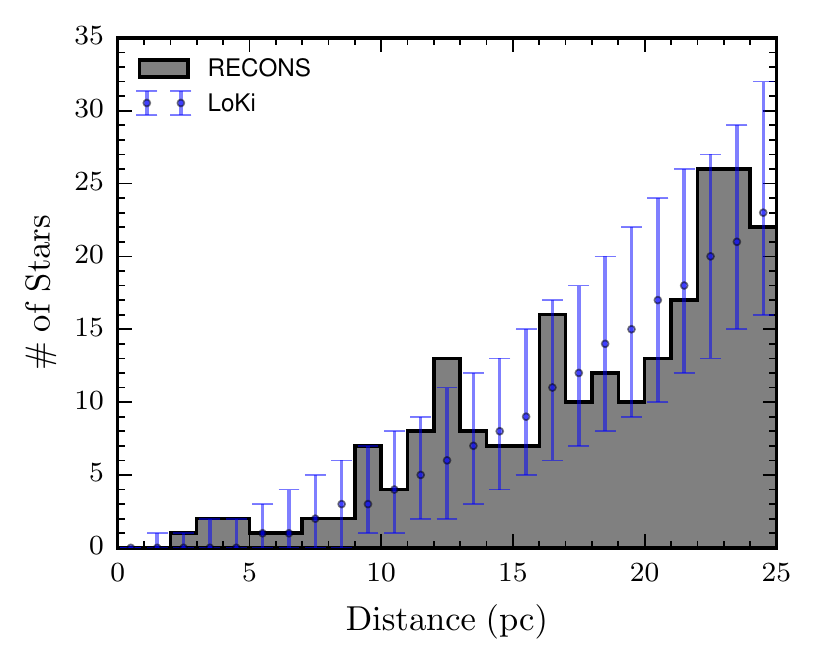}
 \includegraphics[width=\linewidth]{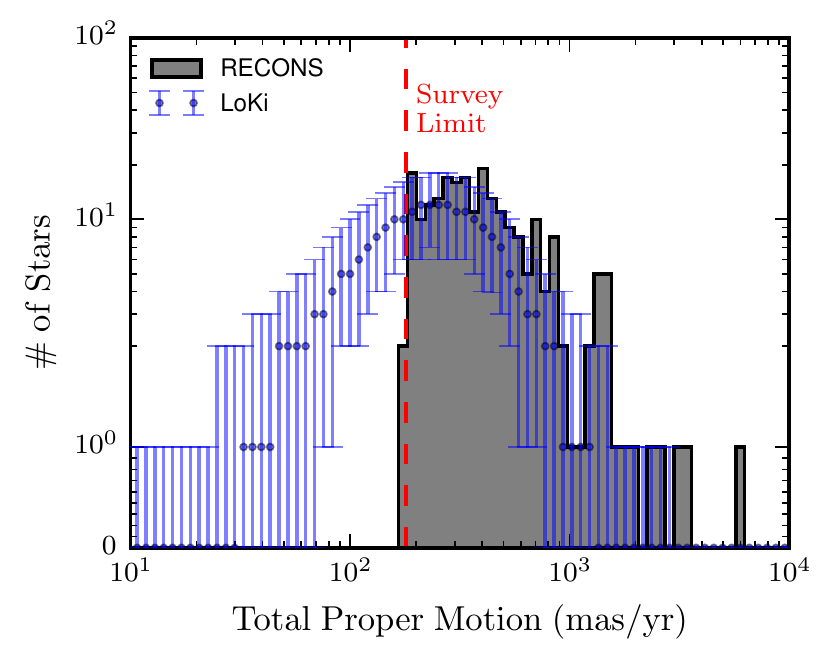}
\caption{Comparison between the model and the RECONS sample \citep{winters:2015:5} for a 3600 deg$^2$ region below the Galactic plane. Points are the 50$^\mathrm{th}$ percentile values and error bars represent the 5$^\mathrm{th}$ and 95$^\mathrm{th}$ percentile values for the 1000 realizations. The red dashed line represents the proper motion limit for the \citet{winters:2015:5} sample. The model is able to reproduce both the distance and proper motion distributions.
\label{fig:recons}}
\end{figure}

	We chose to simulate a 3600 deg$^2$ patch of sky away from the Galactic plane ($0^\circ \leqslant \alpha \leqslant 60^\circ$ and $-60^\circ \leqslant \delta \leqslant 0^\circ$). Since the RECONS sample has parallax measurements with a variety of precisions, we applied a 20\% normal uncertainty to the simulated stars and kept stars within 25 pc. We ran 1000 realizations of the model over the volume listed above using the full density computed from integrating the B10 single-star $r$-band LF. Our results compared to the RECONS sample are shown in Figure~\ref{fig:recons}. Both the model distributions of distances and proper motions follow the observed distributions up to the survey limits. If we use the model to estimate the incompleteness within the volume probed, we estimate the RECONS sample to be 74\% complete using the 95$^\mathrm{th}$ percentile values. The proper motion distribution indicates that the majority of missing stars have small proper motions.
	
\subsection{Model Comparison: SUPERBLINK Sample}\label{comparison3}

	The SUPERBLINK survey \citep[][]{lepine:2002:1190, lepine:2003:921, lepine:2005:1483} is a proper motion and magnitude limited survey. For the comparison, we used the bright M dwarf sub-catalog \citep{lepine:2011:138}. This catalog has a magnitude limit of $J < 10$ and a proper motion limit of $\mu > 40$ mas yr$^{-1}$. The completeness for stars in the northern hemisphere is estimated to be $\approx 90\%$.
	
	To properly simulate this sample, we were required to simulate the magnitude limits in the form of distance limits, and distance uncertainties. The $J < 10$ limit was implemented using the $J$-band LF from B10, and calculating the distance for each $M_J$ bin using a limiting magnitude of $J = 10$. We integrated out to a distance of 200 pc although 80\% of the stars in the \citet{lepine:2011:138} sample have distances $\leqslant 75$ pc. This larger simulated maximum distance was chosen due to the fact that distances were convolved with uncertainties prior to implementing a distance cut of 65 pc (comparing only to the \citealt{lepine:2011:138} stars with $d \leqslant 65$ pc). 
	
	The quoted distance uncertainty in the photometric parallax relationship used in \citet{lepine:2011:138} is between 20\%--50\%. To determine the best uncertainty to fold into the distances, we ran small batches of simulations using different normally distributed uncertainties (between 20\%--50\%), and comparing their distance distributions to SUPERBLINK. We found that 30\% uncertainty gave the expected trends in the distance distributions. 
	
\begin{figure}
\centering
 \includegraphics[width=\linewidth]{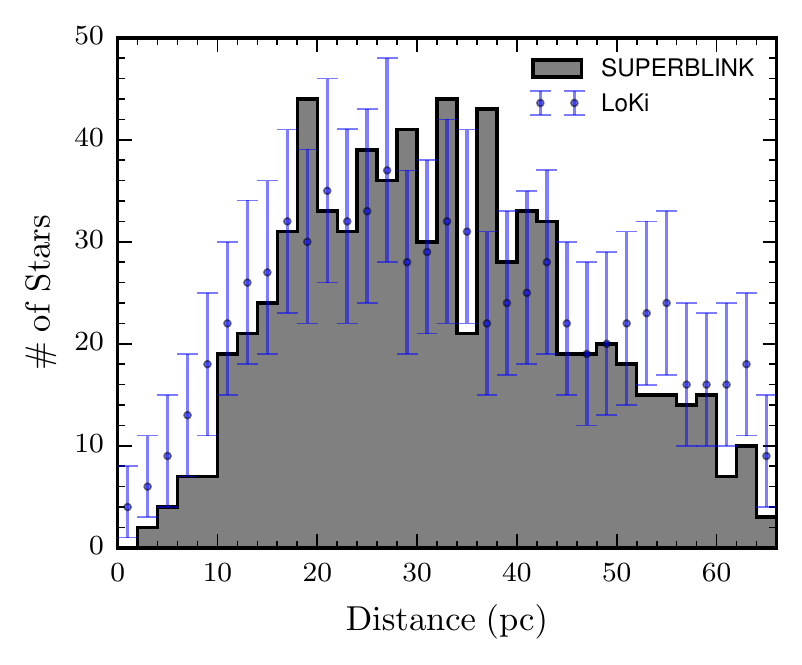}
 \includegraphics[width=\linewidth]{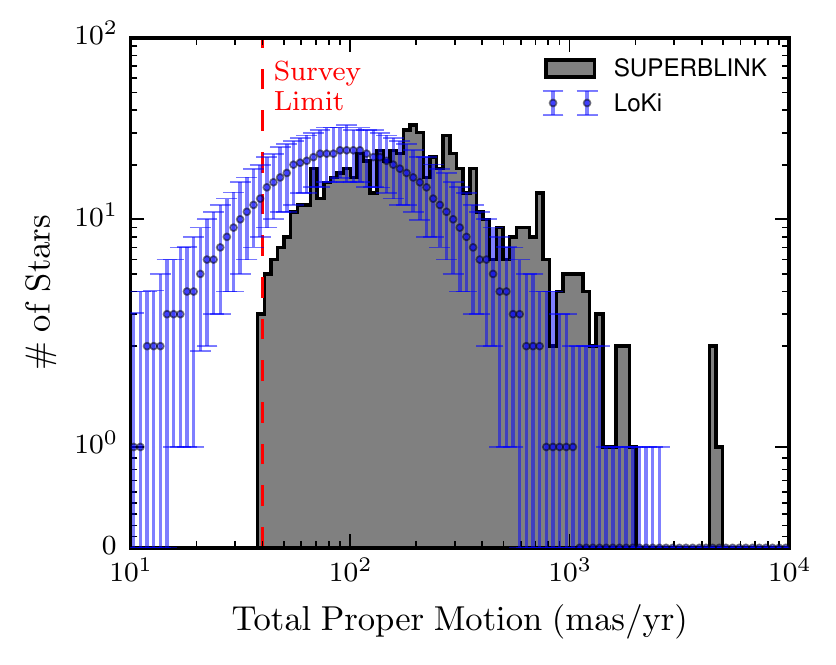}
\caption{Same as Figure~\ref{fig:recons}, but comparing the model against the \citet{lepine:2011:138} sample (M dwarfs with $J < 10$). The model produces similar distributions as the SUPERBLINK sample, but indicates a missing population of stars with small proper motions (similar to the comparison with RECONS).
\label{fig:superblink}}
\end{figure}

	Again, we simulated a 3600 deg$^2$ patch of sky away from the Galactic plane ($160^\circ \leqslant \alpha \leqslant 220^\circ$ and $0^\circ \leqslant \delta \leqslant 60^\circ$) and ran 1000 realizations. Figure~\ref{fig:superblink} shows the SUPERBLINK distributions and the model results, along with the 5$^\mathrm{th}$ and 95$^\mathrm{th}$ percentile confidence intervals. We can again estimate a level of completeness using the 95$^\mathrm{th}$ percentile values, however, caution should be taken as the uncertainties folded into the simulations may be different than the actual uncertainties within the SUPERBLINK survey. The estimated completeness level for the simulated volume is 65\%, with the majority of missing stars at smaller proper motions below the survey limit. 
	
	As is shown in Figure~\ref{fig:superblink}, the completeness of SUPERBLINK should be extremely high for the largest proper motion stars. However, towards the proper motion limit of SUPERBLINK, the completeness drops off. This is to be expected as smaller proper motions are more difficult to measure to high precision. Some of this incompleteness may be accounted for if measurement uncertainty tends to scatter stars towards higher proper motions. However, there still appears to be a large population of nearby stars with small proper motions that has gone relatively undetected due to the requirement of larger proper motions (similar to the comparison with the RECONS sample). The complete SUPERBLINK sample (without the $J < 10$ criterion) will likely resolve much of this incompleteness when some of the fainter stars with smaller proper motions are added to the sample.
	
	The \emph{Gaia} \citep{perryman:2001:339, gaia-collaboration:2016:a1} collaboration recently made Data Release 1 \citep{gaia-collaboration:2016:a2}, which has a proper motion precision of $\sim$1 mas yr$^{-1}$ for non-Hipparcos Tycho-2 stars \citep{lindegren:2016:a4}. However, the final data release for \emph{Gaia} is expected to have a precision better than 0.1 mas yr$^{-1}$. \emph{Gaia} should detect all of the nearby ($\leqslant 60$ pc), earliest-type M dwarfs, and lower-mass objects at closer distances. However, \emph{Gaia} will not be able to detect the lowest-mass M dwarfs out to the distances SDSS, 2MASS, and \emph{WISE} were able to observe them, due to its relatively blue filter \citep{ivezic:2012:251}. The \emph{Gaia} completeness for low-mass dwarfs has been investigated using the LaTE-MoVeRS sample \citep{ theissen:2017:92} and \emph{Gaia} Data Release 1. \citet{ theissen:2017:92} found that \emph{Gaia} was $\sim$70\% complete for low-mass dwarfs with $i < 20$, and less than 30\% complete for dwarfs with $i \geqslant 20$. Although \emph{Gaia} will not be able to probe the entire volume that the MoVeRS sample covers, it will allow us to validate the model across the entire proper motion range and with much smaller simulated distance uncertainties for nearby ($\lesssim 30$ pc) stars. \emph{Gaia} will be especially critical in uncovering the potential population of nearby stars with small proper motions that have been primarily ignored, and resolving the true completeness of the SUPERBLINK sample.

\subsection{Simulating a Galactic Volume within the SDSS Footprint}\label{galacticvolume}

	To properly estimate the level of completeness, we need to simulate the complete volume ($\alpha$, $\delta$, and $d$) from where the sample was extracted. However, due to the time-delay-integrate nature of SDSS, getting the exact outline of the imaging footprint in $\alpha$ and $\delta$ coordinates is extremely complicated. To further complicate matters, some fields observed by SDSS fail processing by the photometric pipeline. This is primarily due to large or bright objects within the frame causing the photometric pipeline to time-out \citep{blanton:2011:31}. To quantify the number of bad fields within the SDSS footprint, we retrieved all the field IDs and number of extracted objects within the field from the \textsc{Field} table using CasJobs\footnote{\url{http://skyserver.sdss.org/casjobs/}}. Of the 938,046 fields in SDSS, 6,239 fields contain zero objects ($\sim$0.67\%). The vast majority of bad fields (4,271) are found in stripes within the Galactic plane ($|b| < 20^\circ$), which we excluded from the sample. Therefore, bad fields were not a concern for the simulated SDSS volume.
	
\begin{figure*}
\centering
 \includegraphics[width=\linewidth]{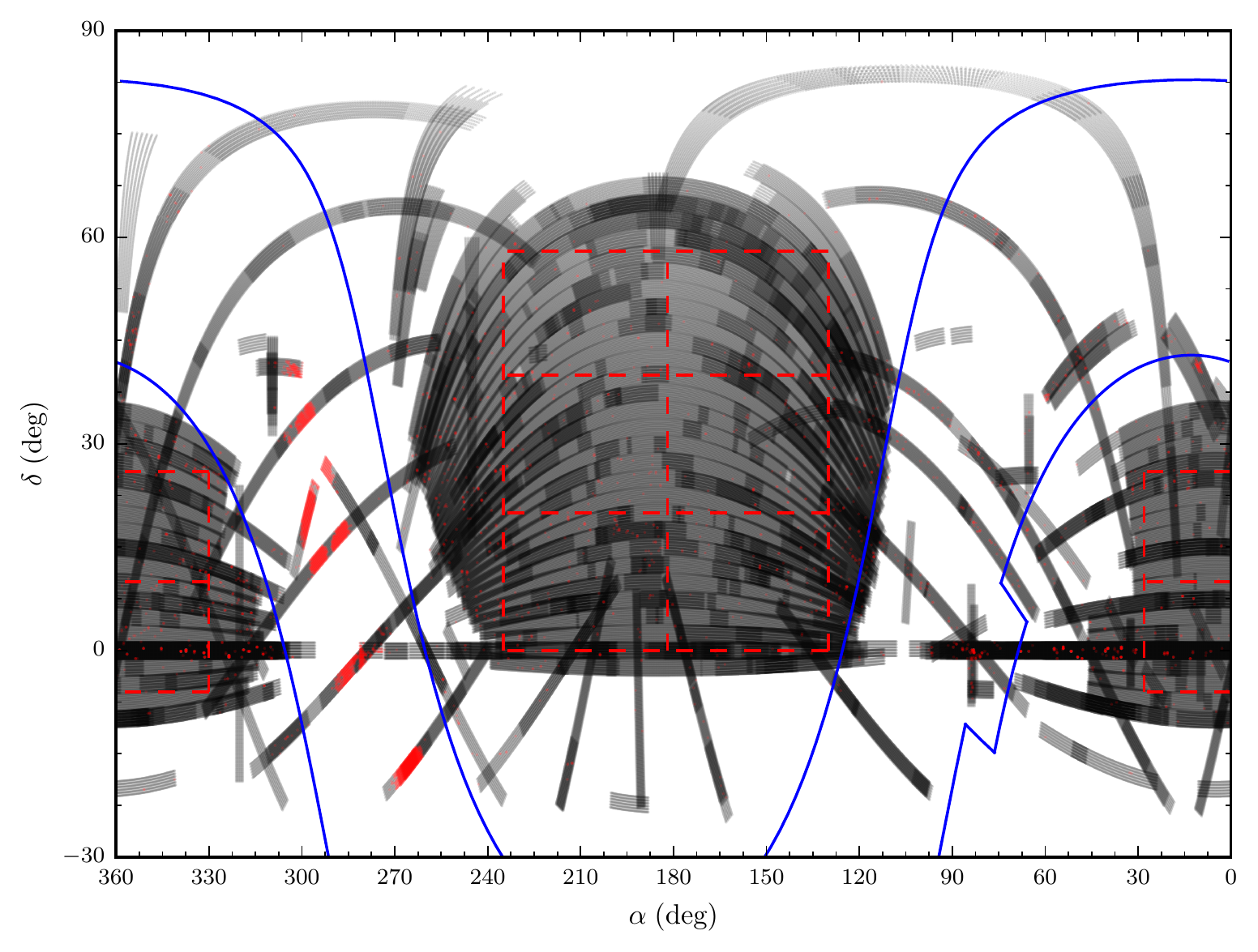}
\caption{The entire SDSS imaging footprint. Individual black points represent the field centers for the good fields (fields with extracted objects), with each point slightly transparent to highlight the areas of high frame density (overlap). Individual red dots represent the field centers for the bad fields (fields with no extracted objects), again, with each point slightly transparent to highlight the areas of high frame density. The region we excluded close to the Galactic plane is outlined in blue. The areas we simulated are outlined in red dashed lines.
\label{fig:footprint}}
\end{figure*}

	Rather than try to simulate the entire SDSS footprint, we chose to simulate large areas within the footprint. Figure~\ref{fig:footprint} shows the fields imaged by SDSS and the selected areas within that footprint. The stripe nature of SDSS is clearly shown, with darker regions indicating heavier coverage. The regions we chose are listed in Table~\ref{tbl:regions}, with larger regions divided into smaller subregions for computational ease and parallelization.
	
\begin{deluxetable}{ccc} 
\tabletypesize{\footnotesize} 
\tablecolumns{3} 
\tablewidth{0pt} 
\tablecaption{Model Simulated Regions\label{tbl:regions}} 
\tablehead{ 
\colhead{Region ID} & \colhead{$\alpha$ Range} & \colhead{$\delta$ Range} \\
 & \colhead{(deg.)} & \colhead{(deg.)} 
}
\startdata 
1	& [0, 28]		& [$-$6, 10] 	 \\ \relax
2	& [0, 28]		& [10, 26] 	 \\ \relax
3	& [130, 182]	& [0, 20] 	 \\ \relax
4	& [130, 182]	& [20, 40] 	 \\ \relax
5	& [130, 182]	& [40, 58] 	 \\ \relax
6	& [182, 235]	& [0, 20] 	 \\ \relax
7	& [182, 235]	& [20, 40] 	 \\ \relax
8	& [182, 235]	& [40, 58] 	 \\ \relax
9	& [330, 360]	& [$-$6, 10] 	 \\ \relax
10	& [330, 360]	& [10, 26] 	 
\enddata 
\end{deluxetable}
	
\subsection{Sampling with the Model to Estimate Completeness}\label{sampling}

	The level of completeness was estimated by simulating stars in regions defined in the previous section. This was done for all stars within the volume, and separately in absolute magnitudes bins defined in Table~\ref{tbl:modelinputs}. The following steps were completed for all simulated regions:

\begin{enumerate}

\item \label{step1} For parallelization, different $r-z$ color ranges (a proxy for stellar mass ranges) were simulated individually. For each $r-z$ color range in Table~\ref{tbl:modelinputs}, we used the B10 color-magnitude relations to obtain the range of absolute magnitudes ($M_r$). 

\item Since the color ranges were continuous, but the B10 $M_r$ LF is given in discrete bins, we chose to interpolate the $M_r$ LF. Using the single-star LF from B10, we interpolated the $M_r$ LF over the $r-z$ color range from the previous step. The B10 LFs are given as median values with asymmetric uncertainties. All three values (median and asymmetric uncertainties) were used to provide a range of possible stellar number densities for the model. Three interpolations were done, one for the median $M_r$ value, one for the upper $M_r$ limit, and one for the lower $M_r$ limit. This step is illustrated in Figure~\ref{fig:LFexample}.

\item A random LF value was drawn for a given $M_r$ value. To do this, the absolute magnitude range (from above) was divided into 10,000 evenly spaced bins. For each bin, a random LF value was drawn from a triangular probability distribution defined by the median value at the apex, and the lower limit and upper limit values as the first and third vertex, respectively. The median, upper limit, and lower limit values were taken from the interpolated LF at the center of each absolute magnitude bin. An example of this step is shown in Figure~\ref{fig:LFexample}.

\begin{figure}
\centering
\includegraphics[width=\linewidth]{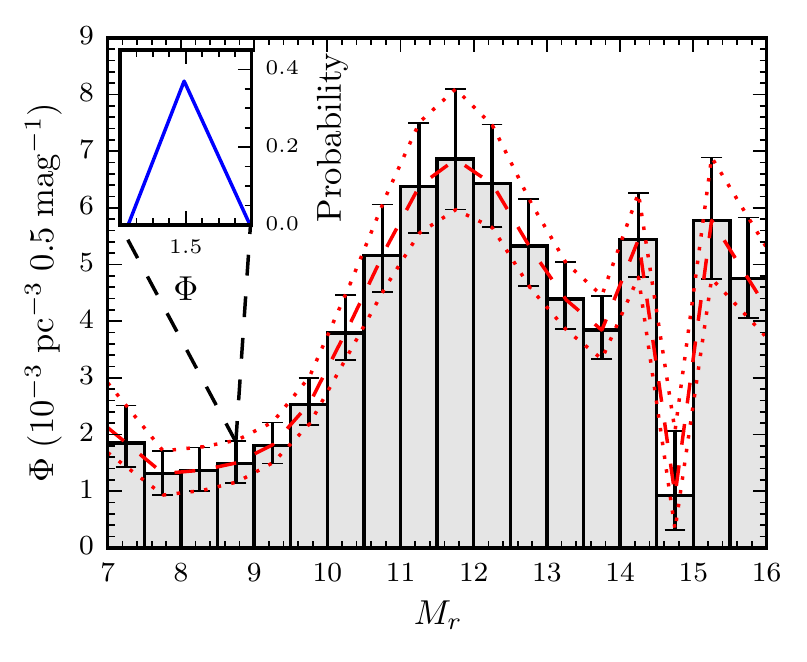}
\caption{The B10 single-star $M_r$ LF. The interpolated functions (Section~\ref{sampling}) between the median values and the upper and lower uncertainties are shown as the red dashed and dotted lines, respectively. The inset plot shows an example of the triangular probability distribution taken at the discrete point $M_r = 8.75$, used to pull a random value from the LF. 
\label{fig:LFexample}}
\end{figure}

\item The LF values from the previous step were then integrated over the absolute magnitude range (from step~\ref{step1}) to produce the local stellar density scaled to the plane, $\rho(R_0,0)$.

\item Using the stellar density from the previous step, we integrated the density profile, Equation~(\ref{eqn:density}), along the LOS in 1 pc deep, discrete pyramidal ``cells." Each cell along the LOS was parameterized by the $\alpha$ and $\delta$ range, and the distance range (defined in Table~\ref{tbl:modelinputs}). Multiplying the volume of the cell by the average stellar density within the cell gave us the total number of stars within each cell. Summing all the cells gave us the total number of stars along the LOS.

\item The next step was distributing stars randomly within the given volume. For the relatively small angular ranges, we assumed that the $\alpha$ and $\delta$ positions for the stars were uniformly random within the range. Distances are more complicated as the distribution of distances is dependent on LOS through the Galaxy. To build a representative distribution of distances along the given LOS, we used the number of stars in each cell, and the distance to the center of each pyramidal cell from the previous step. This distribution was transformed into an inverse cumulative distribution function, which was sampled from in the following step \citep[known as inverse transform sampling;][]{press:1992:}.

\item Stars were then distributed in a three-dimensional space within the defined volume using the rejection method \citep{press:1992:}. This generated uniformly random $\alpha$ and $\delta$ coordinates, and distances randomly chosen through inverse sampling of the distribution created in the previous step.

\item The three dimensional $\alpha$, $\delta$, and distances were converted to Galactic cylindrical coordinates ($R, T, Z$).

\item Each star was then given $V_R$, $V_\theta$, and $V_Z$ velocities dependent on the average $V_R$, $V_\theta$, and $V_Z$ and corresponding dispersion found at each star's Galactic height, based on Equation~(\ref{eqn:dispersion}). These velocities were subsequently converted into $UVW$ velocities.

\item $UVW$ velocities were converted into proper motion components and radial velocities following the inverse of the methods described in \citet{johnson:1987:864}. We disregard the radial velocities as they are not required for the completeness estimates.

\item Lastly, a variable proper motion cut was made based on the minimum proper motion within the MoVeRS sample for the volume and color range simulated. This ensured the simulations only included stars which had distances and tangential motions that would have been detected for the MoVeRS sample.

\item The previous steps were repeated 100 times to build distributions of counts to estimate the random uncertainty in the model.

\end{enumerate}

The \emph{LoKi} Galactic model is available to the community through GitHub\footnote{\url{https://github.com/ctheissen/LoKi}}.

\bibliography{ms}
%\bibliography{../../MyLibrary}
\bibliographystyle{hapj}

\end{document}